\documentclass[11pt,english]{article}
\usepackage[OT1]{fontenc}
\usepackage[latin9]{inputenc}
\usepackage{geometry}
\usepackage{amsmath}
\usepackage{amssymb}

\makeatletter
\usepackage{amsmath}
\usepackage{amsthm}
\usepackage{amssymb}

\theoremstyle{plain}
\newtheorem{theorem}{Theorem}[section]

\theoremstyle{plain}
\newtheorem{theorem-seq}{Theorem}

\newenvironment{myproof}[1][Proof.]{\par
	\pushQED{\qed}%
	\normalfont \topsep6\p@\@plus6\p@\relax
	\trivlist
		\item[\hskip\labelsep
		\bfseries
	#1]\ignorespaces
	}{%
	\popQED\endtrivlist\@endpefalse
}


\newcommand{\eqdef}{\stackrel{\rm{def}}{=}}

\newcommand{\E}{\mathbb{E}}

\newcommand{\N}{\mathbb{N}}
\newcommand{\Z}{\mathbb{Z}}
\newcommand{\R}{\mathbb{R}}

\ifx \C \undefined
\newcommand{\C}{\mathbb{C}}
\else
\renewcommand{\C}{\mathbb{C}}
\fi

\newcommand{\B}{\left\{ 0,1 \right \}}

\usepackage{crossreftools}
\usepackage{cleveref}
\usepackage{aliascnt}
\crefalias{theorem-seq}{theorem}

\let \ref \Cref
\crefname{enumi}{}{}

\newaliascnt{definition-cnt}{theorem}
\theoremstyle{definition}
\newtheorem{definition}[definition-cnt]{Definition}
\crefname{definition-cnt}{definition}{definitions}
\newaliascnt{conjecture-cnt}{theorem}
    \theoremstyle{plain}    
    \newtheorem{conjecture}[conjecture-cnt]{Conjecture} 
\crefname{conjecture-cnt}{conjecture}{conjectures}
\newaliascnt{remark-cnt}{theorem}
    \theoremstyle{definition}
    \newtheorem{remark}[remark-cnt]{Remark}
\crefname{remark-cnt}{remark}{remarks}
\newaliascnt{proposition-cnt}{theorem}
    \theoremstyle{plain}    
    \newtheorem{proposition}[proposition-cnt]{Proposition} 
\crefname{proposition-cnt}{proposition}{propositions}
\newaliascnt{fact-cnt}{theorem}
    \theoremstyle{plain}    
    \newtheorem{fact}[fact-cnt]{Fact}
\crefname{fact-cnt}{fact}{facts}
\newaliascnt{lemma-cnt}{theorem}
\theoremstyle{plain}    
\newtheorem{lemma}[lemma-cnt]{Lemma}  
	\crefname{lemma-cnt}{lemma}{lemmas}
\newaliascnt{notation-cnt}{theorem}
    \theoremstyle{definition}    
    \newtheorem{notation}[notation-cnt]{Notation}
\crefname{notation-cnt}{notation}{notations}
\theoremstyle{plain}
\newtheorem*{restated@theorem}{\rep@title}

\newenvironment{restated}[1]{%
	 \def\rep@title{#1}%
	 \begin{restated@theorem}}%
	{\end{restated@theorem}}

\newaliascnt{claim-cnt}{theorem}
    \theoremstyle{plain}    
    \newtheorem{claim}[claim-cnt]{Claim}
\crefname{claim-cnt}{claim}{claims}

\usepackage{babel}
\usepackage[nomessages]{fp}

\makeatother

\usepackage{babel}
\begin{document}
\title{Lifting with Inner Functions of Polynomial Discrepancy}
\author{Yahel Manor\thanks{Department of Computer Science, University of Haifa, Haifa 3498838,
Israel. Supported by the Israel Science Foundation (grant No. 716/20).}\and Or Meir\thanks{Department of Computer Science, University of Haifa, Haifa 3498838,
Israel. \texttt{ormeir@cs.haifa.ac.il}. Partially supported by the
Israel Science Foundation (grant No. 716/20).}}
\maketitle
\begin{abstract}
\global\long\def\B{\left\{  0,1\right\}  }%
\global\long\def\dcc{D^{\mathrm{cc}}}%
\global\long\def\rcc{R^{\mathrm{cc}}}%
\global\long\def\ddt{D^{\mathrm{dt}}}%
\global\long\def\rdt{R^{\mathrm{dt}}}%
\global\long\def\Err{\mathrm{Err}}%
\global\long\def\disc{{\rm disc}}%
\global\long\def\defeq{\eqdef}%
\global\long\def\cE{\mathcal{E}}%
\global\long\def\cV{\mathcal{V}}%
\global\long\def\D{\Lambda}%
\global\long\def\cX{\mathcal{X}}%
\global\long\def\ent{\mathrm{ent}}%
\global\long\def\cY{\mathcal{Y}}%
\global\long\def\cH{\mathcal{H}}%
\global\long\def\fix{\mathrm{fix}}%
\global\long\def\free{\mathrm{free}}%
\global\long\def\D{\Lambda}%
\global\long\def\cI{\mathcal{I}}%
\global\long\def\cO{\mathcal{O}}%
\global\long\def\cS{\mathcal{S}}%
\global\long\def\b{\mathrm{bias}}%
\global\long\def\Hm{H_{\infty}}%
\global\long\def\Dm{D_{\infty}}%
\global\long\def\KMsg{K_{\text{msg}}}%
\global\long\def\KPrt{K_{\text{prt}}}%
Lifting theorems are theorems that bound the communication complexity
of a composed function~$f\circ g^{n}$ in terms of the query complexity
of~$f$ and the communication complexity of~$g$. Such theorems
constitute a powerful generalization of direct-sum theorems for~$g$,
and have seen numerous applications in recent years.

We prove a new lifting theorem that works for every two functions~$f,g$
such that the discrepancy of~$g$ is at most inverse polynomial in
the input length of~$f$. Our result is a significant generalization
of the known direct-sum theorem for discrepancy, and extends the range
of inner functions~$g$ for which lifting theorems hold.
\end{abstract}

\section{\label{sec:Introduction}Introduction}

The direct-sum question is a fundamental question in complexity theory,
which asks whether computing a function~$g$ on $n$~independent
inputs is $n$~times harder than computing it on a single input.
A related type of result, which is sometimes referred to as an ``XOR
lemma'', says that computing the XOR of the outputs of~$g$ on $n$~independent
inputs is about $n$~times harder than computing~$g$ on a single
coordinate. Both questions received much attention in the communication
complexity literature, see, e.g., \cite{KRW91,FKNN95,KKN92,CSWY01,S01sdpt,JRS03,BPSW05,JRS05,LS09,BBCR10,J11,S11,BR11,B12}.

A lifting theorem is a powerful generalization of both direct-sum
theorems and XOR lemmas. Let $f\colon\B^{n}\to\cO$ and $g\colon\Lambda\times\Lambda\to\B$
be functions (where $\Lambda$ and $\cO$ are some arbitrary sets).
The block-composed function $f\circ g^{n}$ is the function that corresponds
to the following task: Alice gets $x_{1},\ldots,x_{n}\in\Lambda$,
Bob gets $y_{1},\ldots,y_{n}\in\Lambda$, and they wish to compute
the output of $f$ on the $n$-bit string whose $i$-th bit is $g(x_{i},y_{i})$.
Lifting theorems say that the ``natural way'' for computing $f\circ g^{n}$
is more-or-less the best way. In particular, direct-sum theorems and
XOR lemmas can be viewed as lifting theorems for the special cases
where $f$ is the identity function and the parity function respectively.

A bit more formally, observe that there is an obvious protocol for
computing~$f\circ g^{n}$: Alice and Bob jointly simulate a decision
tree of optimal height for solving $f$. Any time the tree queries
the $i$-th bit, they compute $g$ on $(x_{i},y_{i})$ by invoking
the best possible communication protocol for $g$. A (query-to-communication)\emph{
lifting theorem} is a theorem that says that this protocol is roughly
optimal. Specifically, let $\ddt(f)$ and $\dcc(g)$ denote the deterministic
query complexity of~$f$ and communication complexity of~$g$ respectively,
and let $\rdt(f)$ and $\rcc(g)$ denote the corresponding randomized
complexities. Then, a lifting theorem says that
\begin{align}
\dcc(f\circ g^{n}) & =\Omega\left(\ddt(f)\cdot\dcc(g)\right) & \text{(in the deterministic setting)}\label{eq:general-lifting-theorem}\\
\rcc(f\circ g^{n}) & =\Omega\left(\rdt(f)\cdot\rcc(g)\right) & \text{(in the randomized setting)}.\nonumber 
\end{align}
In other words, a lifting theorem says that the communication complexity
of $f\circ g^{n}$ is close to the upper bound that is obtained by
the natural protocol.

In recent years, lifting theorems found numerous applications, such
as proving lower bounds on monotone circuit complexity and proof complexity
(e.g. \cite{RM97,GP14,RPRC16,PR17,GGKS18,PR18,RMNPR20,RMNPRV20}),
the separation of partition number and deterministic communication
complexity~\cite{GPW15}, proving lower bounds on data structures
\cite{CKLM18}, and an application to the famous log-rank conjecture
\cite{HHL16}, to name a few.

For most applications, it is sufficient to prove a lifting theorem
that holds for every outer function~$f$, but only for one particular
choice of the inner function~$g$. Moreover, it is desirable that
the inner function~$g$ would be a simple as possible, and that its
input length~$b=\log\left|\Lambda\right|$ would be a small as possible
in terms of in the input length~$n$ of the outer function~$f$.
For these reasons, the function~$g$ is often referred to as the
``gadget''.

On the other hand, if we view lifting theorems as a generalization
of direct-sum theorems, then it is an important research goal to prove
lifting theorems for as many inner functions~$g$ as possible, including
``complicated'' ones. This goal is not only interesting in its own
right, but might also lead to additional applications. Indeed, this
goal is a natural extension of the long line of research that attempts
to prove direct-sum theorems for as many functions as possible. This
is the perspective we take in this work, following Chattopadhyay et.
al. \cite{CKLM17,CFKMP19}. In particular, we intentionally avoid
the term ``gadget'', since we now view the function~$g$ as the
main object of study.

\paragraph*{Previous work. }

The first lifting theorem, due to Raz and McKenzie~\cite{RM99},
holds only when the inner function~$g$ is the index function. For
a long time, this was the only inner function for which lifting theorems
were known to hold for every outer function~$f$. Then, the works
of Chattopadhyay et. al. \cite{CKLM17} and Wu et. al. \cite{WYY17}
proved a lifting theorem for the case where $g$~is the inner product
function. The work of \cite{CKLM17} went further than that, and showed
that their lifting theorem holds for any inner function~$g$ that
satisfies a certain hitting property. This includes, for example,
the gap-Hamming-distance problem.

All the above results are stated only for the deterministic setting.
In the randomized setting, G\"{o}\"{o}s, Pitassi, and Watson \cite{GPW17}
proved a lifting theorem with the inner function~$g$ being the index
function. In addition, G\"{o}\"{o}s et. al.~\cite{GLMWZ15} proved
a lifting theorem in the non-deterministic setting (as well as several
related settings) with $g$~being the inner product function. 

More recently, Chattopadhyay et. al. \cite{CFKMP19} proved a lifting
theorem that holds for every inner function~$g$ that has logarithmic
input length and exponentially small discrepancy. This theorem holds
in both the deterministic and randomized setting, and includes the
cases where $g$ is the inner product function or a random function.
Since our work builds on the lifting theorem of \cite{CFKMP19}, we
discuss this result in more detail. The \emph{discrepancy} of~$g$,
denoted $\disc(g)$, is a natural and widely-studied property of functions,
and is equal to the maximum bias of~$g$ in any combinatorial rectangle.
Formally, it is defined as follows:
\begin{definition}
\label{def:discrepancy}Let $g:\Lambda\times\Lambda\to\B$ be a function,
and let $U,V$ be independent random variables that are uniformly
distributed over~$\Lambda$. Given a combinatorial rectangle $R\subseteq\Lambda\times\Lambda$,
the \emph{discrepancy of~$g$ with respect to~$R$}, denoted $\disc_{R}(g)$,
is defined as follows:
\[
\disc_{R}(g)=\left|\Pr\left[g(U,V)=0\text{ and }(U,V)\in R\right]-\Pr\left[g(U,V)=1\text{ and }(U,V)\in R\right]\right|.
\]
The \emph{discrepancy of~$g$}, denoted $\disc(g)$, is defined as
the maximum of $\disc_{R}(g)$ over all combinatorial rectangles~$R\subseteq\Lambda\times\Lambda$.
\end{definition}

Informally, the main theorem of \cite{CFKMP19} says that if $b=\log\left|\Lambda\right|$,
$\disc(g)=2^{-\Omega\left(b\right)}$ and $\text{\ensuremath{b\ge c\cdot\log n}}$
for some constant~$c$, then
\[
\dcc(f\circ g^{n})=\Omega\left(\ddt(f)\cdot b\right)\qquad\text{and}\qquad\rcc_{1/3}(f\circ g^{n})=\Omega\left(\rdt_{1/3}(f)\cdot b\right).
\]
We note that when $\disc(g)=2^{-\Omega\left(b\right)}$, it holds
that $\dcc(g)\ge\rcc(g)\ge\Omega(\log\left|\Lambda\right|)$, and
therefore the latter result is equivalent to \ref{eq:general-lifting-theorem}.

\paragraph*{The research agenda of \cite{CFKMP19}.}

As discussed above, we would like to prove a lifting theorem that
holds for as many inner functions~$g$ as possible. Inspired by the
literature on direct-sum theorems, \cite{CFKMP19} conjectured that
lifting theorems should hold for every inner function~$g$ that has
a sufficiently large information cost~$\mathrm{IC}(g)$.
\begin{conjecture}[{special case of \cite[Conj. 1.4]{CFKMP19}}]
There exists a constant $c>0$ such that the following holds. Let
$f:\B^{n}\to\cO$ and $g:\Lambda\times\Lambda\to\B$ be an arbitrary
function such that $\mathrm{IC}(g)\ge c\cdot\log n$. Then 
\[
\rcc(f\circ g^{n})=\Omega\left(\rdt(f)\cdot\mathrm{IC}(g)\right).
\]
\end{conjecture}

Proving this conjecture is a fairly ambitious goal. As an intermediate
goal, \cite{CFKMP19} suggested to prove this conjecture for complexity
measures that are simpler than~$\mathrm{IC}(g)$. In light of their
result, it is natural to start with discrepancy. It has long been
known that the quantity $\Delta(g)\defeq\log\frac{1}{\disc(g)}$ is
a lower bound on~$\rcc(g)$ up to a constant factor. More recently,
it has even been shown that~$\Delta(g)$ is a lower bound on $\mathrm{IC}(g)$
up to a constant factor~\cite{BW12}. Motivated by this consideration,
\cite{CFKMP19} suggested the following natural conjecture: for every
function~$g$ such that $\Delta(g)\ge c\cdot\log n$, it holds that
$\rcc(f\circ g^{n})=\Omega\left(\rdt(f)\cdot\Delta(g)\right)$ (see
Conjecture 1.5 there). The lifting theorem of \cite{CFKMP19} proves
this conjecture for the special case where $\Delta(g)=\Omega(b)$.

\paragraph*{Our result.}

In this work, we prove the latter conjecture of \cite{CFKMP19} in
full, by waiving the limitation of $\Delta(g)=\Omega(b)$ from their
result, where $b=\log\left|\Lambda\right|$. As in previous works,
our result holds even if $f$ is replaced with a general search problem~$\cS$.
In what follows, we denote by $R_{\beta}^{dt}(\cS)$ and $R_{\beta}^{cc}(\cS\circ g^{n})$
the randomized query complexity of~$\cS$ with error~$\beta$ and
the randomized communication complexity of~$\cS\circ g^{n}$ with
error~$\beta$ respectively. We now state our result formally.
\begin{theorem}[Main theorem]
\label{main-theorem}There exists a universal constant~$c$ such
that the following holds: Let $\cS$ be a search problem that takes
inputs from~$\B^{n}$, and let $g:\Lambda\times\Lambda\to\B$ be
an arbitrary function such that $\Delta(g)\ge c\cdot\log n$. Then
\[
\dcc(\cS\circ g^{n})=\Omega\left(\ddt(\cS)\cdot\Delta(g)\right),
\]
and for every $\beta>0$ it holds that
\[
\rcc_{\beta}(\cS\circ g^{n})=\Omega\left(\left(\rdt_{\beta^{\prime}}(\cS)-O(1)\right)\cdot\Delta(g)\right),
\]
where $\beta^{\prime}=\beta+2^{-\Delta(g)/50}$.
\end{theorem}

\paragraph{Discrepancy with respect to product distributions }

We note that \ref{def:discrepancy} is in fact a special case of the
common definition of discrepancy. The general definition refers to
an arbitrary distribution~$\mu$ over~$\Lambda\times\Lambda$. The
\emph{discrepancy of~$g$ with respect to~$\mu$} is defined similarly
to \ref{def:discrepancy} except that the random variables~$U,V$
are distributed according to~$\mu$ rather than the uniform distribution.
We show that \ref{main-theorem} holds even where the discrepancy
is with respect to a product distribution. We do so by reducing the
case of product distribution into the case of uniform distribution,
more details can be found in \ref{sec:Discrepancy-on-product}.
\begin{remark}
It is interesting to note that one of the first direct-sum results
in the randomized setting went along these lines. In particular, the
work of Shaltiel~\cite{S01sdpt} implies that for every function~$g$
such that $\Delta(g)\ge c$ for some universal constant~$c$, it
holds that $R^{cc}(g^{n})=\Omega\left(n\cdot\Delta(g)\right)$. Our
main theorem can be viewed as a generalization of that result.
\end{remark}

\begin{remark}
A natural question is whether the requirement that $\Delta(g)\ge c\cdot\log n$
is necessary. In principle, it is possible that this requirement could
be relaxed. Any such relaxation, however, would imply a lifting theorem
that allows inner functions of smaller input length than is currently
known, which would be considered a significant breakthrough.
\end{remark}

\begin{remark}
In order to facilitate the presentation, we restricted our discussion
on the previous works to lifting theorems that hold for every outer
function~$f$ (and indeed, every search problem~$S$). If one is
willing to make certain assumptions on the outer function~$f$, it
is possible to prove stronger lifting theorems that in particular
allow for a wider variety of inner functions (see, e.g., \cite{S09,SZ09,GP14,HHL16,RMNPRV20,ABK21}).
\end{remark}

\subsection{\label{subsec:Our-techniques}Our techniques}

Following the previous works, we use a ``simulation argument'':
We show that given a protocol that computes $f\circ g^{n}$ with communication
complexity~$C$, we can construct a decision tree that computes~$f$
with query complexity~$O(\frac{C}{\Delta(g)})$. In particular, we
follow the simulation argument of~\cite{CFKMP19} and extend their
main technical lemma. We now describe this argument in more detail,
focusing on the main lemma of \cite{CFKMP19} and our extension of
that lemma. For simplicity, we focus on the deterministic setting,
but the proof in the randomized setting follows similar ideas.

\paragraph*{The simulation argument.}

We assume that we have a protocol~$\Pi$ that computes $f\circ g^{n}$,
and would like to construct a decision tree~$T$ that computes~$f$.
The basic idea is that given an input~$\text{\ensuremath{z\in\B^{n}}}$,
the tree~$T$ uses the protocol~$\Pi$ to find a pair of inputs~$(x,y)\in\Lambda^{n}\times\Lambda^{n}$
such that $(f\circ g^{n})(x,y)=f(z)$, and then returns the output
of~$\Pi$ on~$(x,y)$. 

In order to find the pair~$(x,y)$, the tree~$T$ maintains a pair
of random variables~$(X,Y)$. Initially, the variables $(X,Y)$ are
uniformly distributed over~$\Lambda^{n}\times\Lambda^{n}$. Then,
the tree gradually changes the distribution of~$(X,Y)$ until they
satisfy $(f\circ g^{n})(X,Y)=f(z)$ with probability~$1$, at which
point the tree chooses~$(x,y)$ to be an arbitrary pair in the support
of~$(X,Y)$. This manipulation of the distribution of~$(X,Y)$ is
guided by a simulation of the protocol~$\Pi$ on~$(X,Y)$ (hence
the name ``simulation argument''). Throughout this process, the
decision tree maintains the following structure of~$(X,Y)$:
\begin{itemize}
\item There is a set of coordinates, denoted $F\subseteq$$\left[n\right]$,
such that for every $i\in F$ it holds that~$\text{\ensuremath{g(X_{i},Y_{i})=z_{i}}}$
with probability~$1$.
\item $X_{\left[n\right]\backslash F}$ and $Y_{\left[n\right]\backslash F}$
are \emph{dense} in the following sense: for every $J\subseteq\left[n\right]\backslash F$,
the variables $X_{J}$ and~$Y_{J}$ have high min-entropy.
\end{itemize}
Intuitively, the set~$F$ is the set of coordinates~$i$ for which
the simulation of~$\Pi$ has already computed~$g(X_{i},Y_{i})$,
while for the coordinates~$i\in\left[n\right]\backslash F$ the value~$g(X_{i},Y_{i})$
is unknown. Initially, the set~$F$ is empty, and then it is gradually
expanded until it holds that $(f\circ g^{n})(X,Y)=f(z)$.

\paragraph*{The main lemma of \cite{CFKMP19}.}

Suppose now that as part of the process described above, we would
like expand the set~$F$ by adding a new set of coordinates~$I\subseteq\left[n\right]\backslash F$.
This means that we should condition the distribution of~$(X,Y)$
on the event that $g^{I}(X_{I},Y_{I})=z_{I}$. This conditioning,
however, decreases the min-entropy of~$(X,Y)$, which might cause
$X_{\left[n\right]\backslash(F\cup I)}$ and $Y_{\left[n\right]\backslash(F\cup I)}$
to lose their density.

In order to resolve this issue, \cite{CFKMP19} defined a notion of
``sparsifying\emph{ }values'' of $X$ and~$Y$. Informally, a value~$x$
in the support of~$X$ is called\emph{ sparsifying} if after conditioning
$Y$ on the event $g^{I}(x_{I},Y_{I})=z_{I}$, the variable~$Y_{\left[n\right]\backslash(F\cup I)}$
ceases to be dense. A sparsifying value of~$Y$ is defined similarly.
It is not hard to see that if $X$ and~$Y$ do not have any sparsifying
values in their supports, then the density of~$X_{\left[n\right]\backslash(F\cup I)}$
and~$Y_{\left[n\right]\backslash(F\cup I)}$ is maintained after
the conditioning on~$\text{\ensuremath{g^{I}(X_{I},Y_{I})=z_{I}}}$.
Therefore, \cite{CFKMP19} design their decision tree such that before
the conditioning on the event~$g^{I}(X_{I},Y_{I})=z_{I}$, the tree
first removes the sparsifying values from the supports of~$X$ and~$Y$.

The removal of sparsifying values, however, raises another issue:
when we remove values from the supports of~$X$ and~$Y$, we decrease
the min-entropy of~$X$ and~$Y$. In particular, the removal of
the sparsifying values might cause $X_{\left[n\right]\backslash F}$
and $Y_{\left[n\right]\backslash F}$ to lose their density. This
issue is resolved by the main technical lemma of~\cite{CFKMP19}.
Informally, this lemma says that if $X_{\left[n\right]\backslash F}$
and $Y_{\left[n\right]\backslash F}$ are dense, then the sparsifying
values are very rare. This means that the removal of these values
barely changes the min-entropy of~$X$ and~$Y$, and in particular,
does not violate the density property.

\paragraph*{Our contribution.}

Recall that the lifting theorem of~\cite{CFKMP19} requires that~$\Delta(g)=\Omega(b)$
(where $b=\log\left|\Lambda\right|$), and that our goal is to waive
that requirement. Unfortunately, it turns out that main lemma of~\cite{CFKMP19}
fails when $\Delta(g)$ is very small relatively to~$b$. In fact,
in \ref{sec:Counterexample} we provide an example in which \emph{all
}the values in the support of~$X$ are sparsifying. 

Hence, unlike \cite{CFKMP19}, we cannot afford to remove the sparsifying
values before conditioning on the event $g^{I}(X_{I},Y_{I})=z_{I}$.
Therefore, in our simulation, the variables $X_{\left[n\right]\backslash F}$
and $Y_{\left[n\right]\backslash F}$ sometimes lose their density
after the conditioning. Nevertheless, we observe that even if the
density property breaks in this way, it can often be restored by removing
some more values from the supports of~$X$ and~$Y$. We formalize
this intuition by defining a notion of ``recoverable values''. Informally,
a value~$x$ in the support of~$X$ is called \emph{recoverable}
if after conditioning $Y$ on the event $g^{I}(x_{I},Y_{I})=z_{I}$,
the density of $Y_{\left[n\right]\backslash(F\cup I)}$ can be restored
by discarding some values from its support.

Our main lemma says, informally, that if $X_{\left[n\right]\backslash F}$
and $Y_{\left[n\right]\backslash F}$ are dense, then almost all the
values of~$X$ and~$Y$ are recoverable. In particular, we can afford
to remove the unrecoverable values of~$X$ and~$Y$ without violating
their density. Given our lemma, it is easy to fix the simulation argument
of~\cite{CFKMP19}: whenever our decision tree is about to condition
on an event~$g^{I}(x_{I},Y_{I})=z_{I}$, it first discards the unrecoverable
values of~$X$ and~$Y$; then, after the conditioning, the decision
tree restores the density property by discarding some additional values.
The rest of our argument proceeds exactly as in~\cite{CFKMP19}.

\paragraph*{The proof of our main lemma.}

The definition of a sparsifying value of~$X$ can be stated as follows:
the value~$x$ is sparsifying if there exists a value~$y_{J}$ such
that the probability 
\begin{equation}
\Pr\left[Y_{J}=y_{J}\mid g(x_{I},Y_{I})=z_{I}\right]\label{eq:introduction-y_J-z_I}
\end{equation}
is too high. On the other hand, it can be showed that a value~$x$
is \emph{unrecoverable} if there are \emph{many} such corresponding
values~$y_{J}$. Indeed, if there are only few such values~$y_{J}$,
then we can recover the density of~$Y_{\left[n\right]\backslash(F\cup I)}$
by discarding them.

Very roughly, the main lemma of~\cite{CFKMP19} is proved by showing
that for every~$y_{J}$, there is only a very small number of corresponding~$x$'s
for which the latter probability is too high. Then, by taking union
bound over all possible choices of~$y_{J}$, it follows that there
are only few values~$x$ for which there exists some corresponding~$y_{J}$.
In other words, there are only few sparsifying values.

This argument works in the setting of~\cite{CFKMP19} because they
can prove a very strong upper bound on the number of values~$x$
for a single~$y_{J}$ \textemdash{} indeed, the bound is sufficiently
strong to survive the union bound. In our setting, on the other hand,
the fact that we assume a smaller value of~$\Delta(g)$ translates
to a weaker bound on the number of values~$x$ for a single~$y_{J}$.
In particular, we cannot afford to use the union bound. Instead, we
take a different approach: we observe that, since for every~$y_{J}$
there is only a small number of corresponding~$x$'s, it follows
by an averaging argument that there can only be a small number of~$x$'s
that have \emph{many} corresponding~$y_{J}$'s. In other words, there
can only be a small number of \emph{unrecoverable}~$x$'s.

Implementing this idea is more difficult than it might seem at a first
glance. The key difficulty is that when we say ``values~$x$ that
have many corresponding~$y_{J}$'s'' we do not refer to the absolute
number of~$y_{J}$'s but rather to their probability mass. Specifically,
the probability distribution according to which the $y_{J}$'s should
be counted is the probability distribution of \ref{eq:introduction-y_J-z_I}.
Unfortunately, this means that for every value~$x$, we count the
$y_{J}$'s according to a different distribution, which renders a
simple averaging argument impossible. We overcome this difficulty
by proving a finer upper bound on the number of~$x$'s for each~$y_{J}$
and using a careful bucketing scheme for the averaging argument.

\section{\label{sec:Preliminaries}Preliminaries}

We assume familiarity with the basic definitions of communication
complexity (see, e.g., \cite{KN_book}). For any $n\in\N$, we denote
$\left[n\right]\eqdef\left\{ 1,\ldots,n\right\} $. We denote by~$c\in\N$
some large universal constant that will be chosen later ($c=1000$
will do). For the rest of this paper, we fix some natural number~$n\in\N$,
a finite set~$\Lambda$, and denote~$\text{\ensuremath{b=\log\left|\Lambda\right|}}$.
We fix $g:\D\times\D\to\B$~to be an arbitrary function such that
$\Delta(g)\ge c\cdot\log n$ (where $\Delta(g)\defeq\log\frac{1}{\disc(g)}$),
and abbreviate $\Delta\defeq\Delta(g)$. Since our main theorem holds
trivially when~$n=1$, we assume that $n\ge2$. Furthermore, throughout
this paper,~$X$ and~$Y$ denote independent random variables that
take values from~$\D^{n}$.

Let $I\subseteq\left[n\right]$ be a set of coordinates. We denote
by $\D^{I}$ the set of strings over alphabet~$\D$ of length~$\left|I\right|$
and index the coordinates of the string by~$I$. Given a string~$x\in\D^{n}$,
we denote by~$x_{I}\in\D^{I}$ the projection of~$x$ to the coordinates
in~$I$ (in particular, $x_{\emptyset}$ is defined to be the empty
string). We denote by $g^{I}:\D^{I}\times\D^{I}\to\B^{I}$ the function
that takes as inputs $\left|I\right|$ pairs from~$\D\times\D$ that
are indexed by~$I$, and outputs the string in~$\B^{I}$ whose $i$-th
bit is the output of~$g$ on the $i$-th pair. In particular, we
denote $g^{n}\eqdef g^{\left[n\right]}$, so $g^{n}$ is the direct-sum
function that takes as inputs~$x,y\in\D^{n}$ and outputs the binary
string
\[
g^{n}(x,y)\eqdef\left(g(x_{1},y_{1}),\ldots,g(x_{n},y_{n})\right).
\]
We denote by $g^{\oplus I}:\D^{I}\times\D^{I}\to\B$ the function
that given $x,y\in\D^{I}$ outputs the parity of the string $g^{I}(x,y)$.
The following bound is used throughout the paper.
\begin{proposition}
\label{fact:binomal_sum}Assume that $\beta,l\in\mathbb{R}$ such
that $\beta\leq1$ and $l\geq1$. Then it holds that 
\[
\sum_{S\subseteq\left[n\right],\left|S\right|\geq l}\beta^{\left|S\right|}\cdot\frac{1}{n^{\left|S\right|}}\leq2\beta^{l}
\]
. 
\end{proposition}

\begin{myproof}
It holds that 
\begin{align*}
\sum_{S\subseteq\left[n\right],\left|S\right|\geq l}\beta^{\left|S\right|}\frac{1}{n^{\left|S\right|}} & \leq\sum_{s=\left\lceil l\right\rceil }^{n}\binom{n}{s}\beta^{s}\frac{1}{n^{s}}\\
 & \leq\sum_{s=\left\lceil l\right\rceil }^{n}\beta^{s}\frac{1}{s!} & \text{(\ensuremath{\binom{n}{s}}\ensuremath{\ensuremath{\leq\frac{n^{s}}{s!}}})}\\
 & \leq2\sum_{s=\left\lceil l\right\rceil }^{\infty}\left(\frac{\beta}{2}\right)^{s} & \text{(\ensuremath{s!\geq2^{s-1}})}\\
 & \leq2\frac{\left(\frac{\beta}{2}\right)^{\left\lceil l\right\rceil }}{1-\frac{\beta}{2}} & \text{(geometric sum)}\\
 & \leq2\frac{\left(\frac{\beta}{2}\right)^{l}}{1-\frac{\beta}{2}} & \text{(\ensuremath{\frac{\beta}{2}\leq1} and \ensuremath{l\leq\left\lceil l\right\rceil })}\\
 & =2\beta^{l}\cdot\frac{\left(\frac{1}{2}\right)^{l}}{1-\frac{\beta}{2}} & \text{}\\
 & \leq2\beta^{l} & \text{(\ensuremath{\frac{1}{2}\leq1-\frac{\beta}{2}}).\qedhere}
\end{align*}
\end{myproof}

\paragraph*{Search problems.}

Given a finite set of inputs $\cI$ and a finite set of outputs~$\cO$,
a \emph{search problem}~$\cS$ is a relation between $\cI$ and~$\cO$.
Given $z\in\cI$, we denote by $\cS(z)$ the set of outputs $o\in\cO$
such that $(z,o)\in\cS$. Without loss of generality, we may assume
that $\cS(z)$ is always non-empty, since otherwise we can set $\cS(z)=\left\{ \bot\right\} $
where $\bot$ is some special failure symbol that does not belong
to~$\cO$.

Intuitively, a search problem~$\cS$ represents the following task:
given an input $z\in\cI$, find a solution $o\in\cS(z)$. In particular,
if $\cI=\cX\times\cY$ for some finite sets $\cX,\cY$, we say that
a deterministic protocol~$\Pi$ \emph{solves}~$\cS$ if for every
input $(x,y)\in\cI$, the output of $\Pi$ is in~$\cS(x,y)$. We
say that a randomized protocol~$\Pi$ \emph{solves}~$\cS$ \emph{with
error}~$\beta$ if for every input $(x,y)\in\cI$, the output of~$\Pi$
is in~$\cS(x,y)$ with probability at least~$1-\beta$. We denote
by $\dcc(\cS)$ the deterministic communication complexity of a search
problem~$\cS$. Given $\beta>0$, we denote by $\rcc_{\beta}(\cS)$-th
randomized (public-coin) communication complexity of~$\cS$ with
error~$\beta$. In case that~$\beta$ is omitted one should assume
that~$\beta=\frac{1}{3}$.

Given a search problem~$\cS\subseteq\B^{n}\times\cO$, we denote
by $\cS\circ g^{n}\subseteq(\D^{n}\times\D^{n})\times\cO$ the search
problem that satisfies $\cS\circ g^{n}(x,y)=\cS(g^{n}(x,y))$ for
every $x,y\in\D^{n}$.

\subsection{Decision trees}

Informally, a decision tree is an algorithm that solves a search problem~$\cS\subseteq\B^{n}\times\cO$
by querying the individual bits of its input. The decision tree is
computationally unbounded, and its complexity is measured by the number
of bits it queries.

Formally, a \emph{deterministic decision tree}~$T$ from $\B^{n}$
to~$\cO$ is a binary tree in which every internal node is labeled
with a coordinate in~$\left[n\right]$ (which represents a query),
every edge is labeled by a bit (which represents the answer to the
query), and every leaf is labeled by an output in~$\cO$. Such a
tree computes a function from~$\B^{n}$ to~$\cO$ in the natural
way, and with a slight abuse of notation, we denote this function
by~$T$ as well. The \emph{query complexity} of~$T$ is the depth
of the tree. We say that a tree~$T$ solves a search problem $\cS\subseteq\B^{n}\times\cO$
if for every $z\in\B^{n}$ it holds that $T(z)\in\cS(z)$. The \emph{deterministic
query complexity of}~$\cS$, denoted $\ddt(\cS)$, is the minimal
query complexity of a decision tree that solves~$\cS$.

A \emph{randomized decision tree}~$T$ is a random variable that
takes deterministic decision trees as values. The \emph{query complexity}
of~$T$ is the maximal depth of a tree in the support of~$T$. We
say that~$T$ \emph{solves a search problem $\cS\subseteq\B^{n}\times\cO$
with error~$\beta$} if for every $z\in\B^{n}$ it holds that
\[
\Pr\left[T(z)\in\cS(z)\right]\ge1-\beta.
\]
The \emph{randomized query complexity of~$\cS$ with error~$\beta$},
denoted $\rdt_{\beta}\left(\cS\right)$, is the minimal query complexity
of a randomized decision tree that solves~$\cS$ with error~$\beta$.
In case that~$\beta$ is omitted, one should assume that~$\beta=\frac{1}{3}$.

\subsection{Probability}

Below we recall some standard definitions and facts from probability
theory. Recall that the exponential distribution, denoted $\text{Ex}(\lambda)$,
is defined by the following cumulative probability distribution 
\[
1-e^{-\lambda x}\text{ for }x\ge0.
\]
The Erlang distribution, denoted $\text{Erl}(k,\lambda)$, is defined
as the sum of $k$ exponential variables with parameters $\lambda$
and its cumulative probability distribution of~$\text{Erl}(k,\lambda)$
is
\[
1-e^{-\lambda x}\sum_{i=0}^{k-1}\frac{\left(\lambda x\right)^{i}}{i!}\text{ for }x\ge0.
\]

Given two distributions $\mu_{1},\mu_{2}$ over a finite sample space~$\Omega$,
the \emph{statistical distance} (or \emph{total variation distance})
between $\mu_{1}$ and~$\mu_{2}$ is
\[
\left|\mu_{1}-\mu_{2}\right|\defeq\max_{\cE\subseteq\Omega}\left\{ \left|\mu_{1}(\cE)-\mu_{2}(\cE)\right|\right\} .
\]
We say that $\mu_{1}$ and $\mu_{2}$ are $\varepsilon$\emph{-close}
if $\left|\mu_{1}-\mu_{2}\right|\le\varepsilon$.
\begin{fact}
\label{fact:dist-on-cond}Let $\cE$ be some event and $\mu$ some
distribution. Then 
\[
\left|\mu-\left(\mu\mid\cE\right)\right|\leq1-\Pr\left[\cE\right].
\]
\end{fact}

Let $V$ be a random variable that takes values from a finite set~$\cV$.
The \emph{min-entropy }of~$V$, denoted~$\Hm(V)$, is the largest
number~$k\in\R$ such that for every value~$x$ it holds that $\text{\ensuremath{\Pr\left[V=v\right]\le2^{-k}}}$.
The \emph{deficiency} of~$V$ is defined as
\[
\Dm(V)\defeq\log\left|\cV\right|-H_{\infty}(V).
\]
Intuitively, the deficiency of $V$ measures the amount of information
that is known about $V$ relative to the uniform distribution. Deficiency
has the following easy-to-prove properties.
\begin{fact}
\label{fact:deficiency-nonnegative}For every random variable~$V$
it holds that 
\[
\Dm\left(V\right)\ge0.
\]
\end{fact}

\begin{fact}
\label{fact:deficiency-cond}For every random variable~$V$ and an
event $\mathcal{E}$ with positive probability it holds that 
\[
\Dm\left(V\mid\mathcal{E}\right)\le\Dm\left(V\right)+\log\frac{1}{\Pr\left[\mathcal{E}\right]}.
\]
\end{fact}

\begin{fact}
\label{fact:deficiency-monotone}Let $V_{1},V_{2}$ be random variables.
Then, 
\[
\Dm\left(V_{1}\right)\le\Dm\left(V_{1},V_{2}\right).
\]
\end{fact}

\subsubsection{Vazirani's Lemma}

Given a boolean random variable~$V$, we denote the bias of~$V$
by
\[
\b(V)\defeq\left|\Pr\left[V=0\right]-\Pr\left[V=1\right]\right|.
\]
Vazirani's lemma is a useful result that says that a random string
is close to being uniformly distributed if the parity of every set
of bits in the string has a small bias. We use the following variants
of the lemma.
\begin{lemma}[\cite{GLMWZ15}]
\label{lem:vazirani}Let $\varepsilon>0$, and let $Z$ be a random
variable taking values in~$\B^{m}$. If for every non-empty set $S\subseteq\left[m\right]$
it holds that 
\begin{equation}
\b(\bigoplus_{i\in S}Z_{i})\le\varepsilon\cdot\left(2\cdot m\right)^{-\left|S\right|}\label{eq:bias}
\end{equation}
then for every~$z\in\B^{m}$ it holds that 
\[
\left(1-\varepsilon\right)\cdot\frac{1}{2^{m}}\le\Pr\left[Z=z\right]\le\left(1+\varepsilon\right)\cdot\frac{1}{2^{m}}.
\]
\end{lemma}

The following version of Vazirani's lemma bounds the deficiency of
the random variable via a weaker assumption on the biases
\begin{lemma}[\cite{CFKMP19}]
\label{lem:second vazirani}Let $t\in\mathbb{N}$ be such that $t\geq1$,
and let $Z$ be a random variable taking values in $\B^{m}$. If for
every set $S\subseteq\left[m\right]$ such that $\left|S\right|\geq t$
it holds that 
\[
\b\left(\bigoplus_{i\in S}Z_{I}\right)\leq\left(2\cdot m\right)^{-\left|S\right|}
\]
then $\Dm\left(Z\right)\le t\log m+1$.
\end{lemma}

\subsubsection{Coupling}

Let $\mu_{1},\mu_{2}$ be two distributions over a sample space $\Omega$.
A \emph{coupling} \emph{of $\mu_{1}$ and $\mu_{2}$} is a distribution~$\nu$
over the sample space $\Omega^{2}$ whose marginals over the first
and second coordinates are $\mu_{1}$ and $\mu_{2}$ respectively.
The following standard fact characterizes the statistical distance
between $\mu_{1}$ and~$\mu_{2}$ using couplings.
\begin{fact}
\label{fact:coupling}Let $\mu_{1},\mu_{2}$ be two distributions
over a sample space~$\Omega$. For every coupling $\nu$ of $\mu_{1}$
and~$\mu_{2}$ it holds that 
\[
\left|\mu_{1}-\mu_{2}\right|=\min_{\nu}\Pr_{(V_{1},V_{2})\gets\nu}\left[V_{1}\ne V_{2}\right],
\]
where the minimum is taken over all couplings $\nu$ of $\mu_{1}$
and $\mu_{2}$. In particular, any coupling gives an upper bound on
the statistical distance. 
\end{fact}

\subsection{Prefix-free codes}

A set of strings $C\subseteq\B^{*}$ is called a \emph{prefix-free
code} if no string in~$C$ is a prefix of another string in~$C$.
Given a string $w\in\B^{*}$, we denote its length by~$\left|w\right|$.
We use the following simple corollary of Kraft's inequality.
\begin{fact}[Corollary of Kraft's inequality]
\label{Fact:prefix-code}Let $C\subseteq\B^{*}$ be a finite prefix-free
code, and let~$W$ be a random string that takes values from $C$.
Then, there exists a string $w\in C$ such that $\text{\ensuremath{\Pr\left[W=w\right]\geq2^{-\left|w\right|}}}$.
\end{fact}

A simple proof of \ref{Fact:prefix-code} can be found in \cite{CFKMP19}.

\subsection{Properties of discrepancy}

Recall that $\Delta\defeq\Delta(g)=\log\frac{1}{\disc(g)}$, that
$X,Y$ are independent random variables that take values from~$\D^{n}$,
and that $g^{\oplus S}:\D^{S}\times\D^{S}\to\B$ is the function that
given $x_{S},y_{S}\in\D^{S}$ outputs the parity of the string~$g^{S}(x_{S},y_{S})$.
We use the following properties of discrepancy. In what follows, the
parameter $\lambda$ controls $\b\left(g^{\oplus S}(x_{S},Y_{S})\right)$
and the parameter~$\gamma$ controls the error probability.
\begin{lemma}[{see, e.g., \cite[Cor. 2.12]{CFKMP19}}]
\label{discrepancy-XOR-extractor}Let $\lambda>0$ and let $S\subseteq\left[n\right]$.
If
\[
\Dm(X_{S})+\Dm(Y_{S})\le(\Delta(g)-6-\lambda)\cdot\left|S\right|
\]
then
\[
\b\left(g^{\oplus S}(X_{S},Y_{S})\right)\le2^{-\lambda\left|S\right|}.
\]
\end{lemma}

\begin{lemma}[{see, e.g., \cite[Cor. 2.13]{CFKMP19}}]
\label{discrepancy-XOR-sampling}Let $\gamma,\lambda>0$ and let
$S\subseteq\left[n\right]$. If 
\[
\Dm(X_{S})+\Dm(Y_{S})\le(\Delta(g)-7-\gamma-\lambda)\cdot\left|S\right|
\]
then the probability that $X$ takes a value~$x\in\D^{n}$ such that
\[
\b\left(g^{\oplus S}(x_{S},Y_{S})\right)>2^{-\lambda\left|S\right|}
\]
is less than $2^{-\gamma\left|S\right|}$.
\end{lemma}

\subsection{Background from \cite{CFKMP19}}

In this section, we review some definitions and results from~\cite{CFKMP19}
that we use in our proofs. We present those definitions and results
somewhat differently than~\cite{CFKMP19} in order to streamline
the proofs in the setting where $\Delta\ll b$. The most significant
deviation from the presentation of~\cite{CFKMP19} is the following
definition of a $\sigma$-sparse random variable, which replaces the
notion of a $\delta$-dense random variable from~\cite{GLMWZ15,GPW17,CFKMP19}.
Both notions are aimed to capture a random variable over $\D^{n}$
on which very little information is known. 
\begin{definition}
Let $X$ be a random variable taking values from~$\Lambda^{n}$,
and let $\sigma>0$. We say that~$X$ is \emph{$\sigma$-sparse }if
for every set $S\subseteq\left[n\right]$ it holds that $\Dm(X_{S})\le\sigma\cdot\Delta\cdot\left|S\right|.$
\end{definition}

\begin{remark}
The relation between the above definition to the notion of $\delta$-dense
random variable of~\cite{GLMWZ15,GPW17,CFKMP19} is the following:
$X$ is $\sigma$-sparse if and only if it is $(1-\frac{\Delta}{b}\cdot\sigma)$-dense.
\end{remark}

As explained in \ref{subsec:Our-techniques}, our proof relies on
a simulation argument that takes a protocol~$\Pi$ for~$\cS\circ g^{n}$
and constructs a decision tree~$T$ for~$\cS$. We use the following
notion of restriction to keep track of the queries that the tree makes
and their answers.
\begin{definition}
A \emph{restriction~$\rho$} is a string in $\left\{ 0,1,*\right\} ^{n}$.
We say that a coordinate~$i\in\left[n\right]$ is \emph{free} in~$\rho$
if $\rho_{i}=*$, and otherwise we say that $i$~is \emph{fixed}.
Given a restriction $\rho\in\left\{ 0,1,*\right\} ^{n}$, we denote
by $\free(\rho)$ and $\fix(\rho)$ the sets of free and fixed coordinates
of~$\rho$ respectively. We say that a string~$z\in\B^{n}$ is \emph{consistent}
with $\rho$ if $z_{\fix(\rho)}=\rho_{\fix(\rho)}$.
\end{definition}

Intuitively, $\fix(\rho)$ represents the queries that have been made
so far, and $\free(\rho)$ represents the coordinates that have not
been queried yet. As explained in \ref{subsec:Our-techniques}, the
decision tree maintains a pair of random variables $X,Y$ with a certain
structure, which is captured by the following definition.
\begin{definition}
\label{structured-RVs}Let $\rho\in\left\{ 0,1,*\right\} ^{n}$ be
a restriction, let $\sigma_{X},\sigma_{Y}>0$, and let $X,Y$ be independent
random variables that take values from $\D^{n}$. We say that $X$~and~$Y$
are $(\rho,\sigma_{X},\sigma_{Y})$\emph{-structured} if there exist
$\sigma_{X},\sigma_{Y}>0$ such that $X_{\free(\rho)}$ and $Y_{\free(\rho)}$
are $\sigma_{X}$-sparse and $\sigma_{Y}$-sparse respectively and
\[
g^{\fix(\rho)}\left(X_{\fix(\rho)},Y_{\fix(\rho)}\right)=\rho_{\fix(\rho)}.
\]
\end{definition}

Intuitively, this structure says that $(X_{\fix(\rho)},Y_{\fix(\rho)})$
must be consistent with the queries that the decision tree has made
so far, and that the simulated protocol~$\Pi$ does not know much
about the free coordinates. The following proposition formalizes the
intuition that the simulated protocol does not know the value of~$g$
on the free coordinates. In what follows, the parameter~$\gamma$
controls how close are the values of~$g$ on the free coordinates
to being uniformly distributed.
\begin{proposition}[{\cite[Prop. 3.10]{CFKMP19}}]
\label{multiplicative-uniformity}Let $\gamma\geq0$. Let $X,Y$
be random variables that are $(\rho,\sigma_{X},\sigma_{Y})$-structured
for $\sigma_{X}+\sigma_{Y}\leq1-\frac{8}{c}-\gamma$, and let $I=\free\left(\rho\right)$.
Then, 
\[
\forall z_{I}\in\B^{I}:\Pr\left[g^{I}\left(X_{I},Y_{I}\right)=z_{I}\right]\in\left(1\pm2^{-\gamma\Delta}\right)\cdot2^{-\left|I\right|}.
\]
\end{proposition}

Next, we state the uniform marginals lemma of~\cite{CFKMP19} (which
generalized an earlier lemma of \cite{GPW17}). Intuitively, this
lemma says that the simulated protocol~$\Pi$ cannot distinguish
between the distribution~$(X,Y)$ and the same distribution conditioned
on~$g^{n}(X,Y)=z$. In what follows, the parameter~$\gamma$ controls
the indistinguishability.
\begin{lemma}[{Uniform marginals lemma, \cite[Lemma 3.4]{CFKMP19}}]
\label{lem:uniform-marginal-lemma}Let $\gamma\geq0$, let $\rho$
be a restriction, and let $z\in\B^{n}$ be a string that is consistent
with $\rho$. Let $X,Y$ be $\left(\rho,\sigma_{X},\sigma_{Y}\right)$-structured
random variables that are uniformly distributed over sets $\mathcal{X},\mathcal{Y}\subseteq\Lambda^{n}$
respectively such that $\text{\ensuremath{\sigma_{X}+\sigma_{Y}\leq1-\frac{10}{c}-\gamma}}$
. Let $\left(X',Y'\right)$ be uniformly distributed over $\left(g^{n}\right)^{-1}\left(z\right)\cap\left(\cX\times\cY\right)$.
Then,~$X$ and $Y$ are $2^{-\gamma\Delta}$-close to $X'$ and $Y'$
respectively.
\end{lemma}

The following folklore fact allows us to transform an arbitrary variable
over~$\D^{n}$ into a $\sigma$-sparse one by fixing some of its
coordinates.
\begin{proposition}[{see, e.g., \cite[Prop. 3.6]{CFKMP19}}]
\label{pro:density-restoring-fixing}Let $X$ be a random variable,
let $\sigma_{X}>0$, and let $I\subseteq\left[n\right]$ be a maximal
subset of coordinates such that $\Dm(X_{I})>\sigma_{X}\cdot\Delta\cdot\left|I\right|$.
Let $x_{I}\in\D^{I}$ be a value such that 
\[
\Pr\left[X_{I}=x_{I}\right]>2^{\sigma_{X}\cdot\Delta\cdot\left|I\right|-b\cdot\left|I\right|}.
\]
Then, the random variable $X_{\left[n\right]-I}\mid X_{I}=x_{I}$
is $\sigma_{X}$-sparse.
\end{proposition}

\ref{pro:density-restoring-fixing} is useful in the deterministic
setting, since in this setting the decision tree is free to condition
the distributions of $X,Y$ in any way that does not increase their
sparsity. In the randomized setting, however, the decision tree is
more restricted, and cannot condition the inputs on events such as
$X_{I}=x_{I}$ which may have very low probability. In \cite{GPW17},
this issue was resolved by observing that the probability space can
be partitioned to disjoint events of the form~$X_{I}=x_{I}$, and
that the randomized simulation can use such a partition to achieve
the same effect of \ref{pro:density-restoring-fixing}. This leads
to the following lemma, which we use as well.
\begin{lemma}[Density-restoring partition \cite{GPW17}]
\label{lem:density-restoring-partition}Let $\cX\subseteq\D^{n}$
denote the support of~$X$, and let $\sigma_{X}>0$. Then, there
exists a partition 
\[
\cX\defeq\cX^{1}\cup\dots\cup\cX^{\ell}
\]
where each $\cX^{j}$ is associated with a set $I_{j}\subseteq\left[n\right]$
and a value $x_{j}\in\D^{I_{j}}$ such that:
\begin{itemize}
\item $X_{I_{j}}\mid X\in\cX^{j}$ is fixed to~$x_{j}$.
\item $X_{\left[n\right]-I_{j}}\mid X\in\cX^{j}$ is $\sigma_{X}$-sparse.
\end{itemize}
Moreover, if we denote $p_{\ge j}\defeq\Pr\left[X\in\cX^{j}\cup\ldots\cup\cX^{\ell}\right]$,
then it holds that

\[
\Dm(X_{\left[n\right]-I_{j}}\mid X\in\cX^{j})\leq\Dm(X)+\sigma_{X}\cdot\Delta\cdot\left|I_{j}\right|+\log\frac{1}{p_{\ge j}}.
\]
\end{lemma}

In what follows, we recall some definitions and the main lemma from
\cite{CFKMP19}. Those definitions and lemma are not used to prove
our main result, but are given here so they can be compared to our
definitions and main lemma.
\begin{definition}[\cite{CFKMP19}]
\label{def:cfkmp-dangerous}Let $Y$ be a random variable taking
values from $\Lambda^{n}$. We say that a value $x\in\Lambda^{n}$
is \emph{leaking} if there exists a set $I\subseteq\left[n\right]$
and an assignment $z_{I}\in\B^{I}$ such that 
\[
\Pr\left[g^{I}\left(x_{I},Y_{I}\right)=z_{I}\right]<2^{-\left|I\right|-1}.
\]
Let $\sigma_{Y},\varepsilon>0$, and suppose that $Y$ is $\sigma_{Y}$-sparse.
We say that a value $x\in\Lambda^{n}$ is $\varepsilon$\emph{-sparsifying}
if there exists a set $I\subseteq\left[n\right]$ and an assignment
$z_{I}\in\B^{I}$ such that the random variable $\mbox{\ensuremath{Y_{\left[n\right]-I}\mid g^{I}\left(x_{I},Y_{I}\right)=z_{I}}}$
is not $(\sigma_{Y}+\varepsilon)$-sparse. We say that a value $x\in\Lambda^{n}$
is $\varepsilon$\emph{-dangerous} if it is either leaking or $\varepsilon$-sparsifying.
\end{definition}

\begin{lemma}[main lemma of \cite{CFKMP19}]
\label{main-cfkmp}There exists universal constants $h,c$ such that
the following holds: Let $b$ be some number such that $b\geq c\cdot\log n$
and let $\gamma,\varepsilon,\sigma_{X},\sigma_{Y}>0$ be such that
$\varepsilon\geq\frac{4}{\Delta}$, and $\sigma_{X}+\sigma_{Y}\leq1-\frac{h\cdot b\cdot\log n}{\Delta^{2}\cdot\varepsilon}-\gamma$.
Let $X,Y$ be $\left(\rho,\sigma_{X},\sigma_{Y}\right)$-structured
random variables. Then, the probability that $X_{\free\left(\rho\right)}$
takes a value that is $\varepsilon$-dangerous for $Y_{\free\left(\rho\right)}$
is at most $2^{-\gamma\Delta}$. 
\end{lemma}

\section{\label{sec:Main-lemma}The main lemma}

\edef\epsCns{4}
\FPeval{\epsCnsPropRed}{clip(3+\epsCns{})}
\edef\cnsRefinedWeaklySparsifying{15}
\FPeval{\cnsProbabilityOfRedeemable}{clip(\cnsRefinedWeaklySparsifying{}+\epsCnsPropRed{})}
\FPeval{\cnsAlmostAlwaysNot}{clip(\cnsProbabilityOfRedeemable{}+2)}
\FPeval{\cnsMainLemma}{clip(max(\cnsAlmostAlwaysNot{},11)+1)}In this section, we state and prove our main lemma. As discussed in
the introduction, our simulation argument maintains a pair of random
variables $X,Y\in\D^{n}$. A crucial part of the simulation consists
of removing certain ``unsafe'' values from the supports of these
variables. Our main lemma says that almost all values are safe. 

There are two criteria for a value $x\in\Lambda^{n}$ to be ``safe'':
First, it should hold that after we condition $Y$ on an event of
the form $g^{I}\left(x_{I},Y_{I}\right)=z_{I}$, the density of $Y$
can be recovered by conditioning on a high probability event (such
values are called \emph{recoverable}). Second, it holds that $g^{n}(x,Y)$
is distributed almost uniformly (such values are called \emph{almost
uniform}). This guarantees that from the point of view of Alice, who
knows~$X$, the value of $g^{n}(X,Y)$ is distributed almost uniformly.
For the simplicity of notation we denote:
\begin{notation}
For $x\in\Lambda^{n}$ we define the random variable $Z^{x}\defeq g^{n}\left(x,Y\right)$.
\end{notation}

We now formally define the notions of safe, recoverable and almost
uniform values.
\begin{definition}[safe values]
\label{dangerous}Let $\alpha\geq0$. Let $Y$ be a random variable
in~$\D^{n}$, let $\sigma_{Y}>0$ be the minimal value for which
$Y$ is $\sigma_{Y}$-sparse and let $x\in\D^{n}$. We say that $x$~is
\emph{almost uniform} (for $Y$) if for any assignment~$z\in\B^{n}$
it holds that 
\[
\Pr\left[Z^{x}=z\right]\in2^{-n}\left(1\pm2^{-\frac{\Delta}{10}}\right).
\]
We say that $x$~is\textbf{ $\alpha$}-\emph{recoverable} (for $Y$)
if for all $I\subseteq\left[n\right]$ and $z_{I}$ the following
holds: there exists an event~$\cE$ that only depend on $Y$ such
that $\Pr\left[\cE\mid Z_{I}^{x}=z_{I}\right]\ge1-2^{-\alpha\Delta}$
and such that the random variable
\[
Y_{\left[n\right]-I}\mid\cE\text{ and }Z_{I}^{x}=z_{I}
\]
is $(\sigma_{Y}+\frac{\epsCns}{c})$-sparse. We say that $x$~is
$\alpha$\emph{-safe} (for $Y$) if it is both almost uniform and
\textbf{$\alpha$}-recoverable. Almost uniform, recoverable, and safe
values for $X$ are defined analogously.
\end{definition}

We turn to state our main lemma.
\begin{lemma}[Main lemma]
\label{lem:main-lemma}For every $c>0$ the following holds. Let
$\alpha\ge\frac{1}{\Delta}$ and $\gamma>0$, and let $X$ and~$Y$
be independent random variables such that $X$ is $\sigma_{X}$-sparse.
Let $\sigma_{Y}>0$ be the minimal value for which $Y$ is $\sigma_{Y}$-sparse.
If $\sigma_{X}+2\sigma_{Y}\leq\frac{9}{10}-\frac{\cnsMainLemma}{c}-\gamma-\alpha$,
then 
\[
\Pr_{x\sim X}\left[x\text{ is not \ensuremath{\alpha}-safe for }Y\right]\le2^{-\gamma\cdot\Delta}.
\]
\end{lemma}

In the rest of this section, we prove the main lemma. Let $\alpha,\sigma_{X},\sigma_{Y}$
be as in the lemma. Let~$X,Y$ be independent random variables such
that $X$ is $\sigma_{X}$-sparse and $Y$ is $\sigma_{Y}$-sparse.
The following two propositions, which are proved in \ref{subsec:Proof-of leaking,subsec:Proof-of recoverable},
upper bound the probabilities that~$X$ takes a value that is not
almost uniform or not recoverable respectively. \ref{lem:main-lemma}
follows immediately from the following propositions using the union
bound.
\begin{proposition}
\label{leaking-upper-bound}Let $\gamma>0$ be a real number such
that $\sigma_{X}+\sigma_{Y}\leq\frac{9}{10}-\frac{11}{c}-\gamma$.
The probability that $X$ takes a value that is not almost uniform
is at most~$2^{-\gamma\cdot\Delta}$.
\end{proposition}

\begin{proposition}
\label{pro:almost always not strongly sparsifying}Let $\gamma>0$
and assume that $\sigma_{X}+2\cdot\sigma_{Y}\leq1-\frac{\cnsAlmostAlwaysNot}{c}-\gamma-\alpha$.
Then, the probability that $X$ takes an almost uniform value~$x$
that is not $\alpha$-recoverable is at most $2^{-\gamma\cdot\Delta}$. 
\end{proposition}

\begin{myproof}[Proof of \ref{lem:main-lemma} from \ref{leaking-upper-bound,pro:almost always not strongly sparsifying}]
Any value that is not $\alpha$-safe is either not almost uniform
or almost uniform but not $\alpha$-recoverable. By applying \ref{leaking-upper-bound}
with $\gamma=\gamma+\frac{1}{c}$, it follows that the probability
that $X$ takes a value that is not almost uniform is at most $2^{-(\gamma+\frac{1}{c})\cdot\Delta}\le2^{-\gamma\Delta-1}$.
By applying \ref{pro:almost always not strongly sparsifying} with
$\gamma=\gamma+\frac{1}{c}$, and $\alpha=\alpha$, it follows that
the probability that $X$ takes an almost uniform value that is not
$\alpha$-recoverable is at most~$2^{-(\gamma+\frac{1}{c})\cdot\Delta}\le2^{-\gamma\Delta-1}$.
Therefore, the probability that $X$ takes a value that is not $\alpha$-safe
value is at most $2^{-\gamma\Delta-1}+2^{-\gamma\Delta-1}=2^{-\gamma\Delta}$.
\end{myproof}
\begin{remark}
We now compare our notion of safe values to the notion of dangerous
values in \cite{CFKMP19}. 
\begin{itemize}
\item Our requirement that safe values would be almost uniform corresponds
to the requirement of \cite{CFKMP19} that safe values would be non-leaking.
However, our definition is stronger: Both definitions bound the probability
of $\Pr\left[Z_{I}^{x}=z_{I}\right]$. The definition of almost-uniform
values bounds the probability tightly from both above and below while
the definition of non-leaking values bounds it only from below. We
use our stronger requirement to bound the probability of a certain
event in \ref{subsec:Proof-Of-Correctness-Without-Halting}. 
\item Our requirement that safe values would be recoverable corresponds
to the requirement of \cite{CFKMP19} that safe values would be non-sparsifying.
Our requirement is weaker: Both definitions regard the sparsity of
the random variable $Y_{\left[n\right]-I}\mid Z_{I}^{x}=z_{I}$. While
the definition of \cite{CFKMP19} requires the random variable to
have low sparsity, our definition only requires that the sparsity
of the random variable can be made low by conditioning on an additional
event. As discussed in the introduction, this weakening is necessary
as when $\Delta\ll b$, there might not be enough values $x$ that
satisfy the stronger requirement (see \ref{sec:Counterexample}). 
\end{itemize}
\end{remark}

\subsection{\label{subsec:Proof-of leaking}Proof of \ref{leaking-upper-bound}}

In this section we prove \ref{leaking-upper-bound}, restated next,
following the ideas of~\cite{CFKMP19}. Essentially, the proof uses
Vazirani's lemma to reduce the problem to bounding $\b(g^{\oplus S}(x_{S},Y_{S}))$
for most values of $x$. This is done using the fact that $X$ and
$Y$ have low sparsity together with the low discrepancy of~$g$. 
\begin{restated}{\ref{leaking-upper-bound}}
Let $\gamma>0$ be a real number such that $\sigma_{X}+\sigma_{Y}\leq\frac{9}{10}-\frac{11}{c}-\gamma$.
The probability that~$X$ takes a value that is not almost uniform
is at most~$2^{-\gamma\cdot\Delta}$.
\end{restated}

\begin{myproof}
We start by observing that for every $x\in\D^{n}$, if it holds that
\[
\b(g^{\oplus S}(x_{S},Y_{S}))\le2^{-\frac{\Delta}{10}}\cdot{(2n)}^{-\left|S\right|}
\]
for every non-empty set~$S\subseteq\left[n\right]$, then by first
variant of Vazirani's lemma (\ref{lem:vazirani}) we get that~$x$
is almost uniform.

It remains to show that with probability at least $1-2^{-\gamma\cdot\Delta}$
the random variable~$X$ takes a value~$x$ that satisfies the latter
condition on the biases. We start by lower bounding the probability
that $\b(g^{\oplus S}(x_{S},Y_{S}))\le2^{-\frac{\Delta}{10}}\cdot{(2n)}^{-\left|S\right|}$
for a specific set~$S\subseteq\left[n\right]$. Fix a non-empty set
$S\subseteq\left[n\right]$. By assumption, it holds that 
\begin{align*}
\Dm(X_{S})+\Dm(Y_{S}) & \le(1-\frac{11}{c}-\gamma-\frac{1}{10})\cdot\Delta\cdot\left|S\right|\\
 & =\left(\Delta-\frac{7\Delta}{c}-\gamma\Delta-\frac{\Delta}{10}-\frac{4\Delta}{c}\right)\cdot\left|S\right|\\
 & \le\left(\Delta-7-\gamma\Delta-\frac{\Delta}{10}-2\log n-2\right)\cdot\left|S\right|. & \text{(\ensuremath{\Delta\geq c\log n})}
\end{align*}
By applying \ref{discrepancy-XOR-sampling} with~$\gamma=\gamma\Delta+\log n+1$
and~$\lambda=\log n+1+\frac{\Delta}{10}$ it follows that with probability
at least~$1-2^{-\gamma\Delta-1}\cdot\frac{1}{n^{\left|S\right|}}$,
the random variable~$X$ takes a value~$x$ such that
\[
\b\left(g^{\oplus S}(x_{S},Y_{S})\right)\le2^{-\frac{\Delta}{10}}\cdot\left(2n\right)^{-\left|S\right|}.
\]
Next, by taking the union bound over all non-empty sets~$S\subseteq\left[n\right]$,
it follows that 
\[
\Pr\left[\exists S\neq\emptyset:\b\left(g^{\oplus S}(x_{S},Y_{S})\right)>2^{-\frac{\Delta}{10}}\cdot{(2n)}^{-\left|S\right|}\right]\leq\sum_{S\subseteq\left[n\right]:S\ne\emptyset}2^{-\gamma\Delta-1}\cdot\frac{1}{n^{\left|S\right|}}<2^{-\gamma\Delta-1}\cdot2=2^{-\gamma\Delta}
\]
by \ref{fact:binomal_sum}. Therefore with probability at least $1-2^{-\gamma\Delta}$,
the random variable~$X$ takes a value~$x$ such that $\b(g^{\oplus S}(x_{S},Y_{S}))\le2^{-\frac{\Delta}{10}}\cdot{(2n)}^{-\left|S\right|}$
for all non-empty sets~$S\subseteq\left[n\right]$, as required.
\end{myproof}

\subsection{\label{subsec:Proof-of recoverable}Proof of \ref{pro:almost always not strongly sparsifying}}

In the following subsection we prove \ref{pro:almost always not strongly sparsifying}.
To this end, we introduce the notion of light values. A value $x$
is \emph{light} if after conditioning on $x$ the distribution $Y_{J}\mid Z_{I,x_{I}}=z_{I}$
is ``mostly'' $\left(\sigma_{Y}+\frac{\epsCns}{c}\right)$-sparse,
that is, for almost all values $y_{J}$ it holds that
\[
\Pr\left[Y_{J}=y_{J}\mid Z_{I}^{x}=z_{I}\right]\leq2^{\left(\sigma_{Y}+\frac{\epsCns}{c}\right)\cdot\Delta\cdot\left|J\right|-b\cdot\left|J\right|}.
\]
The term ``light'' is motivated by the intuitive idea that the values~$y_{J}$
that violate the sparsity requirement are ``heavy'' in terms of
their probability mass, so a ``light'' value~$x$ is one that does
not have many ``heavy'' values~$y_{J}$. We show that when a values
$x$ is light then it also recoverable. Intuitively, the density of
$Y_{J}$ can be restored by conditioning on the high probability event
that the heavy values are not selected. We now formally define light
values. In the following definition of heavy value there is an additional
parameter $t$, which measures by how much the heavy value~$y_{J}$
violates the $(\sigma_{Y}+\frac{\epsCns}{c})$-sparsity. This parameter
is used later in the proof. 
\begin{definition}
Let $x\in\D^{n},J\subseteq\left[n\right]$ and $t\in\mathbb{R}$.
We say that $y_{J}$ is $t$-\emph{heavy} (for $x$ and $z_{I}$)
if
\[
\Pr\left[Y_{J}=y_{J}\mid Z_{I}^{x}=z_{I}\right]>2^{(\sigma_{Y}+\frac{\epsCns}{c})\cdot\Delta\cdot\left|J\right|-b\cdot\left|J\right|+t-1}
\]
We say that a value $y_{J}$ is heavy if it is $0$-heavy. We say
that a value $x$ is $\alpha$\emph{-light} with respect to~$J$
if for every $I\subseteq\left[n\right]-J$ and $z_{I}\in\B^{I}$ it
holds that
\[
\Pr\left[Y_{J}\text{ is heavy for \ensuremath{x} and \ensuremath{z_{I}}}\mid Z_{I}^{x}=z_{I}\right]\le2^{-\alpha\Delta}\cdot\left(\frac{1}{2n}\right)^{\left|J\right|}.
\]
A value $x$ is $\alpha$\emph{-light} if it is $\alpha$\emph{-light}
with respect to all sets~$J$.
\end{definition}

We now state the relationship between light values and recoverable
values that was mentioned earlier.
\begin{proposition}
\label{redeemable-implies-non-sparsifying}If $x\in\D^{n}$ is $\alpha$\textup{\emph{-light}}
then it is $\alpha$-recoverable.
\end{proposition}

We postpone the proof of \ref{redeemable-implies-non-sparsifying}
to the end of this subsection. To prove \ref{pro:almost always not strongly sparsifying},
we consider values $x$ that are not light with respect to some specific
$J$. The following proposition bounds the probability of such values.
We then complete the proof by taking union bound over all sets $J$.
\begin{proposition}
\label{probability-of-redeemable}Assume that $\sigma_{X}+2\cdot\sigma_{Y}\leq1-\frac{\cnsProbabilityOfRedeemable}{c}-\gamma-\alpha$.
For every $J\subseteq\left[n\right]$, the probability that $X$ takes
an almost uniform value~$x$ that is not $\alpha$-light with respect
to~$J$ is at most $2^{-\gamma\cdot\Delta\cdot\left|J\right|}$.
\end{proposition}

The proof \ref{probability-of-redeemable} is provided in \ref{subsec:proof-of-light}.
We now prove \ref{pro:almost always not strongly sparsifying} restated
below. 
\begin{restated}{\ref{pro:almost always not strongly sparsifying}}
Assume that $\sigma_{X}+2\cdot\sigma_{Y}\leq1-\frac{\cnsAlmostAlwaysNot}{c}-\gamma-\alpha$.
Then, the probability that $X$ takes an almost uniform value~$x$
that is not $\alpha$-recoverable is at most $2^{-\gamma\cdot\Delta}$. 
\end{restated}

\begin{myproof}
By applying \ref{probability-of-redeemable} with $\gamma=\gamma+\frac{2}{c}$,
we obtain that for every set~$J\subseteq\left[n\right]$, the probability
that $X$ takes an almost uniform value~$x$ that is not $\alpha$-light
with respect to~$J$ is at most~$2^{-\gamma\cdot\Delta}\cdot\frac{1}{\left(2n\right)^{\left|J\right|}}$.
By the union bound and \ref{fact:binomal_sum}, we obtain that with
probability at least~$1-2^{-\gamma\cdot\Delta}$, the random variable~$X$
takes a value~$x$ that is $\alpha$-light with respect to all~$J\subseteq\left[n\right]$.
Such a value~$x$ is $\alpha$-recoverable by \ref{redeemable-implies-non-sparsifying},
so the required result follows.
\end{myproof}
\begin{myproof}[Proof of \ref{redeemable-implies-non-sparsifying}]
Let $\alpha\ge\frac{1}{\Delta}$ and let $x\in\D^{n}$ be $\alpha$-light.
We show that $x$ is $\alpha$-recoverable by showing that for every
$I\subseteq\left[n\right]$ and $z_{I}\in\D^{I}$ there exists an
event~$\cE$ such that the following random variable is $(\sigma_{Y}+\frac{\epsCns}{c})$-sparse:
\[
Y_{\left[n\right]-I}\mid\cE\text{ and }Z_{I}^{x}=z_{I}.
\]
We choose $\cE$ to be the event such that $y_{J}$ are not heavy
for some non-empty set $J\subseteq\left[n\right]-I$. We first prove
that $\Pr\left[\neg\mathcal{E}\mid Z_{I,x_{I}}=z_{I}\right]<2^{-\alpha\Delta}$.
By the union bound, it holds that 
\begin{align*}
\Pr\left[\neg\mathcal{E}\mid Z_{I,x_{I}}=z_{I}\right] & =\Pr\left[\exists J\neq\emptyset\subseteq\left[n\right]-I:Y_{J}\text{ is heavy for \ensuremath{x} and \ensuremath{z_{I}}}\mid Z_{I}^{x}=z_{I}\right]\\
 & \leq\sum_{\emptyset\ne J\subseteq\left[n\right]-I}\Pr\left[Y_{J}\text{ is heavy for \ensuremath{x} and \ensuremath{z_{I}}}\mid Z_{I}^{x}=z_{I}\right]\\
 & \leq\sum_{\emptyset\ne J\subseteq\left[n\right]-I}2^{-\alpha\Delta}\cdot\left(\frac{1}{2n}\right)^{\left|J\right|} & \text{(\ensuremath{x} is \ensuremath{\alpha}-light)}\\
 & \leq2^{-\alpha\Delta} & \text{(\ref{fact:binomal_sum})}
\end{align*}
It remain to prove that the random variable
\[
Y_{\left[n\right]-I}\mid\cE\text{ and }Z_{I}^{x}=z_{I}
\]
is $\left(\sigma_{Y}+\frac{\epsCns}{c}\right)$-sparse. For every
$J\subseteq\left[n\right]-I$, let $y_{J}$ be some value of $Y_{J}$
then it holds that 
\begin{align*}
\Pr\left[Y_{J}=y_{J}\mid\cE\text{ and }Z_{I,x_{I}}=z_{I}\right] & =\frac{\Pr\left[Y_{J}=y_{J}\text{ and }Y\in\cE\mid Z_{I}^{x}=z_{I}\right]}{\Pr\left[\cE\mid Z_{I}^{x}=z_{I}\right]}\\
 & \leq\frac{2^{(\sigma_{Y}+\frac{\epsCns}{c})\cdot\Delta\cdot\left|J\right|-b\cdot\left|J\right|-1}}{\Pr\left[\cE\mid Z_{I}^{x}=z_{I}\right]} & \text{(\ensuremath{y_{J}} is not heavy)}\\
 & \leq\frac{2^{(\sigma_{Y}+\frac{\epsCns}{c})\cdot\Delta\cdot\left|J\right|-b\cdot\left|J\right|-1}}{1-2^{-\alpha\Delta}}\\
 & \leq\frac{2^{(\sigma_{Y}+\frac{\epsCns}{c})\cdot\Delta\cdot\left|J\right|-b\cdot\left|J\right|-1}}{1-\frac{1}{2}} & \text{(since \ensuremath{\alpha\ge\frac{1}{\Delta}})}\\
 & \leq2^{(\sigma_{Y}+\frac{\epsCns}{c})\cdot\Delta\cdot\left|J\right|-b\cdot\left|J\right|},
\end{align*}
and therefore the above random variable is $(\sigma_{Y}+\frac{\epsCns}{c})$-sparse,
as required.
\end{myproof}

\subsection{\label{subsec:proof-of-light}Proof of \ref{probability-of-redeemable}}

The proof of \ref{probability-of-redeemable} consist of two parts.
In the first part, we prove that for any $y_{J}$ there are only few
$x$ such that $y_{J}$ is heavy for that $x$. In the second part,
we complete the proof using an averaging and bucketing argument. The
first part is encapsulated in the following result.
\begin{proposition}
\label{refined-weakly-sparsifying-upper-bound}Let $\gamma\geq\frac{2}{c}$
and assume that $\sigma_{X}+\sigma_{Y}\leq1-\frac{\cnsRefinedWeaklySparsifying}{c}-\gamma$.
Then, for every $J\subseteq\left[n\right]$ and for every~$y_{J}\in\D^{J}$,
the probability that $X$ takes an almost uniform value~$x$ such
that $y_{J}$ is $t$-heavy for $x$ is at most $2^{-\gamma\cdot\Delta\cdot\left|J\right|-2t}$.
\end{proposition}

The proof of \ref{refined-weakly-sparsifying-upper-bound} is given
in \ref{subsec:Proof-of-sparsifying}, and the rest of this section
is devoted to proving \ref{probability-of-redeemable} using \ref{refined-weakly-sparsifying-upper-bound}.
An important point about the proposition is that we get stronger bounds
on the probability for higher values of $t$, that is, for values
$y$ that violate the sparsity more strongly.

It is tempting to try and deduce \ref{probability-of-redeemable}
directly from \ref{refined-weakly-sparsifying-upper-bound} by an
averaging argument. Such an argument would first say that, for all
$y$, there is only a small probability that $X$ takes a value for
which $y$ is heavy by \ref{refined-weakly-sparsifying-upper-bound}.
Thus, for most values of $x$, the probability of choosing $y$ such
that $y$ is heavy for $x$ should be small as well. There is, however,
a significant obstacle here: in order to prove \ref{probability-of-redeemable},
we need to bound probability that $Y_{J}$ is heavy for~$x$ with
respect to a distribution that \emph{depends on~$x$}, namely, the
distribution $Y_{J}$ conditioned on the value of $g^{I}\left(x_{I},Y_{I}\right)$.
In contrast, the naive averaging argument assumes that $X$ and~$Y$
are independent. This complication renders a naive averaging argument
impossible. 

In order to overcome this obstacle, we note that conditioning on $Z_{I}^{x}=z_{I}$
can increase the probabilities of getting some $y_{J}$. This increase
in probability is characterize by the maximal $t$ such that $y_{J}$
is $t$-heavy for~$x$. Thus, for each $x$ we consider all $y_{J}$
such that $y_{J}$ is $t$-heavy for~$x$, and place them into buckets
according to the value of~$t$. Then, we bound the weight of each
bucket separately, while making use of the fact that \ref{refined-weakly-sparsifying-upper-bound}
provides a stronger upper bound for larger values of~$t$. Using
this bucketing scheme turns out to be sufficient for the averaging
argument to go through. We now prove \ref{probability-of-redeemable}
restated below. 
\begin{restated}{\ref{probability-of-redeemable}}
Assume that $\sigma_{X}+2\cdot\sigma_{Y}\leq1-\frac{\cnsProbabilityOfRedeemable}{c}-\gamma-\alpha$.
For every $J\subseteq\left[n\right]$, the probability that $X$ takes
an almost uniform value~$x$ that is not $\alpha$-light with respect
to~$J$ is at most $2^{-\gamma\cdot\Delta\cdot\left|J\right|}$.
\end{restated}

\begin{myproof}
Let $J\subseteq\left[n\right]$, and let $\cX$ and $\cY_{J}$ denote
the supports of~$X$ and $Y_{J}$ respectively. For every $x\in\cX$
and $y_{J}\in\cY_{J}$, let $t(x,y_{J},I,z_{I})\in\mathbb{R}$ be
the number for which 
\[
\Pr\left[Y_{J}=y_{J}\mid Z_{I}^{x}=z_{I}\right]=2^{(\sigma_{Y}+\frac{\epsCns}{c})\cdot\Delta\cdot\left|J\right|-b\cdot\left|J\right|+t_{x,y_{J}}-1}.
\]
We define $t_{x,y_{J}}$ as the maximum of $t(x,y_{J},I,z_{I})$ over
all possible $I,z_{I}$. Next, consider a table whose rows and columns
are indexed by $\cX$ and~$\cY_{J}$ respectively. For every row~$x\in\cX$
and column~$y_{J}\in\cY_{J}$, we set the corresponding entry to
be
\[
\ent(x,y_{J})\defeq\begin{cases}
2^{(\sigma_{Y}+\frac{\epsCns}{c})\cdot\Delta\cdot\left|J\right|-b\cdot\left|J\right|+t_{x,y_{J}}-1} & t_{x,y_{J}}>0\text{ and \ensuremath{x} is almost uniform}\\
0 & \text{otherwise}.
\end{cases}
\]
Now we use a bucketing argument. We take each pair $(x,y_{J})$ and
put it in the bucket that is labeled by $\left\lceil t_{x,y_{J}}\right\rceil $.
Then, for each $y_{J}$, we upper bound the probability of $(X,y_{J})$
to be in each bucket. Let $\gamma'=\gamma+\sigma_{Y}+\alpha+\frac{\epsCnsPropRed}{c}$.
Note that when $\left\lceil t_{X,y_{J}}\right\rceil =t$, we can bound
$t_{X,y_{J}}$ from below by $t-1$. Then, by applying \ref{refined-weakly-sparsifying-upper-bound}
with $\gamma=\gamma'$, we get that for every~$y_{J}$ and every~$t\in\Z_{>0}$
it holds that 
\begin{equation}
\Pr\left[\left\lceil t_{X,y_{J}}\right\rceil =t\text{ and }X\text{ is almost uniform}\right]\le2^{-\gamma'\cdot\Delta\cdot\left|J\right|-2(t-1)}.\label{eq:Bucket-bound}
\end{equation}
Therefore, for every $y_{J}\in\cY_{J}$, the expected entry in the
$y_{J}$-th column (over the random choice of~$X$) is
\begin{align*}
\E\left[\ent(X,y_{J})\right]= & \sum_{t=1}^{\infty}\left(\Pr\left[\left\lceil t_{X,y_{J}}\right\rceil =t\text{ and }X\text{ is almost uniform}\right]\cdot\right.\\
 & \left.\E\left[\ent(X,y_{J})\mid\left\lceil t_{X,y_{J}}\right\rceil =t\text{ and }X\text{ is almost uniform}\right]\right)\\
\leq & 2^{(\sigma_{Y}+\frac{\epsCns}{c})\cdot\Delta\cdot\left|J\right|-b\cdot\left|J\right|-1}\cdot\sum_{t=1}^{\infty}\Pr\left[\left\lceil t_{X,y_{J}}\right\rceil =t\text{ and }X\text{ is almost uniform}\right]\cdot2^{t}\\
\le & 2^{(\sigma_{Y}+\frac{\epsCns}{c})\cdot\Delta\cdot\left|J\right|-b\cdot\left|J\right|-1}\cdot\sum_{t=1}^{\infty}2^{-\gamma'\cdot\Delta\cdot\left|J\right|-2(t-1)}\cdot2^{t} & \text{(\ref{eq:Bucket-bound})}\\
= & 2^{(\sigma_{Y}+\frac{\epsCns}{c})\cdot\Delta\cdot\left|J\right|-b\cdot\left|J\right|-1}\cdot2^{-\gamma'\cdot\Delta\cdot\left|J\right|+2}\cdot\sum_{t=1}^{\infty}2^{-t}\\
\le & 2^{(\sigma_{Y}+\frac{\epsCns}{c}-\gamma'+\frac{1}{c})\cdot\Delta\cdot\left|J\right|-b\cdot\left|J\right|}\\
= & 2^{(\sigma_{Y}+\frac{\epsCns}{c}-\gamma-\sigma_{Y}-\alpha-\frac{\epsCnsPropRed}{c}+\frac{1}{c})\cdot\Delta\cdot\left|J\right|-b\cdot\left|J\right|} & \text{(definition of \ensuremath{\gamma'})}\\
\leq & 2^{-(\gamma+\alpha+\frac{2}{c})\cdot\Delta\cdot\left|J\right|-b\cdot\left|J\right|}.
\end{align*}
It follows that the expected sum of a random row of the table (over
the random choice of~$X$) is
\begin{align*}
\E\left[\sum_{y_{J}\in\cY_{J}}\ent(X,y_{J})\right] & =\sum_{y_{J}\in\cY_{J}}\E\left[\ent(X,y_{J})\right]\\
 & \le\sum_{y_{J}\in\cY_{J}}2^{-(\gamma+\alpha+\frac{2}{c})\cdot\Delta\cdot\left|J\right|-b\cdot\left|J\right|}\\
 & =2^{-(\gamma+\alpha+\frac{2}{c})\cdot\Delta\cdot\left|J\right|}.
\end{align*}
By Markov's inequality, the probability that the sum of the $X$-th
row is more than $2^{-\left(\alpha+\frac{2}{c}\right)\cdot\Delta\left|J\right|}$
is upper bounded by $2^{-\gamma\cdot\Delta\cdot\left|J\right|}$.
We now prove that if a value~$x\in\cX$ is almost uniform and the
sum in the $x$-th row is at most $2^{-\left(\alpha+\frac{2}{c}\right)\cdot\Delta\left|J\right|}$,
then $x$ is $\alpha$-light with respect to~$J$, and this will
finish the proof of the proposition.

Let $x\in\cX$ be such a value. We prove that $x$ is $\alpha$-light
with respect to~$J$. Let $I\subseteq\left[n\right]-J$ and let $z_{I}\in\B^{I}$.
We would like to prove that
\[
\Pr\left[Y_{J}\text{ is heavy for \ensuremath{x} and \ensuremath{z_{I}}}\mid Z_{I}^{x}=z_{I}\right]\le2^{-\alpha\Delta}\cdot\left(\frac{1}{2n}\right)^{\left|J\right|}.
\]
Observe that for every heavy $y_{J}$, it holds that $\Pr\left[Y_{J}=y_{J}\mid Z_{I}^{x}=z_{I}\right]\le\ent(x,y_{J})$.
It follows that
\begin{align*}
 & \Pr\left[Y_{J}\text{ is heavy for \ensuremath{x} and \ensuremath{z_{I}}}\mid Z_{I}^{x}=z_{I}\right]\\
\le & \sum_{y_{J}\text{ is heavy for \ensuremath{x} and \ensuremath{z_{I}}}}\Pr\left[Y_{J}=y_{J}\mid Z_{I}^{x}=z_{I}\right]\\
\le & \sum_{y_{J}}\ent(x,y_{J})\\
\le & 2^{-\left(\alpha+\frac{2}{c}\right)\cdot\Delta\left|J\right|}\\
\le & 2^{-\alpha\Delta}\cdot\left(\frac{1}{2n}\right)^{\left|J\right|} & \text{(\ensuremath{\left|J\right|\geq1,\Delta\ge c\log n})}
\end{align*}
as required.
\end{myproof}

\subsection{\label{subsec:Proof-of-sparsifying}Proof of \ref{refined-weakly-sparsifying-upper-bound}}

In this section, we prove \ref{refined-weakly-sparsifying-upper-bound},
restated next.
\begin{restated}{\ref{refined-weakly-sparsifying-upper-bound}}
Let $\gamma\geq\frac{2}{c}$ and assume that $\sigma_{X}+\sigma_{Y}\leq1-\frac{\cnsRefinedWeaklySparsifying}{c}-\gamma$.
Then, for every $J\subseteq\left[n\right]$ and for every~$y_{J}\in\D^{J}$,
the probability that $X$ takes an almost uniform value~$x$ such
that $y_{J}$ is $t$-heavy for $x$ is at most $2^{-\gamma\cdot\Delta\cdot\left|J\right|-2t}$.
\end{restated}

\ref{refined-weakly-sparsifying-upper-bound} is essentially a more
refined version of the analysis in~\cite{CFKMP19}. An important
point about this proposition is that it gives stronger bounds for
larger values of~$t$. The proof of \ref{refined-weakly-sparsifying-upper-bound}
follows closely the equivalent proof in \cite{CFKMP19} while taking
the parameter $t$ into consideration. The proof consists of three
main steps: first, we use Bayes' formula to reduce the upper bound
of \ref{refined-weakly-sparsifying-upper-bound} to upper bounding
the probability of a related type of values, called \emph{skewing}~values;
then, we use Vazirani's lemma to reduce the latter task to the task
of the upper bounding the biases of $Z_{I}^{x}$. Finally, we upper
bound the biases of $Z_{I}^{x}$ using the low deficiency of~$X$
and~$Y$ and the discrepancy of~$g$. We start by formally defining
skewing values,\emph{ }and then establish their connection to heavy
values.
\begin{definition}
Let $J\subseteq\left[n\right],y_{J}\in\mathrm{supp}(Y_{J})$, and
$I\subseteq\left[n\right]-J$. Let $e(y_{J})$ be the real number
that satisfies. 
\[
\Pr\left[Y_{J}=y_{J}\right]=2^{\sigma_{Y}\cdot\Delta\cdot\left|J\right|-b\cdot\left|J\right|-e(y_{J})}
\]
 We note that $e\left(y_{J}\right)$ is non-negative as we assume
that $Y$ is $\sigma_{Y}$-sparse. We say that $x$ is $t$-\emph{skewing}
for $y_{J}$ (with respect to $I$) if
\[
\Dm\left(Z_{I}^{x}\mid Y_{J}=y_{J}\right)>\epsCns\cdot\log n\cdot\left|J\right|+e\left(y_{J}\right)+t-2
\]
\end{definition}

\begin{proposition}
\label{sparsifying to skewing}Let $x\in\D^{n}$, $J\subseteq\left[n\right]$,
and $y_{J}\in\D^{J}$. If $y_{J}$ is $t$-heavy for $x$ and is almost
uniform then $x$ is $t$-skewing for $y_{J}$ (with respect to some
set~$I$).
\end{proposition}

\begin{myproof}
By the assumption that $y_{J}$ is $t$-heavy for $x$, there exist
$I\subseteq\left[n\right]$ and $z_{I}\in\D^{I}$ such that
\begin{equation}
\Pr\left[Y_{J}=y_{J}\mid Z_{I}^{x}=z_{I}\right]>2^{(\sigma_{Y}+\frac{\epsCns}{c})\cdot\Delta\cdot\left|J\right|+t-b\cdot\left|J\right|-1}.\label{eq:sparsifying-to-skewing}
\end{equation}
We show that $x$ is $t$-skewing for $y_{J}$ with respect to $I$.
It follows that
\begin{align*}
\Pr\left[Z_{I}^{x}=z_{I}\mid Y_{J}=y_{J}\right] & =\frac{\Pr\left[Y_{J}=y_{J}\mid Z_{I}^{x}=z_{I}\right]\cdot\Pr\left[Z_{I}^{x}=z_{I}\right]}{\Pr\left[Y_{J}=y_{J}\right]} & \text{(Bayes' formula)}\\
 & >\frac{2^{(\sigma_{Y}+\frac{\epsCns}{c})\cdot\Delta\cdot\left|J\right|+t-b\cdot\left|J\right|-1}\cdot\Pr\left[Z_{I}^{x}=z_{I}\right]}{\Pr\left[Y_{J}=y_{J}\right]} & \text{(\ref{eq:sparsifying-to-skewing})}\\
 & \ge\frac{2^{(\sigma_{Y}+\frac{\epsCns}{c})\cdot\Delta\cdot\left|J\right|+t-b\cdot\left|J\right|-1}\cdot2^{-\left|I\right|-1}}{\Pr\left[Y_{J}=y_{J}\right]} & \text{(\ensuremath{x} is almost uniform)}\\
 & =\frac{2^{(\sigma_{Y}+\frac{\epsCns}{c})\cdot\Delta\cdot\left|J\right|+t-b\cdot\left|J\right|-1}\cdot2^{-\left|I\right|-1}}{2^{\sigma_{Y}\cdot\Delta\cdot\left|J\right|-b\cdot\left|J\right|-e(y_{J})}} & \text{(definition of \ensuremath{e(y_{J})})}\\
 & =2^{\epsCns\cdot\log n\cdot\left|J\right|+e\left(y_{J}\right)+t-2-\left|I\right|} & \text{(\ensuremath{\Delta\ge c\cdot\log n}).}
\end{align*}
The latter inequality implies that
\[
\Dm\left(Z_{I}^{x}\mid Y_{J}=y_{J}\right)>\epsCns\cdot\log n\cdot\left|J\right|+e\left(y_{J}\right)+t-2.
\]
as required.
\end{myproof}
Next, we formally define biasing values and relate them to skewing
values via the usage of Vazirani lemma, thus allowing us to focus
on the biases. Informally, biasing values are values~$x$ such that
when conditioning on $Y_{J}=y_{J}$, the bias of $g^{\oplus S}\left(x_{S},Y_{S}\right)=\bigoplus_{i\in S}Z_{i}^{x}$
is too high for some~$S\subseteq I$.
\begin{definition}
Let $J\subseteq\left[n\right]$ and let $y_{J}\in\D^{J}$. We say
that $x$ is \emph{$t$-biasing} for $y_{J}$ if there exists a set~$S\subseteq\left[n\right]-J$
such that $\left|S\right|\geq\epsCns\cdot\left|J\right|+\frac{t+e(y_{J})-3}{\log n}$
and 
\[
\b\left(g^{\oplus S}\left(x_{S},Y_{S}\right)\mid Y_{J}=y_{J}\right)>\left(2n\right)^{-\left|S\right|}.
\]
\end{definition}

\begin{proposition}
\label{not biasing to not skewing}Let $x\in\D^{n}$, let $J\subseteq\left[n\right]$,
and let $y_{J}\in\D^{J}$. If $x$ is not $t$-biasing for~$y_{J}$
then~$x\text{ is not }t$-skewing for $y_{J}$. 
\end{proposition}

\begin{myproof}
Let $x,J,y_{J}$ be as in the proposition, and assume that $x$ is
not $t$-biasing for~$y_{J}$. We prove that $x\text{ is not }t$-skewing
for $y_{J}$. To this end, we prove that for every $I\subseteq\left[n\right]-J$
it holds that
\[
\Dm\left(Z_{I}^{x}\mid Y_{J}=y_{J}\right)\le\epsCns\cdot\log n\cdot\left|J\right|+e\left(y_{J}\right)+t-2.
\]
Let $I\subseteq\left[n\right]-J$. By assumption, $x$ is not $t$-biasing
for~$y_{J}$. Therefore we can apply the second variant of Vazirani's
lemma (\ref{lem:second vazirani}) to the random variable $Z_{I,x_{I}}\mid Y_{J}=y_{J}$
and obtain that
\begin{align*}
\Dm\left(Z_{I}^{x}\mid Y_{J}=y_{J}\right) & \le\left(\epsCns\cdot\left|J\right|+\frac{t+e(y_{J})-3}{\log n}\right)\cdot\log\left|I\right|+1\\
 & \le\left(\epsCns\cdot\left|J\right|+\frac{t+e(y_{J})-3}{\log n}\right)\cdot\log n+1\\
 & =\epsCns\cdot\log n\cdot\left|J\right|+t+e(y_{J})-2,
\end{align*}
as required.
\end{myproof}
We finally prove \ref{refined-weakly-sparsifying-upper-bound}, restated
next.
\begin{restated}{\ref{refined-weakly-sparsifying-upper-bound}}
Let $\gamma\geq\frac{2}{c}$ and assume that $\sigma_{X}+\sigma_{Y}\leq1-\frac{\cnsRefinedWeaklySparsifying}{c}-\gamma$.
Then, for every $J\subseteq\left[n\right]$ and for every~$y_{J}\in\D^{J}$,
the probability that $X$ takes an almost uniform value~$x$ such
that $y_{J}$ is $t$-heavy for $x$ is at most $2^{-\gamma\cdot\Delta\cdot\left|J\right|-2t}$.
\end{restated}

\begin{myproof}
Let $J\subseteq\left[n\right]$ and let $y_{J}\in\D^{J}$. We first
observe that it suffices to prove that with probability at least $1-2^{-\gamma\cdot\Delta\cdot\left|J\right|-2t}$,
the random variable~$X$ takes a value~$x$ that is not $t$-biasing
for~$y_{J}$. Indeed, if $x$ is a value that is not $t$-biasing
for~$y_{J}$, then by \ref{not biasing to not skewing} it is not
$t$-skewing for~$y_{J}$, and then by \ref{sparsifying to skewing}
it cannot be the case that $y_{J}$ is $t$-heavy for $x$ and that
$x$ is almost uniform. It remains to upper bound the probability
that $x$ is $t$-biasing for~$y_{J}$.

We start by upper bounding the probability that $X$ takes a value~$x$
such that
\[
\b\left(g^{\oplus S}\left(x_{S},Y_{S}\right)\mid Y_{J}=y_{J}\right)>\left(2n\right)^{-\left|S\right|}
\]
for some fixed non-empty set~$S\subseteq\left[n\right]-J$ such that
$\left|S\right|\geq\epsCns\cdot\left|J\right|+\frac{t+e(y_{J})-3}{\log n}$
(the case where $S$ is empty is trivial). Let $S$ be such a set.
In order to upper bound the latter probability, we use \ref{discrepancy-XOR-sampling},
which in turn requires us to upper bound the deficiencies $\Dm(X_{S})$
and $\Dm(Y_{S}\mid Y_{J}=y_{J})$. By assumption, we know that~$\Dm(X_{S})\le\sigma_{X}\cdot\Delta\cdot\left|S\right|$.
We turn to upper bound $\Dm(Y_{S}\mid Y_{J}=y_{J})$. For every $y_{S}\in\D^{S}$,
it holds that
\begin{align*}
\Pr\left[Y_{S}=y_{S}\mid Y_{J}=y_{J}\right] & =\frac{\Pr\left[Y_{S\cup J}=y_{S\cup J}\right]}{\Pr\left[Y_{J}=y_{J}\right]}\\
 & =\frac{\Pr\left[Y_{S\cup J}=y_{S\cup J}\right]}{2^{\sigma_{Y}\cdot\Delta\cdot\left|J\right|-b\cdot\left|J\right|-e(y_{J})}} & \text{(Definition of \ensuremath{e(y_{J})})}\\
 & \le\frac{2^{\sigma_{Y}\cdot\Delta\cdot(\left|S\right|+\left|J\right|)-b\cdot(\left|S\right|+\left|J\right|)}}{2^{\sigma_{Y}\cdot\Delta\cdot\left|J\right|-b\cdot\left|J\right|-e(y_{J})}} & \text{(\ensuremath{Y} is \ensuremath{\sigma_{Y}}-sparse)}\\
 & =2^{\sigma_{Y}\cdot\Delta\cdot\left|S\right|+e(y_{J})-b\cdot\left|S\right|}
\end{align*}
and thus
\[
\Dm(Y_{S}\mid Y_{J}=y_{J})\le\sigma_{Y}\cdot\Delta\cdot\left|S\right|+e(y_{J}).
\]
By our assumption on the size of~$S$, it follows that
\[
e(y_{J})\le\log n\cdot\left|S\right|+3\le4\cdot\log n\cdot\left|S\right|\le\frac{4}{c}\cdot\Delta\cdot\left|S\right|
\]
and therefore
\begin{align*}
D(X_{S})+\Dm(Y_{S}\mid Y_{J}=y_{J}) & \le(\sigma_{X}+\sigma_{Y}+\frac{4}{c})\cdot\Delta\cdot\left|S\right|\\
 & \le(1-\frac{11}{c}-\gamma)\cdot\Delta\cdot\left|S\right| & \text{(\ensuremath{\sigma_{X}+\sigma_{Y}\leq1-\frac{\cnsRefinedWeaklySparsifying}{c}-\gamma})}\\
 & =\left(\Delta-\frac{7\Delta}{c}-\gamma\Delta-\frac{4\Delta}{c}\right)\cdot\left|S\right|\\
 & \le\left(\Delta-7-\gamma\Delta-2\log n-2\right)\cdot\left|S\right|.
\end{align*}
Now, by applying \ref{discrepancy-XOR-sampling} with $\gamma=\gamma\Delta+\log n+1$
and $\lambda=\log n+1$, it follows that the probability that $X$
takes a value~$x$ such that 
\[
\b\left(g^{\oplus S}\left(x_{S},Y_{S}\right)\mid Y_{J}=y_{J}\right)>\left(2n\right)^{-\left|S\right|}
\]
is at most 
\[
2^{-\gamma\cdot\Delta\cdot\left|S\right|}\cdot\frac{1}{\left(2n\right)^{\left|S\right|}},
\]
where the inequality holds since $S$ is assumed to be non-empty.
By taking union bound over all relevant sets~$S$, it follows that
the probability that $X$~takes a value~$x$ that is $t$-biasing
for~$y_{J}$ is at most
\begin{align*}
 & \sum_{S\subseteq\left[n\right]:\left|S\right|\ge\epsCns\cdot\left|J\right|+\frac{t+e(y_{J})-3}{\log n}}2^{-\gamma\cdot\Delta\cdot\left|S\right|}\cdot\frac{1}{\left(2n\right)^{\left|S\right|}}\\
 & \le\sum_{S\subseteq\left[n\right]:\left|S\right|\ge\epsCns\cdot\left|J\right|+\frac{t-3}{\log n}}2^{-\gamma\cdot\Delta\cdot\left|S\right|}\cdot\frac{1}{\left(2n\right)^{\left|S\right|}} & \text{(\ensuremath{e(y_{J})\ge0})}\\
 & \le2\cdot2^{-\gamma\cdot\Delta\cdot\left(\epsCns\cdot\left|J\right|+\frac{t-3}{\log n}\right)}\cdot\frac{1}{2} & \text{(\ref{fact:binomal_sum})}\\
 & \le2^{-\gamma\cdot\Delta\cdot\left(\epsCns\left|J\right|+\frac{t-3}{\log n}\right)}\\
 & \le2^{-\gamma\cdot\Delta\left|J\right|-\gamma\cdot\Delta\frac{t}{\log n}} & \text{(\ensuremath{\left|J\right|\ge1,\log n\geq1})}\\
 & \leq2^{-\gamma\cdot\Delta\cdot\left|J\right|-2t} & \text{(\ensuremath{\frac{\gamma\cdot\Delta}{\log n}\geq\frac{2\Delta}{c\log n}\geq2})}
\end{align*}
as required.
\end{myproof}

\section{\label{sec:deterministic-lifting}The deterministic lifting theorem}

In this section we prove the deterministic part of our main theorem,
restated below. 
\begin{theorem}[Deterministic lifting theorem]
There exists a universal constant~$c\in\N$ such that the following
holds. Let $\mathcal{S}$ be a search problem that takes inputs from
$\B^{n}$ and let $g:\Lambda\times\Lambda\to\B$ be an arbitrary function
such that $\Delta(g)\ge c\cdot\log n$. Then there is a deterministic
decision tree T that solves $\mathcal{S}$ with complexity $O\left(\frac{D^{cc}\left(\mathcal{S}\circ g^{n}\right)}{\Delta(g)}\right)$.
\end{theorem}

In what follows, we fix~$\Pi$ to be an optimal deterministic protocol
for $\cS\circ g^{n}$ and let~$\dcc(\cS\circ g^{n})$ be its complexity.
We construct a decision tree $T$ that solves~$\cS$ using $O(\frac{\dcc(\cS\circ g^{n})}{\Delta})$
queries. We construct the decision tree~$T$ in \ref{subsec:deterministic-construction},
prove its correctness in \ref{subsec:deterministiccorrectness}, and
analyze its query complexity in \ref{subsec:deterministic-query-complexity}.

\paragraph*{High-level idea of the proof.}

Intuitively, given an input~$z\in\B^{n}$, the decision tree~$T$
attempts to construct a full transcript~$\pi$ of~$\Pi$ that is
consistent with some input $(x,y)$ such that $g^{n}(x,y)=z$. Such
a transcript must output a solution in
\[
(\cS\circ g^{n})(x,y)=\cS\left(g^{n}(x,y)\right)=\cS(z)
\]
thus solving~$\cS$ on~$z$. The tree works iteratively, constructing
the transcript~$\pi$ message-by-message. Throughout this process,
the tree maintains random variables $X,Y$ that are distributed over
the inputs of Alice and Bob such that the input~$(X,Y)$ is consistent
with~$\pi$. In particular, the tree preserves the invariant that
the variables~$(X,Y)$ are $(\rho,\sigma_{X},\sigma_{Y})$-structured
where $\rho$ is a restriction that is consistent with $z$ and for
some values $\sigma_{X}$ and $\sigma_{Y}$. The restriction $\rho$
keep track which of the input bits of $z$ have been queried, and
is maintained accordingly by the tree. By the uniform marginals lemma
(\ref{lem:uniform-marginal-lemma}), we get that the protocol~$\Pi$
cannot distinguish between the distribution of~$(X,Y)$ and the same
distribution conditioned on~$g^{n}(X,Y)=z$. Hence, when the protocol
ends, $\pi$ must be consistent with an input in~$\left(g^{n}\right)^{-1}(z)$.

We now describe how the tree preserves the foregoing invariant. Suppose
without loss of generality that the invariant is violated because
$X_{I}$ is not $\sigma_{X}$-sparse (the case where $Y$ is not $\sigma_{Y}$-sparse
is treated similarly). Specifically, assume that the deficiency of~$X_{I}$
is too large for some set of coordinates~$I\subseteq\free(\rho)$.
To restore the structure, the tree queries~$z_{I}$ and conditions~$X$
and~$Y$ on $\text{\ensuremath{g^{I}(X_{I},Y_{I})=z_{I}}}$. The
conditioning on $\text{\ensuremath{g^{I}(X_{I},Y_{I})=z_{I}}}$, however,
could increase the deficiency of $X$ and $Y$, which might violate
their structure. In order to avoid this, the tree conditions $X$
in advance on taking a safe value. Moreover, after the conditioning
on~ $g^{I}(X_{I},Y_{I})=z_{I}$ the tree recovers the density of
$Y$ by conditioning it on a high-probability event, as in the the
definition of a recoverable value (\ref{dangerous}).

In order to upper bound the query complexity of the tree, we keep
track of the deficiency $\Dm\left(X_{\free\left(\rho\right)},Y_{\free\left(\rho\right)}\right)$
throughout the simulation of the protocol. We prove that the messages
that the tree sends in the simulation increase the deficiencies by
at most~$O(\dcc(\cS\circ g^{n}))$ overall. On the other hand, whenever
the tree queries a set of coordinates~$I$, the deficiency decreases
by at least~$\Omega(\left|I\right|\cdot\Delta)$. As a result we
get that the tree cannot make more then $O\left(\frac{\dcc(\cS\circ g^{n})}{\Delta}\right)$
queries in total.

\subsection{\label{subsec:deterministic-construction}The construction of the
deterministic decision tree}

\FPeval{\epsCnsInEqOnSigma}{clip(\cnsMainLemma{}+2)}
\edef\cCns{200}
\edef\sigmaDeterministic{0.25}
\FPeval{\cnsLHSdeterminstic}{round(3*\sigmaDeterministic{}+\epsCns{}/\cCns{},2)}
\FPeval{\cnsRHSdeterminstic}{round(1-\epsCnsInEqOnSigma{}/\cCns{},3)}
\FPeval{\cnsQueyrCompDeterministic}{round(\sigmaDeterministic{}-3/\cCns{},3)}

In this section we describe the construction of the deterministic
decision tree $T$. For the construction we set $\sigma\defeq\frac{1}{4},\alpha\defeq\frac{1}{c}$
and $c=\cCns$.

\paragraph{\label{par:Invariants-and-assertions.}Invariants.}

The tree maintains a partial transcript $\pi$, a restriction $\rho$,
and two independent random variables $X,Y$ over~$\Lambda^{n}$ such
that the input $(X,Y)$ is consistent with~$\pi$. The tree works
iteratively. In each iteration the tree simulates a single round of
the protocol $\Pi$. The tree maintains the invariant that at the
beginning of each iteration, if it is Alice's turn to speak in the
simulated protocol, then $X$ and $Y$ are $(\rho,\sigma+\frac{\epsCns}{c},\sigma)$-structured.
If it is Bob's turn to speak, it is the other way around.

\paragraph{The algorithm of the tree.}

When $T$ starts the simulation, the tree sets the transcript $\pi$
to be empty , the restriction $\rho$ to $\left\{ *\right\} ^{n}$,
and $X,Y$ to uniform random variables over~$\Lambda^{n}$. It is
easy to verify that the invariant holds at the beginning of the simulation.
We now describe a single iteration of the tree. Without lose of generality,
we assume that this is Alice's turn. The tree performs the following
steps:
\begin{enumerate}
\item \label{enu:condition-on-not-dangerous}The tree conditions $X_{\free\left(\rho\right)}$
on taking an $\alpha$-safe value for $Y_{\free\left(\rho\right)}$. 
\item \label{enu:msg-choosing}Let $M\left(x,\pi\right)$ be the message
that Alice sends on partial transcript $\pi$ and input $x$. The
tree chooses a message $m$ such that $P\left[M\left(X,\pi\right)=m\right]\ge2^{-\left|m\right|}$,
adds it to $\pi$, and conditions $X$ on the event that $M\left(X,\pi\right)=m$.
We note that such a message $m$ must exist (see explanation below).
\item \label{enu:Density-violates}Let $I\subseteq\free\left(\rho\right)$
be a maximal set that violates the $\sigma$-sparsity of~$X_{\free\left(\rho\right)}$,
and let $x_{I}$ be a value such that $\Pr\left[X_{I}=x_{I}\right]>2^{\sigma\cdot\Delta\cdot\left|I\right|-b\cdot\left|I\right|}$.
The tree conditions~$X$ on $X_{I}=x_{I}$.
\item \label{enu:query-z}The tree query $z_{I}$ and sets $\rho_{I}$ to
$z_{I}$.
\item \label{enu:conditions-on-equale-z}The tree conditions $Y$ on $g^{I}\left(x_{I},Y_{I}\right)=z_{I}$.
\item \label{enu:discard-y}The tree conditions $Y$ on an event $\cE$
such that $Y\mid\cE$ is $\left(\sigma+\frac{\epsCns}{c}\right)$-sparse
and $\Pr\left[\cE\right]\geq\frac{1}{2}$. Such an event must exist
since the value $x$ is safe, and in particular, recoverable
\end{enumerate}
When the protocol~$\Pi$ halts, the tree $T$ halts as well and returns
the output of the transcript~$\pi$. A few additional comments are
in order:
\begin{itemize}
\item We need to explain why we never condition $X$ nor $Y$ on an event
with zero probability. Regarding Step~\ref{enu:condition-on-not-dangerous},
we need to prove that there exist $\alpha$-safe values. To do so
we apply \ref{lem:main-lemma} with $\gamma=\frac{1}{c}$ (and $\alpha$
as defined above). As required by the lemma, it holds that 
\[
\sigma_{X}+2\sigma_{Y}\leq3\sigma+\frac{\epsCns}{c}=\cnsLHSdeterminstic\leq\cnsRHSdeterminstic=1-\frac{\epsCnsInEqOnSigma}{c}=1-\frac{\cnsMainLemma}{c}-\gamma-\alpha.
\]
Therefore we know that the conditioning is on an event with probability
of at least $\text{\ensuremath{1-2^{-\frac{\Delta}{c}}>0}}$. Regarding
the conditioning at Step~\ref{enu:conditions-on-equale-z}, we note
that the value~$x_{I}$ is safe. Hence, $x_{I}$ is almost uniform
and the event $g^{I}\left(x_{I},Y_{I}\right)=z_{I}$ is not empty.
\item In Step~\ref{enu:msg-choosing}, we claimed that there must be a
message $m$ such that $\Pr\left[M\left(X,\pi\right)=m\right]\geq2^{-\left|m\right|}$.
To this end, we note that the set of possible messages of Alice must
be a prefix-free code. Otherwise, there would be two messages $m_{1},m_{2}$
such that $m_{1}$ is prefix of $m_{2}$. In case that Bob got the
message $m_{1}$ from Alice, Bob will not know whether Alice sent
$m_{2}$ but the remaining bits have not arrived yet, or Alice sent
the message $m_{1}$. We complete the proof of the claim by using
\ref{Fact:prefix-code} that ensures that such a message exists.
\item We need to prove that at the end of Step~\ref{enu:discard-y}, the
variable $X$ is $\sigma$-sparse and $Y$ is $\left(\sigma+\frac{\epsCns}{c}\right)$-sparse
as required by the invariants. It holds that $X$ is $\sigma$-sparse
by \ref{pro:density-restoring-fixing}, and $Y$ is $\left(\sigma+\frac{\epsCns}{c}\right)$-sparse
by the definition of the event $\cE$ at end of Step~\ref{enu:discard-y}.
\end{itemize}

\subsection{\label{subsec:deterministiccorrectness}Correctness of the deterministic
decision tree}

We now prove that the tree always returns a correct answer, that is,
given an input $z$ the tree outputs an answer $o$ such that $\left(z,o\right)\in\cS$.
Let~$X,Y$ and~$\pi$ be the distributions and transcript at the
end of the simulation. Let $o$ be the output associated with the
transcript $\pi$. As asserted in \ref{par:Invariants-and-assertions.},
every $\left(x,y\right)\in\text{supp}\left(X,Y\right)$ is consistent
with the transcript $\pi$. By the fact that $\Pi$ solves $\cS\circ g^{n}$,
for every pair $\left(x,y\right)\in\text{supp}\left(X,Y\right)$ it
holds that
\[
\left(g^{n}\left(x,y\right),o\right)\in\mathcal{S}.
\]
To complete the proof of correctness, we show that there exists a
pair $\left(x,y\right)\in\text{supp}\left(X,Y\right)$ such that $g^{n}\left(x,y\right)=z$
and therefore $\left(z,o\right)\in\mathcal{S}$. As $X,Y$ are $\left(\rho,\sigma_{X},\sigma_{Y}\right)$-structured,
it is guaranteed that 
\[
g^{\text{fix}\left(\rho\right)}\left(X_{\text{fix}\left(\rho\right)},Y_{\text{fix}\left(\rho\right)}\right)=z_{\text{fix}\left(\rho\right)}
\]
 with probability $1$. Regarding the free part, we apply \ref{multiplicative-uniformity}
with $\gamma=\frac{1}{c}$. (so the requirement $2\sigma+\frac{\epsCns}{c}\leq1-\frac{8}{c}-\gamma$
holds) and we get that
\[
\Pr\left[g^{\free\left(\rho\right)}\left(X_{\free\left(\rho\right)},Y_{\free\left(\rho\right)}\right)=z_{\free\left(\rho\right)}\right]>0.
\]
Therefore there must be some pair $\left(x,y\right)\in\text{supp}\left(X,Y\right)$
such that $g^{\free\left(\rho\right)}\left(x_{\free\left(\rho\right)},y_{\free\left(\rho\right)}\right)=z_{\free\left(\rho\right)}$.
Both facts together ensure us that there exist a pair $\left(x,y\right)\in\text{supp}\left(X,Y\right)$
such that $g^{n}\left(x,y\right)=z$, as required.

\subsection{\label{subsec:deterministic-query-complexity}Query complexity of
the deterministic decision tree}

In this section, we upper bound the query complexity of the tree by
$O\left(\frac{D^{cc}\left(\cS\circ g^{n}\right)}{\Delta}\right)$.
The upper bound is proven using a potential argument. We define our
potential function to be the deficiency of $X,Y$, i.e.,
\[
\Dm\left(X_{\free\left(\rho\right)},Y_{\free\left(\rho\right)}\right)=\Dm\left(X_{\free\left(\rho\right)}\right)+\Dm\left(Y_{\free\left(\rho\right)}\right).
\]
We prove that whenever the simulated protocol sends a message $m$,
the deficiency increases by at most~$O\left(\left|m\right|\right)$,
and that whenever the tree makes a query, the deficiency decreased
by at least~$\Omega\left(\Delta\right)$. The deficiency is equal
to zero at the beginning of the simulation and it is always non-negative
by \ref{fact:deficiency-nonnegative}. The length of the transcript
is bounded by $D^{cc}\left(\mathcal{S}\circ g^{n}\right)$, and therefore
we get a bound of $O\left(\frac{D^{cc}\left(\mathcal{S}\circ g^{n}\right)}{\Delta}\right)$
queries. 

We now analyze in detail the changes in the deficiency during a single
iteration of the tree step-by-step:
\begin{enumerate}
\item In Step~\ref{enu:condition-on-not-dangerous} the tree conditions
on the event that $X_{\free\left(\rho\right)}$ is safe. By applying
\ref{lem:main-lemma} with $\gamma=\frac{1}{c}$, we obtain that $\Pr\left[X_{\free\left(\rho\right)}\text{ is not safe}\right]\leq2^{-\frac{\Delta}{c}}\leq\frac{1}{2}$.
Therefore, it follows that $\Dm\left(X_{\free\left(\rho\right)}\right)$
increases at most by $1$ by \ref{fact:deficiency-cond}.
\item In Step~\ref{enu:msg-choosing}, the deficiency $\Dm\left(X_{\free(\rho)}\right)$
increases by at most $\log\frac{1}{\Pr\left[m\right]}$ by \ref{fact:deficiency-cond}.
As $m$ is chosen such that $\Pr\left[m\right]\ge2^{-\left|m\right|}$,
it holds that $\Dm\left(X_{\free(\rho)}\right)$ increases by at most
$\left|m\right|$
\item In Step~\ref{enu:Density-violates}, the deficiency $\Dm\left(X_{\free(\rho)}\right)$
increases by at most $b\cdot\left|I\right|-\sigma\cdot\Delta\cdot\left|I\right|$
by \ref{fact:deficiency-cond}.
\item In Step~\ref{enu:query-z}, we reduce $\free\left(\rho\right)$ by
$\left|I\right|$. The deficiency $\Dm\left(X_{\free\left(\rho\right)}\right)$
is decreased by $b\cdot\left|I\right|$ since~$X_{I}$ is a constant,
while $\Dm\left(Y_{\free\left(\rho\right)}\right)$ does not increase
by \ref{fact:deficiency-monotone}.
\item In Step~\ref{enu:conditions-on-equale-z}, we get that $\Dm\left(Y_{\free\left(\rho\right)}\right)$
increases by at most $\log\frac{1}{\Pr\left[g\left(x_{I},Y_{I}\right)=z_{I}\right]}$
by \ref{fact:deficiency-cond}. Since~$x_{I}$ is almost uniform,
we know that $\Pr\left[g\left(x_{I},Y_{I}\right)=z_{I}\right]\geq2^{-\left|I\right|-1}$,
and therefore $\Dm\left(Y_{\free\left(\rho\right)}\right)$ increases
by at most $\left|I\right|+1$.
\item In Step~\ref{enu:discard-y} the tree conditions $Y$ on the event
$\cE$ such that $\Pr\left[Y\in\cE\right]\geq\frac{1}{2}$. By \ref{fact:deficiency-cond}
we get that $\Dm(Y_{\free})$ increases by at most~$1$.
\end{enumerate}
At the end, we have that any message $m$ increase the deficiency
by at most $\left|m\right|+1\in O\left(\left|m\right|\right)$, and
for any set of queries $I$ the deficiency decreases by at least
\begin{align*}
b\cdot\left|I\right|-(b\cdot\left|I\right|-\sigma\cdot\Delta\cdot\left|I\right|)-\left|I\right|-2 & \geq\left(\sigma-\frac{3}{\Delta}\right)\cdot\Delta\cdot\left|I\right| & \text{(rearranging,\ensuremath{\left|I\right|}\ensuremath{\ensuremath{\geq}1})}\\
 & \geq\left(\sigma-\frac{3}{c}\right)\cdot\Delta\cdot\left|I\right| & \text{(\ensuremath{\Delta>c\log n})}\\
 & \in\Omega\left(\Delta\cdot\left|I\right|\right) & \text{(\ensuremath{\sigma-\frac{3}{c}=\cnsQueyrCompDeterministic} is a positive constant).}
\end{align*}

\section{\label{sec:randomized-lifting}The randomized lifting theorem}

\edef\sigmaCRandCns{2}
\edef\KpDeltaCns{2}
\edef\cnsCrand{1000}
\edef\sigmaMainCnsRand{0.1}
\edef\alphaCnsRand{0.1}
\FPeval{\sigmaCnsXAfterSafe}{clip(1+\epsCns{})}
\FPeval{\sigmaCnsOneXTwoY}
{clip(3*\sigmaCRandCns{}+\epsCns{})}
\FPeval{\cnsLHSrandom}{round(3*\sigmaMainCnsRand{}+\sigmaCnsOneXTwoY{}/\cnsCrand{},3)}
\FPeval{\cnsRHSrandom}{round(0.7-\cnsMainLemma{}/\cnsCrand{},3)}
\edef\KCnsPart{5}
\FPeval{\KCns}{clip(1+\KCnsPart{})}
\FPeval{\MaxDefTimes}{clip(2+\KCns{})}
\FPeval{\KDeltaCns}{clip(\KpDeltaCns{}+1)}
\FPeval{\MaxDefTimesTen}{clip(max(\MaxDefTimes{},\KDeltaCns{})/\sigmaMainCnsRand{})}In this section we prove the randomized lifting theorem. We start
by stating the following simulation result, which implies the lifting
theorem as a simple consequence. 
\begin{notation}
Let $\Pi$ be some protocol that takes inputs in~$\D^{n}\times\D^{n}$.
For every $z\in\B^{n}$ we let $\Pi_{z}^{\prime}$ denote the distribution
of transcripts of the protocol~$\Pi$ on uniformly random inputs
$X,Y$ conditioned on the event $g^{n}\left(X,Y\right)=z$. 
\end{notation}

\begin{theorem}
\label{thm:rand-sim}Let $g:\D\times\D\to\B$ be a function such that
$\Delta\left(g\right)\geq1000\log n$. Let $\Pi$ be a public-coin
randomized protocol that takes inputs from $\D^{n}\times\D^{n}$ and
uses at most $C$ bits of communication. Then, there is a randomized
decision tree given an input $z\in\B^{n}$, samples from a distribution
that is $\left(2^{-\frac{\Delta\left(g\right)}{20}}\cdot\left(1+C\right)\right)$-close
to the distribution $\Pi_{z}^{\prime}$ and makes at most $\MaxDefTimesTen\left(\frac{C}{\Delta\left(g\right)}+1\right)$
queries.
\end{theorem}

Before we prove \ref{thm:rand-sim}, we show how \ref{thm:rand-sim}
implies \ref{main-theorem}, restated next.
\begin{restated}{\ref{main-theorem}}[Randomized part]
 There exists a universal constant~$c$ such that the following
holds: Let $\cS$ be a search problem that takes inputs from~$\B^{n}$,
let $g:\Lambda\times\Lambda\to\B$ be an arbitrary function such that
$\Delta(g)\ge c\cdot\log n$, and let $\beta>0$. Then, it holds that
\[
\rcc_{\beta}(\cS\circ g^{n})\in\Omega\left(\left(\rdt_{\beta^{\prime}}(\cS)-\MaxDefTimesTen\right)\cdot\Delta(g)\right),
\]
where $\beta^{\prime}=\beta+2^{-\Delta(g)/50}$.
\end{restated}

\begin{myproof}
We choose $c=\cnsCrand$. Let $\Pi$ be an optimal protocol that solves
$S\circ g^{n}$ with error probability $\beta$, and denote by~$C$
the complexity of $\Pi$. In the case that $C\ge n\cdot\Delta\left(g\right)$,
the lower bound holds trivially, and we therefore assume that $C<n\cdot\Delta\left(g\right)$.
Let $T$ be the tree obtained by applying \ref{thm:rand-sim} to $\Pi$.
We construct a decision tree~$T^{\prime}$ for $\cS$ as follows:
on input $z$, the tree~$T^{\prime}$ simulates the tree $T$ on
$z$, thus obtaining a transcript $\pi$ of $\Pi$, and returns the
output associated with this transcript. For a transcript $\pi$ of~$\Pi$,
we denote by $\mathcal{O}\left(\pi\right)$ the output associated
with this transcript. By assumption, for every inputs $\left(x,y\right)\in\Lambda^{n}\times\Lambda^{n}$
such that $g^{n}\left(x,y\right)=z$, it holds that the output of
$\Pi$ on $x,y$ is in $\cS\left(z\right)$ with probability $1-\beta$
as $\cS\left(z\right)=\left(\cS\circ g^{n}\right)\left(x,y\right)$.
The error probability of $T^{\prime}$ on $z$ is

\begin{align*}
\Pr_{o\leftarrow T^{\prime}\left(z\right)}\left[\left(z,o\right)\notin\mathcal{S}\right] & =\Pr_{\pi\leftarrow T}\left[\left(z,\mathcal{O}\left(\pi\right)\right)\notin\mathcal{S}\right]\\
 & \leq\Pr_{\pi\leftarrow\Pi_{z}^{\prime}}\left[\left(z,\mathcal{O}\left(\pi\right)\right)\notin\mathcal{S}\right]+2^{-\frac{\Delta\left(g\right)}{20}}\cdot\left(1+C\right) & \text{(by \ref{thm:rand-sim})}\\
 & \leq\beta+2^{-\frac{\Delta\left(g\right)}{20}}\cdot\left(1+C\right)\\
 & \leq\beta+2^{-\frac{\Delta\left(g\right)}{20}}\cdot n\cdot\Delta\left(g\right)\\
 & \leq\beta+2^{-\frac{\Delta\left(g\right)}{20}}\cdot2^{\frac{\Delta\left(g\right)}{c}}\cdot\Delta\left(g\right) & \text{(\ensuremath{\Delta\left(g\right)\ge c\cdot\log n})}\\
 & \leq\beta+2^{-\frac{\Delta\left(g\right)}{50}}. & \text{(as \ensuremath{\Delta\left(g\right)\geq c=\cnsCrand})}
\end{align*}
where the second to last transition hold for sufficiently large (but
constant) $c$. Therefore, the tree~$T^{\prime}$ solves $\mathcal{S}$
with error probability at most $\beta^{\prime}$. Note that the query
complexity of $T^{\prime}$ is the same as the query complexity of
$T$. Therefore, by \ref{thm:rand-sim}, the query complexity of $T^{\prime}$
is at most $\MaxDefTimesTen\left(\frac{\rcc_{\beta}(\cS\circ g^{n})}{\Delta\left(g\right)}+1\right)$.
Together with the fact that $T^{\prime}$ solves $\cS$ we get that
\begin{align*}
R_{\beta^{\prime}}^{dt}\left(\cS\right) & \leq\MaxDefTimesTen\left(\frac{\rcc_{\beta}(\cS\circ g^{n})}{\Delta\left(g\right)}+1\right)
\end{align*}
and therefore
\begin{align*}
R_{\beta^{\prime}}^{dt}\left(\cS\right)-\MaxDefTimesTen & \leq\MaxDefTimesTen\left(\frac{\rcc_{\beta}(\cS\circ g^{n})}{\Delta\left(g\right)}\right)\\
\left(R_{\beta^{\prime}}^{dt}\left(\cS\right)-\MaxDefTimesTen\right)\cdot\Delta\left(g\right) & \leq\MaxDefTimesTen\left(\rcc_{\beta}(\cS\circ g^{n})\right)\\
\rcc_{\beta}(\cS\circ g^{n}) & \in\Omega\left(\left(R_{\beta^{\prime}}^{dt}\left(\cS\right)-\MaxDefTimesTen\right)\cdot\Delta\left(g\right)\right).\qedhere
\end{align*}
\end{myproof}
In the rest of this section we prove \ref{thm:rand-sim}. The proof
is organized as follow: We first introduce the algorithm of tree $T$
in \ref{subsec:Construction-rand-tree} and prove some basic facts
about the tree. Then, in \ref{subsec:Correctness-of-rand} we prove
the correctness of the tree, that is, we prove that the distribution
of the outputs of the tree given $z$ is indeed close to $\Pi_{z}^{\prime}$.
We end the section with proving the bound on the query complexity
in \ref{subsec:Query-complexity-rand-tree}.

\subsection{\label{subsec:Construction-rand-tree}Construction of the decision
tree}

The construction of the randomized decision tree is similar to the
deterministic tree but has several important differences. The key
difference is in the goal of the tree: in the randomized case, we
should sample a transcript from~$\Pi_{z}'$, whereas in the deterministic
case we only need to find some transcript in the support of~$\Pi_{z}'$.

As in the deterministic case, the tree maintains two random inputs
$X$ and $Y$. As a result of the key difference described above,
the tree cannot condition on low-probability events, as this can drastically
change the distribution of the returned transcript. This constraint
results in number of changes to the way the tree samples messages
and restores density. We now describe these changes:
\begin{itemize}
\item When the deterministic tree chooses a message for Alice in the simulation,
it is sufficient for the tree to choose some high probability message
$m$ and to condition $X$ on sending it. In contrast, the randomized
tree needs to sample a message that is close to the distribution of
the next message in~$\Pi_{z}^{\prime}$. In order to do so in the
case that Alice speaks, the randomized decision tree samples the message
$m$ by first sampling $x$ from $X$ and then simulating the protocol
on~$x$ to obtain~$m$. The case where it is Bob's turn to speak
is handled similarly. In \ref{subsec:Correctness-of-rand} we prove
that this distribution of the message is sufficiently close to its
distribution in~$\Pi_{z}'$ using the uniform marginals lemma. 
\item The change to the selection of messages in the previous item creates
a new problem. Denote by~$M$ be the random message that sampled
by the simulation. In the deterministic case, when the tree conditions
$X$ on $M=m$, the deficiency of $X$ grows by at most $\left|m\right|$,
as the choice of~$m$ guarantees that $\Pr\left[M=m\right]\geq2^{-\left|m\right|}$.
In the randomized case, on the other hand, this does not hold anymore,
since the chosen message~$m$ may have an arbitrarily low probability.
In order to resolve this issue, we maintain a counter~$\KMsg$ that
keep track of the sum $\sum_{m\in\pi}\log\frac{1}{\Pr\left[M=m\right]}$,
that is, the total increase in deficiency caused by sending messages.
The tree halts if the counter~$\KMsg$ surpasses~$C+\Delta$ at
any point. This way we ensure that sending messages contributes at
most~$C+\Delta$ to the deficiency. In \ref{subsec:Correctness-of-rand},
we prove that the tree halts in this way only with small probability,
and therefore this modification does not contribute much to the error
probability. 
\item In the deterministic case, the tree restores the density of the variable
$X$ by conditioning it on an event of the form $X_{I}=x_{I}$ for
some value $x_{I}$. Unfortunately, the event $X_{I}=x_{I}$ might
have too low probability. Instead, we use the density-restoring partition
(\ref{lem:density-restoring-partition}). Specifically the tree sample
some class $\cX_{j}$ of the partition and conditions on~$X\in\cX_{j}$. 
\item The change to the density-restoration procedure create a new problem.
Let $J$ be the a random variable of the partition class, and let
$j$ be the index of the class partition that was chosen. Using the
density-restoring partition increases the deficiency by an additional
term of $\log\frac{1}{\Pr\left[J\ge j\right]}$. As with the messages,
we create a counter $\KPrt$ that keeps track of the sum $\sum\log\frac{1}{\Pr\left[J\ge j\right]}$
and halt if $\KPrt>\KCnsPart C+\KpDeltaCns\Delta$. We show that $\KPrt>\KCnsPart C+\KpDeltaCns\Delta$
occurs with only with negligible probability.
\end{itemize}
The changes described above follow the previous works \cite{GPW17,CFKMP19}.
In particular, our construction closely follows the construction in
\cite{CFKMP19}. As in the deterministic case, the main difference
is the addition of density-restoring step for $Y$ at the end of the
iteration. This, in turn, requires a non-trivial addition to the proof
of correctness (see \ref{subsec:Proof-Of-Correctness-Without-Halting}).
We turn to formally describe the decision tree.

\paragraph*{Parameters.}

Let set $c=\cnsCrand,\sigma\defeq\frac{1}{10}+\frac{\sigmaCRandCns}{c}$
and $\alpha\defeq\frac{1}{10}$. Let $\Delta\defeq\log\frac{1}{\disc\left(g\right)}$
be such that $\Delta>c\log n$. Denote by $C$ the worst-case complexity
of the protocol $\Pi$.

\paragraph*{Assertions and Invariants.}

Throughout the simulation the tree maintains two independent random
variables~$X,Y\in\Lambda^{n}$ that are uniformly distributed over
some subsets~$\mathcal{X},\mathcal{Y}\subseteq\D^{n}$ respectively.
The tree also maintains a restriction $\rho$. It holds that the set
$\fix\left(\rho\right)$ is the set of all queried bits, and $\rho_{\fix\left(\rho\right)}=z_{\fix\left(\rho\right)}$.
The tree simulates $\Pi$ iteratively, where at each iteration the
tree simulates single round. At the beginning of each iteration, if
it is Alice's (respectively Bob's) turn to speak then $X,Y$ are $\left(\rho,\sigma+\frac{\epsCns}{c},\sigma\right)$-structured
(respectively $\left(\rho,\sigma,\sigma+\frac{\epsCns}{c}\right)$-structured).
Throughout the simulation, the tree maintains some partial transcript
$\pi$. At any point in the simulation, it holds that all pairs of
inputs $\left(x,y\right)$ in $\mathcal{X\times Y}$ are consistent
with the partial transcript $\pi$.

\paragraph*{The Algorithm Of The Tree.}

The tree starts by sampling public coins for the randomized protocol
$\Pi$ and fixing them. From this point on, the protocol can be thought
of as a deterministic protocol. The tree then sets initial values
to its variables. The distributions $X,Y$ are initialized to be uniform
over $\Lambda^{n}$. The transcript $\pi$ is initialized to the empty
transcript, the restriction $\rho$ to~$\left\{ *\right\} ^{n}$,
and the counters $\KMsg,\KPrt$ to $0$. We now describe a single
iteration of the tree. Without lose of generality we assume here that
it is Alice's turn to speak in $\pi$. An iteration where it is Bob's
turn to speak in $\pi$ is the same with the exception of swapping
the roles $X$ and $Y$.
\begin{enumerate}
\item \label{enu:condition on uniformity and revealing}The tree conditions
$X_{\free\left(\rho\right)}$ on taking a value $x$ that is $\alpha$-safe
for $Y_{\free\left(\rho\right)}$. 
\item \label{enu:sample-msg}Let the random variable $M$ be the message
that Alice sends on input $X$ and the current partial transcript
$\pi$. The tree samples a message $m$ according to $M$ and appends
it to $\pi$. Then, the tree adds $\log\frac{1}{\Pr\left[M=m\right]}$
to $\KMsg$ and conditions $X$ on $M=m$,. 
\item \label{enu:partion-of-X}Let $\mathcal{X}_{\free\left(\rho\right)}=\mathcal{X}^{1}\cup\dots\cup\mathcal{X}^{l}$
be the density-restoring partition we get from \ref{lem:density-restoring-partition}.
The tree chooses a random class $\cX^{j}$ such that it choose the
$i$-th class with probability $\Pr\left[X_{\free\left(\rho\right)}\in\mathcal{X}^{i}\right]$.
The tree then conditions $X$ on $X_{\free\left(\rho\right)}\in\mathcal{X}^{j}$.
Let $I_{j}$ and $x_{j}$ be the set of coordinates and value associated
with $\mathcal{X}^{j}$ as defined by \ref{lem:density-restoring-partition}.
\item \label{enu:adding-to-K_prt}Recall that
\[
p_{\geq i}\eqdef\Pr\left[X_{\free\left(\rho\right)}\in\mathcal{X}^{i}\cup\dots\cup\mathcal{X}^{l}\right],
\]
where the random variable $X$ here refers to the random variable
as in the start of Step \ref{enu:partion-of-X}. The tree adds $\log\frac{1}{p_{\geq j}}$
to $\KPrt$. 
\item \label{enu:halting-msg-and-prt}If $\KPrt>\KCnsPart C+\KpDeltaCns\Delta$
or $\KMsg>C+\Delta$ then the tree halts. 
\item \label{enu:query-I_j}The tree queries the coordinates in $I_{j}$
and sets $\rho_{I_{j}}=z_{I_{j}}$.
\item \label{enu:condion-on-input}The tree conditions $Y$ on $g^{I_{j}}\left(x_{I_{j}},Y_{I_{J}}\right)=\rho_{I_{j}}$. 
\item \label{enu:restoring-Y}The tree conditions $Y$ on an event $\cE$
such that $\Pr\left[\cE\right]>1-2^{-\alpha\Delta}$ and $Y_{\free\left(\rho\right)}\mid\cE$
is $\left(\sigma+\frac{\epsCns}{c}\right)$-sparse. Such an event
must exist as $x$ is $\alpha$-safe, and in particular, $\alpha$-recoverable.
For the sake of the analysis, we assume that there is a canonical
choice of such the event~$\cE$.
\end{enumerate}
When the simulation reaches the end of the protocol, the decision
tree halts and outputs the transcript~$\pi$. In order for the algorithm
to be well-defined, it remains to explain two points.
\begin{itemize}
\item We explain why all the conditionings done in the tree are on non-empty
events. Regarding Step \ref{enu:condition on uniformity and revealing},
the probability that $X_{\free\left(\rho\right)}$ is safe is lower
bounded by the main lemma. We know that $X_{\free\left(\rho\right)},Y_{\free\left(\rho\right)}$
are $\left(\sigma+\frac{\epsCns}{c}\right)$-sparse and $\sigma$-sparse
respectively. We now apply the main lemma with $\gamma=\frac{1}{10}$,
which is possible since
\[
\sigma_{X}+2\sigma_{Y}=\frac{3}{10}+\frac{\sigmaCnsOneXTwoY}{\cnsCrand}=\cnsLHSrandom\leq\cnsRHSrandom=\frac{7}{10}-\frac{\cnsMainLemma}{\cnsCrand}=\frac{9}{10}-\frac{\cnsMainLemma}{c}-\gamma-\alpha.
\]
Thus the probability is lower bounded by $1-2^{-\frac{1}{10}\Delta}>0$
and we can conclude that the event is not empty. In Step \ref{enu:condion-on-input},
the tree conditions on $g^{I_{j}}\left(x_{I_{j}},Y_{I_{J}}\right)=\rho_{I_{j}}$,
and it holds that $\Pr\left[g^{I_{j}}\left(x_{I_{j}},Y_{I_{J}}\right)=\rho_{I_{j}}\right]>2^{-\left|I\right|-1}$
as $x$ is almost uniform. 
\item Earlier we asserted that if at the beginning of the iteration it is
Alice's turn to speak then $X_{\free\left(\rho\right)},Y_{\free\left(\rho\right)}$
are $\left(\rho,\sigma+\frac{\epsCns}{c},\sigma\right)$-structured,
and they are $\left(\rho,\sigma,\sigma+\frac{\epsCns}{c}\right)$-structured
when it is Bob's turn to speak. Assume that in the start of the iteration
it is the case that $X_{\free\left(\rho\right)},Y_{\free\left(\rho\right)}$
are $\left(\rho,\sigma+\frac{\epsCns}{c},\sigma\right)$-structured.
After Step \ref{enu:partion-of-X} it is guaranteed that $X$ is $\sigma$-sparse,
and $Y$ is $\left(\sigma+\frac{\epsCns}{c}\right)$-sparse by Step
\ref{enu:restoring-Y}. Therefore $X,Y$ swap rules and the assertion
holds.
\end{itemize}

\subsection{\label{subsec:Correctness-of-rand}Correctness of the decision tree}

In this section we prove the correctness of the decision tree, that
is, that on every input $z$ the output distribution of the tree given
$z$ is $\left(2^{-\frac{\Delta\left(g\right)}{20}}\cdot\left(1+C\right)\right)$-close
to $\Pi_{z}^{\prime}$ (recall that $\Pi_{z}^{\prime}$ denotes the
distribution of transcripts of the protocol~$\Pi$ on uniformly random
inputs conditioned on the event $g^{n}\left(X,Y\right)=z$). For the
rest of the proof we fix $z$ to be some input. For simplicity, we
will assume that the protocol always runs for exactly $C$ rounds.
In case that the protocol finishs earlier, we consider all the remaining
rounds as containing empty messages.

We now set up some additional notation. Let $\Pi^{\dagger}$ be a
random transcript constructed by $T$ when invoked on the input~$z$.
In the case that $T$ halts early we let $\Pi^{\dagger}$ to be some
special symbol~$\bot$. Instead of bounding the distance between~$\Pi^{\dagger}$
and~$\Pi_{z}^{\prime}$ directly, it is easier to bound the distance
of each of them to an intermediate transcript. In order to define
the intermediate transcript, we introduce an intermediate decision
tree $T^{*}$, which is the same as the tree $T$ except that Step
\ref{enu:halting-msg-and-prt} is removed. We now define the intermediate
transcript $\Pi^{*}$ to be the transcript generated by~$T^{*}$
when invoked on the input~$z$. The correctness of~$T$ now follows
immediately from the next two claims, that are proved in \ref{subsec:Proof-Of-Correctness-Without-Halting,subsec:proof-of-correctness-halting}
respectively.
\begin{claim}
\label{clm:pi' close to pi*}$\Pi_{z}^{\prime}$ is $C\cdot2^{-\frac{\Delta}{20}}$-close
to $\Pi^{*}$.
\end{claim}

\begin{claim}
\label{clm:pi is close to pi*}$\Pi^{*}$ is $2\cdot2^{-\Delta}$-close
to $\Pi^{\dagger}$.
\end{claim}

\subsubsection{\label{subsec:Proof-Of-Correctness-Without-Halting}Proof Of \ref{clm:pi' close to pi*}}

In order to prove \ref{clm:pi' close to pi*}, we define a notion
called the \emph{extended protocol}, which is an augmented version
of~$\Pi$ that imitates the action of the tree~$T^{*}$. The extended
protocol works in iterations, such that each iteration corresponds
to a single round of the original protocol, and is analogous to a
single iteration of the tree $T^{*}$.

In the following description of the protocol, we denote by $\cX$
and~$\cY$ the set of all inputs of Alice and Bob that are compatible
with the current (partial) transcript of the extended protocol. Additionally,
let $X$ and $Y$ be uniform variables over $\cX$ and $\cY$ respectively.
Throughout the protocol, Alice and Bob maintain a restriction $\rho$,
which is initialized with $\left\{ *\right\} ^{n}$ and is kept consistent
with $g^{n}\left(x,y\right)$. Alice and Bob also maintain shared
partial transcript $\pi$ of the original protocol.

We now describe a single iteration of the extended protocol. Without
loss of generality we assume that it is Alice's turn to speak in this
round. In the case where it is Bob's turn to speak, the roles of Alice
(respectively $X$) and Bob (respectively $Y$) are swapped. In a
single iteration, the extended protocol performs the following steps:
\begin{enumerate}
\item Alice sends $0$ if $x_{\free\left(\rho\right)}$ is $\alpha$-safe
for $Y_{\free\left(\rho\right)}$, and $1$ otherwise. 
\item Alice sends the message $m$ that Alice would send in protocol $\Pi$
on the input $x$ in this round. The message $m$ is appended to $\pi$
by both Alice and Bob. 
\item Let $\cX^{j}$ be the density-restoring partition of $\cX_{\free(\rho)}$.
Alice sends the index $j$ such that $x_{\free(\rho)}\in\cX^{j}$. 
\item Bob sends $g^{I_{j}}\left(x_{I_{j}},y_{I_{j}}\right)$. Both Alice
and Bob update $\rho_{I_{j}}=g^{I_{j}}\left(x_{I_{j}},y_{I_{j}}\right)$. 
\item Bob sends $0$ if $y\in\cE$, where $\cE$ is the event from Step
\ref{enu:restoring-Y} of the tree, and $1$ otherwise. 
\end{enumerate}
The tree $T^{*}$ can be naturally modified to simulate not only the
protocol but also the extended protocol. We call this modified tree
the \emph{extended tree}. More formally, we change~$T^{*}$ such
that it maintains extended transcript $\pi^{e}$ as follows:
\begin{itemize}
\item in Step \ref{enu:condition-on-not-dangerous} the tree appends $0$
to $\pi^{e}$.
\item in Step \ref{enu:msg-choosing} the tree appends the same message
$m$ to $\pi^{e}$ as it appends to $\pi$.
\item in Step \ref{enu:partion-of-X} the tree appends the index $j$ of
the partition class to $\pi^{e}$.
\item in Step \ref{enu:condion-on-input} the tree appends the queried bits
to $\pi^{e}$.
\item in Step \ref{enu:restoring-Y} the tree appends $0$ to $\pi^{e}$.
\end{itemize}
After those changes, the sets $\cX=\mathrm{supp}(X)$ and $\cY=\mathrm{supp}(Y)$
that are maintained by the tree are equal to the set of all inputs
that are consistent with the partial transcript $\pi^{e}$. 

Let $E^{*}$ be a random extended transcript constructed by $T^{*}$
when invoked on the input~$z$. Let $E^{\prime}$ be a transcript
of the extended protocol on random inputs $X^{\prime},Y^{\prime}$
uniformly chosen from $\left(g^{n}\right)^{-1}\left(z\right)$. We
prove \ref{clm:pi' close to pi*} by bounding the statistical distance
between the transcripts $E^{*}$ and $E^{\prime}$. Then, as the transcripts
of random transcript of $\Pi^{*}$ and $\Pi_{z}^{\prime}$ can be
extracted from $E^{*}$ and~$E^{\prime}$, the statistical distance
$\left|\Pi^{*}-\Pi_{z}^{\prime}\right|$ is bounded by $\left|E^{*}-E^{\prime}\right|$. 

We bound the distance $\left|E^{*}-E^{\prime}\right|$ by constructing
a coupling for $E^{*}$ and $E^{\prime}$. Let $E_{\leq i}^{*}$ be
the prefix of~$E^{*}$ that corresponds for the first $i$ iterations
(the same goes for $E_{\leq i}^{\prime}$). The coupling is constructed
round-by-round, that is, we iteratively construct couplings of $E_{\leq i}^{\prime}$
and $E_{\leq i}^{*}$ from a coupling of $E_{\leq i-1}^{\prime}$
and $E_{\leq i-1}^{*}$. We formally state this in the following claim. 
\begin{claim}
For every $i$ exists a coupling of $E_{\leq i}^{\prime}$ and $E_{\leq i}^{*}$
such that 
\[
\Pr\left[E_{\leq i}^{\prime}\neq E_{\leq i}^{*}\right]\leq2^{-\frac{\Delta}{20}}\cdot i.
\]
\end{claim}

\begin{myproof}
We prove the claim by induction on $i$. In the base case of $i=0$
we set both $E'$ and $E^{*}$ to the empty transcript , and then
$\Pr\left[E_{0}^{\prime}\neq E_{0}^{*}\right]=0$ as desired. Assume
by induction that there exists a coupling of $E_{\le i-1}^{\prime}$
and $E_{\le i-1}^{*}$. We describe an algorithm that samples a coupling
of $E_{\le i}^{\prime}$ and $E_{\le i}^{*}$ using the coupling of
$E_{\le i-1}^{\prime}$ and $E_{\le i-1}^{*}$. The algorithm maintains
two partial transcripts~$e^{*}$ and~$e'$, which constitute the
parts of $E_{\le i}'$ and $E_{\le i}^{*}$ that were constructed
until the current step. We will sometimes say that the algorithm fails,
in which case the output for $E_{\leq i}^{*}$ is sampled by sampling
from the distribution $\left(E_{\leq i}^{*}\mid e^{*}\text{ is a prefix of }E_{\leq i}^{*}\right)$
and the output for $E_{\le i}^{\prime}$ is sampled similarly and
independently.

We assume without loss of generality that it is Alice's turn to speak
in the original protocol. The algorithm first samples transcripts
$e'$ and~$e^{*}$ from the coupling of of $E_{\leq i-1}^{\prime}$
and $E_{\leq i-1}^{*}$ respectively. If $e^{\prime}\neq e^{*}$ the
algorithm fails immediately. Let $X$ be the uniform distribution
over the set $\mathcal{X}$ of Alice's inputs that are consistent
with the partial transcript $e^{*}$. The random variable $Y$ is
defined analogously for $\cY$ and Bob. Let $X'$ and~$Y'$ be random
variables that are jointly distributed like~$X$ and~$Y$ conditioned
on $g^{n}\left(X,Y\right)=z$ (respectively). The algorithm now follows
the steps of the extended tree and extended protocol and constructs
$e'$ and~$e^{*}$ by appending messages corresponding to those steps.
\begin{itemize}
\item The algorithm appends~$0$ to~$e^{*}$ since $T^{*}$ always appends~$0$
to its transcript. The algorithm also appends $0$ to the transcript
$e^{\prime}$ with probability $\Pr\left[X_{\free\left(\rho\right)}\text{ is \ensuremath{\alpha}-safe}\mid g^{n}\left(X,Y\right)=z\right]$,
and otherwise it appends $1$ to $e^{\prime}$ and fails. To bound
the failure probability, we use the uniform marginals lemma (\ref{lem:uniform-marginal-lemma})
with $\gamma=\frac{1}{10}$ and get that 
\[
\Pr\left[X_{\free\left(\rho\right)}\text{ is not \ensuremath{\alpha}-safe}\mid g^{n}\left(X,Y\right)=z\right]\leq\Pr\left[X_{\free\left(\rho\right)}\text{ is not \ensuremath{\alpha}-safe}\right]+2^{-\frac{\Delta}{10}}.
\]
Then, by applying the main lemma (\ref{lem:main-lemma}) with $\gamma=\frac{1}{10}$,
we get that $\Pr\left[X_{\free\left(\rho\right)}\text{ is not \ensuremath{\alpha}-safe}\right]$
is at most $2^{-\frac{\Delta}{10}}.$ Note that we can apply the main
lemma and the uniform marginals lemma as $X$ and $Y$ are $(\sigma+\frac{\epsCns}{c})$-sparse
and $\sigma$-sparse, respectively. Thus, the coupling fails at this
step with probability at most $2\cdot2^{-\frac{\Delta}{10}}$.
\item Let $M\left(x\right)$ be the next message in $\Pi$ on input $x$
and let $J\left(x\right)$ be the partition class of the density restoring
partition that contains $x$. We define the random variables $M^{*}=M\left(X\right)$,$J^{*}=J\left(X\right)$,$M^{\prime}=M\left(X^{\prime}\right)$,
and $J^{\prime}=J\left(X^{\prime}\right)$. As we will explain momentarily,
there exists a coupling of the pairs $\left(M^{*},J^{*}\right)$ and~$\left(M^{\prime},J^{\prime}\right)$
s.t $\text{\ensuremath{\Pr\left[\left(M^{*},J^{*}\right)\ne\left(M^{\prime},J^{\prime}\right)\right]\leq2^{-\frac{\Delta}{10}}}}.$
The algorithm samples from this coupling and appends the resulting
samples to $e^{*}$ and $e^{\prime}$. If the samples are different,
the algorithm fails. \\
To show that the required coupling exists, we use the uniform marginals
lemma (\ref{lem:uniform-marginal-lemma}) with $\gamma=\frac{1}{10}$
to show that $X$ and $X^{\prime}$ are $2^{-\frac{\Delta}{10}}$-close.
Therefore $M\left(X\right),J\left(X\right)$ and $M\left(X^{\prime}\right),J\left(X\right)$
are $2^{-\frac{\Delta}{10}}$-close , which implies the existence
of the desired coupling by \ref{fact:coupling}. In order to apply
the uniform marginals lemma, we need to show that $X$ and~$Y$ are
sufficiently sparse. The random variable~$Y$ is $\sigma$-sparse
by the invariant of the tree. To see why $X$ is sufficiently sparse,
recall that $X$ was $(\sigma+\frac{\epsCns}{c})$-sparse at the begining
at the beginning of the iteration, and the last step only conditioned
it on an event of probably $\ge1-2^{-\frac{\Delta}{10}}$.
\item The algorithm appends $z_{I}$ to both $e^{*}$ and $e^{\prime}$,
where $I$ is the set associated with the class~$J^{\prime}=J^{*}$. 
\item For the last message the algorithm appends $0$ to $e^{*}$(since
in the last step of the extended tree it always appends~$0$). The
algorithm appends $0$ to $e^{\prime}$ with probability $\Pr\left[Y^{\prime}\in\cE\right]$,
otherwise it appends $1$ and fails. \\
We turn to bound the probability to fail in this step. As $\Pr\left[Y\notin\cE\right]$
is at most $2^{-\alpha\Delta}$, it is tempting to apply the uniform
marginals lemma to relate $\Pr\left[Y^{\prime}\in\cE\right]$ and
$\Pr\left[Y\in\cE\right]$. Unfortunately the uniform marginals lemma
cannot be applied as $Y$ is not necessarily sparse at this point.
Yet, we are still able relate $\Pr\left[Y^{\prime}\in\cE\right]$
and $\Pr\left[Y\in\cE\right]$ as follows:
\begin{align*}
\Pr\left[Y^{\prime}\in\cE\right]= & \Pr\left[Y\in\cE\mid g^{n}\left(X,Y\right)=z\right]\\
= & \frac{\Pr\left[g^{n}\left(X,Y\right)=z\mid Y\in\cE\right]}{\Pr\left[g^{n}\left(X,Y\right)=z\right]}\cdot\Pr\left[Y\in\cE\right] & \text{(Bayes' rule)}\\
= & \frac{\Pr\left[g^{\free\left(\rho\right)}\left(X_{\free\left(\rho\right)},Y_{\free\left(\rho\right)}\right)=z_{\free\left(\rho\right)}\mid Y\in\cE\right]}{\Pr\left[g^{\free\left(\rho\right)}\left(X_{\free(\rho)},Y_{\free\left(\rho\right)}\right)=z_{\free\left(\rho\right)}\right]}\cdot\Pr\left[Y\in\cE\right]\\
\ge & \frac{\Pr\left[g^{\free\left(\rho\right)}\left(X_{\free\left(\rho\right)},Y_{\free\left(\rho\right)}\right)=z_{\free\left(\rho\right)}\mid Y\in\cE\right]}{2^{-\left|\free\left(\rho\right)\right|}\cdot\left(1+2^{-\frac{\Delta}{10}}\right)}\cdot\Pr\left[Y\in\cE\right] & \text{(\ensuremath{X} is almost uniform)}
\end{align*}
Regarding the last inequality, recall that the tree removes all unsafe
values from the support of $X$, thus all the values in the support
of $X$ are safe and thus almost uniform. Observe that by the definition
of $\cE$, it hold that $Y\mid Y\in\cE$ is $\left(\sigma+\frac{\epsCns}{c}\right)$-sparse.
Thus, we can bound the numerator from below using \ref{multiplicative-uniformity}.
It follows that
\begin{align*}
 & \Pr\left[Y\in\cE\mid g^{n}\left(X,Y\right)=z\right]\\
\ge & \frac{2^{-\left|\free\left(\rho\right)\right|}\cdot\left(1-2^{-\frac{\Delta}{10}}\right)}{2^{-\left|\free\left(\rho\right)\right|}\cdot\left(1+2^{-\frac{\Delta}{10}}\right)}\cdot\Pr\left[Y\in\cE\right] & \text{(By }\text{\ref{multiplicative-uniformity}}\text{)}\\
\geq & \left(1-2\cdot2^{-\frac{\Delta}{10}}\right)\cdot\Pr\left[Y\in\cE\right]\\
\geq & \Pr\left[Y\in\cE\right]-2\cdot2^{-\frac{\Delta}{10}}\geq1-\left(2\cdot2^{-\frac{\Delta}{10}}+2^{-\alpha\Delta}\right).
\end{align*}
Thus, the probability that the algorithm fails at this step is at
most $\left(2\cdot2^{-\frac{\Delta}{10}}+2^{-\alpha\Delta}\right)$. 
\end{itemize}
At last, the sum of all the failure probabilities is
\begin{align*}
 & \overset{\text{First Message}}{\overbrace{2\cdot2^{-\frac{\Delta}{10}}}}+\overset{\text{Second Message}}{\overbrace{2^{-\frac{\Delta}{10}}}}+\overset{\text{Third Message}}{\overbrace{2^{-\frac{\Delta}{10}}}}+\overset{\text{Fifth message}}{\overbrace{2\cdot2^{-\frac{\Delta}{10}}+2^{-\alpha\Delta}}}\\
\leq & 7\cdot2^{-\frac{\Delta}{10}} & \text{(\ensuremath{\alpha=\frac{1}{10}})}\\
\leq & 2^{-\frac{\Delta}{20}}. & \text{(\ensuremath{\Delta\ge c=\cnsCrand})}
\end{align*}
The probability that $E_{\leq i}^{\prime}\neq E_{\leq i}^{*}$ is
at most the failure probability of the algorithm, which is at most
\begin{align*}
\Pr\left[E_{\leq i}^{\prime}\neq E_{\leq i}^{*}\right] & \leq\Pr\left[E_{\leq i-1}^{\prime}\neq E_{\leq i-1}^{*}\right]+2^{-\frac{\Delta}{20}}\\
 & \leq2^{-\frac{\Delta}{20}}\left(i-1\right)+2^{-\frac{\Delta}{20}}\\
 & =2^{-\frac{\Delta}{20}}\cdot i
\end{align*}
thus completing the proof. 
\end{myproof}
As the number of rounds is bounded by the communication complexity
of the protocol we get that there exists a coupling such that 
\[
\Pr\left[E^{\prime}\neq E^{*}\right]\leq2^{-\frac{\Delta}{20}}\cdot C.
\]
Using \ref{fact:coupling} and this coupling, we get that the statistical
distance between $E^{\prime}$ and $E^{*}$ is upper bounded by~$2^{-\frac{\Delta}{20}}\cdot C$.
\begin{remark}
We now explain why we use the notion of almost-uniform values which
was not present in \cite{CFKMP19}. Recall the work of \cite{CFKMP19}
did not use the notion of ``almost uniform'' and instead used the
notion of ``non-leaking''. The notion of ``almost uniform'' bounds
the probability 
\[
\Pr\left[g^{\free\left(\rho\right)}\left(x_{\free\left(\rho\right)},Y_{\free\left(\rho\right)}\right)=z_{\free\left(\rho\right)}\right]
\]
from both below and above, while the notion of ``non-leaking'' only
bounds the probability from below. In the above proof, in order to
bound $\Pr\left[Y\in\cE\mid g^{n}\left(X,Y\right)=z\right]$ we need
to bound the latter probability from above, and this is the reason
that we use the stronger notion of ``almost uniform'' instead of
the weaker notion of ``non-leaking''. The reason that this issue
did not come up in \cite{CFKMP19} is that the step of restoring the
density of~$Y$(Step \ref{enu:restoring-Y}) does not exist in \cite{CFKMP19},
and therefore their analysis does not require to bound $\Pr\left[Y\in\cE\mid g^{n}\left(X,Y\right)=z\right]$.
\end{remark}

\subsubsection{\label{subsec:proof-of-correctness-halting}Proof of \ref{clm:pi is close to pi*}}

We now prove that the transcripts $\Pi^{\dagger}$ and $\Pi^{*}$
are sufficiently close. By definition $\Pi^{\dagger}$ differs from~$\Pi^{*}$
if and only if the tree $T$ halts on Step \ref{enu:halting-msg-and-prt}.
For the ease of notation, we denote the event that the tree halts
due to $\KMsg$ by $\mathcal{\cH}_{\text{msg}}$, and the event that
the tree halts due to $\KPrt$ by $\mathcal{\cH}_{\text{prt}}$. Then,
by \ref{fact:dist-on-cond} and the union bound it hold that 
\[
\left|\Pi^{\dagger}-\Pi^{*}\right|\le\Pr\left[\mathcal{\cH}_{\text{msg}}\text{ or }\mathcal{\cH}_{\text{prt}}\right]\leq\Pr\left[\mathcal{\cH}_{\text{msg}}\right]+\Pr\left[\mathcal{\cH}_{\text{prt}}\right].
\]
In what follows we upper bound the probabilities of $\mathcal{\cH}_{\text{msg}}$
and~$\mathcal{\cH}_{\text{prt}}$ separately.

It suffices to upper bound the probabilities of~$\cH_{\text{msg}}$
and $\mathcal{\cH}_{\text{prt}}$ conditioned on every fixed choice
of the random coins of the protocol, since $\Pr\left[\mathcal{\cH}_{\text{msg}}\right]$
and $\Pr\left[\mathcal{\cH}_{\text{prt}}\right]$ are a convex combinations
of those probabilities. For the rest of this section, we fix a choice
of the random coins , and assume that all the probabilities below
are conditioned on this choice.

We first set up some notation. Let $\bar{M}=\left(M_{1},\ldots,M_{r}\right)$
be the random messages that are chosen by the tree and let $\bar{J}=\left(J_{1},\dots,J_{r}\right)$
be the random variable of the indices of the partition class chosen
by the tree. We denote by $\bar{m}=\left(m_{1},\dots,m_{r}\right)$
and $\bar{j}=\left(j_{1},\dots,j_{r}\right)$ some specific values
of $\bar{M}$ and $\bar{J}$ respectively. 

\paragraph{Bound on the probability of $\mathcal{\protect\cH}_{\text{msg}}$.}

Before formally proving the bound on $\Pr\left[\mathcal{\cH}_{\text{msg}}\right]$
we first give a high-level description of the argument. For every
transcript let $\KMsg$ be the value of $\KMsg$ at the end of the
simulation of this transcript. Then, it holds that the probability
of getting this transcript is at most~$2^{-\KMsg}$. Recall that
the tree halts if $\KMsg>C+\Delta$, so every specific halting transcript
has a probability of at most $2^{-C-\Delta}$. By taking a union bound
over all of them, we get a probability of $2^{-\Delta}$ as required.
In the formal proof, we also need to take into account in the union
bound the different choices of the partition classes, see below. Note
that the following proof is identical to the corresponding proof in
\cite{CFKMP19} and we provide it here for completeness.

We define $\mathcal{B}$ to be the set of all pairs $\left(\bar{m},\bar{j}\right)$
for which $T$ halts on Step \ref{enu:halting-msg-and-prt} because
of $\KMsg$, then it holds that

\begin{align*}
\Pr\left[\mathcal{\cH}_{\text{msg}}\right] & =\sum_{\left(\bar{m},\bar{j}\right)\in\mathcal{B}}\Pr\left[\bar{M}=\bar{m},\bar{J}=\bar{j}\right]\\
 & =\sum_{\left(\bar{m},\bar{j}\right)\in\mathcal{B}}\prod_{i=1}^{r}\Pr\left[M_{i}=m_{i}\mid\bar{M}_{<i}=\overline{m}_{<i},\bar{J}_{<i}=\bar{j}_{<i}\right]\cdot\Pr\left[J_{i}=j_{i}\mid\bar{M}_{\leq i}=\overline{m}_{\leq i},\bar{J}_{<i}=\bar{j}_{<i}\right].
\end{align*}
Now, recall that conditioned on the event $\mathcal{\cH}_{\text{msg}}$,
we know that at the end of the simulation we have 
\[
\KMsg=\sum\log\frac{1}{\Pr\left[M_{i}=m_{i}\mid\bar{M}_{<i}=\overline{m}_{<i},\bar{J}=\bar{j}_{<i}\right]}>C+\Delta.
\]
In other words, 
\[
\prod_{i=1}^{r}\Pr\left[M_{i}=m_{i}\mid\bar{M}_{<i}=\overline{m}_{<i},\bar{J}_{<i}=\bar{j}_{<i}\right]<2^{-C-\Delta},
\]
and therefore we get

\begin{align*}
\Pr\left[\mathcal{\cH}_{\text{msg}}\right] & =\sum_{\left(\bar{m},\bar{j}\right)\in\mathcal{B}}\prod_{i=1}^{r}\Pr\left[M_{i}=m_{i}\mid\bar{M}_{\leq j}=\overline{m}_{<i},\bar{J}_{<i}=\bar{j}_{<i}\right]\cdot\Pr\left[J_{i}=j_{i}\mid\bar{M}_{\leq i}=\bar{m}_{\leq i},\bar{J}_{<i}=\bar{j}_{<i}\right]\\
 & <\sum_{\left(\bar{m},\bar{j}\right)\in\mathcal{B}}2^{-C-\Delta}\cdot\prod_{i=1}^{r}\Pr\left[J_{i}=j_{i}\mid\bar{M}_{\leq j}=\overline{m}_{\leq i},\bar{J}_{<i}=\bar{j}_{<i}\right]\\
 & \leq2^{-C-\Delta}\sum_{\left(\bar{m},\bar{j}\right)}\prod_{i=1}^{r}\Pr\left[J_{i}=j_{i}\mid\bar{M}_{\leq j}=\overline{m}_{\leq i},\bar{J}_{<i}=\bar{j}_{<i}\right].
\end{align*}
We claim that the sum $\sum_{\left(\bar{m},\bar{j}\right)}\prod_{i=1}^{r}\Pr\left[J_{i}=j_{i}\mid\bar{M}_{\leq i}=\bar{m}_{\leq i},\bar{J}_{<i}=\bar{j}_{<i}\right]$
is equal to $\sum_{\bar{m}}1$, which in our case is upper bounded
by $2^{C}$. To see it, observe that for every round $t$ it holds
that 
\begin{align*}
 & \sum_{\bar{m},\bar{j}_{\leq t}}\prod_{i=1}^{t}\Pr\left[J_{i}=j_{i}\mid\bar{M}_{\leq i}=\bar{m}_{\leq i},\bar{J}_{<i}=\bar{j}_{<i}\right]\\
= & \sum_{\bar{m},\bar{j}_{<t}}\left(\left(\prod_{i=1}^{t-1}\Pr\left[J_{i}=j_{i}\mid\bar{M}_{\leq i}=\bar{m}_{\leq i},\bar{J}_{<i}=\bar{j}_{<i}\right]\right)\cdot\sum_{j_{t}}\Pr\left[J_{t}=j_{t}\mid M_{\leq t}=\bar{m}_{\leq t},\bar{J}_{<i}=\bar{j}_{<t}\right]\right)\\
= & \sum_{\bar{m},\bar{j}_{<t}}\prod_{i=1}^{t-1}\Pr\left[J_{i}=j_{i}\mid\bar{M}_{\leq i}=\bar{m}_{\leq i},\bar{J}_{<i}=\bar{j}_{<i}\right],
\end{align*}
where the last transition hold as 
\[
\sum_{j_{t}}\Pr\left[J_{t}=j_{t}\mid\bar{M}_{\leq t}=\bar{m}_{\leq t},\bar{J}_{<i}=\bar{j}_{<t}\right]=1.
\]
By induction, we get that
\[
\sum_{\left(\bar{m},\overline{j}\right)}\prod_{i=1}^{r}\Pr\left[J_{i}=j_{i}\mid\bar{M}_{\leq i}=\bar{m}_{\leq i},\bar{J}_{<i}=\bar{j}_{<i}\right]=\sum_{\bar{m}}1\leq2^{C}.
\]
We now get that
\[
\Pr\left[\mathcal{\cH}_{\text{msg}}\right]<2^{-C-\Delta}\sum_{\bar{m},J}\prod_{i=1}^{r}\Pr\left[J_{i}=j_{i}\mid\bar{M}_{\leq i}=\bar{m}_{\leq i},\bar{J}_{<i}=\bar{j}_{<i}\right]\leq2^{-\Delta},
\]
as required.

\paragraph{Bound on the probability of $\mathcal{\protect\cH}_{\text{prt}}$.}

This part of the proof is a variant of the analysis of \cite{GPW17}.
Let $p^{(i)}$ be the probability $p_{\geq j}$ in the $i$-th iteration.
Using this notation, we can write $\KPrt=\sum_{i=1}^{C}\log\frac{1}{p_{\ge}^{(i)}}$.
For the next claim, we will need the notion of \emph{stochasitc domination.
}Given two real-valued random variables~$X,Y$, we say that $X$
is \emph{stochastically dominant} over $Y$ if for all $t$ it hold
that
\[
\Pr\left[X\ge t\right]\ge\Pr\left[Y\ge t\right].
\]
We will also use the following fact, whose proof is defered to the
end of this section.
\begin{fact}
\label{fct:Dominance-of-sum}Let $A_{1}\dots A_{n}$ and $B_{1}\dots B_{n}$
be random variables over $\mathbb{R}$ such that $A_{i}$ are $i.i.d$
and independent from $B_{1}\dots B_{n}$. Assume that for all $i\leq n$
and $b_{1},\ldots,b_{i-1}\in\mathbb{R}$ the random the random variable
$B_{i}$ is stochastically dominant $A_{i}$ conditioned on $B_{1}=x_{1},\ldots,B_{i-1}=x_{i-1}$.
Then, $\sum_{i=1}^{n}B_{i}$ is stochastically dominant over $\sum_{i=1}^{n}A_{i}$. 
\end{fact}

\begin{claim}
The Erlang distribution $\text{Erl}\left(C,\ln2\right)$ is stochastically
dominant over $\KPrt$. 
\end{claim}

\begin{myproof}
Let $\left(U^{(1)},\dots,U^{(C)}\right)$ be independent uniform random
variables over $\left[0,1\right]$. We start by proving that $\Pr\left[\forall i:p^{(i)}\ge t_{i}\right]\ge\Pr\left[\forall i:U^{(i)}\ge t_{i}\right]$
for every~$t_{i}\in\left[0,1\right]$. We do so by analyzing the
distribution of 
\[
p^{(i)}\mid M_{<i}=\bar{m}_{<i},J_{<i}=\bar{j}_{<i}.
\]
Let $p_{1}\dots p_{l}$ be the probabilities assigned to the different
partition classes in this round. Then, the probability $\Pr\left[p^{(i)}=p_{\ge j}\mid M_{<i}=\bar{m}_{<i},J_{<i}=\bar{j}_{<i}\right]$
is equal to $p_{\ge j}-p_{\ge\left(j+1\right)}$. Thus, it easy to
see that 
\[
\Pr\left[p^{(i)}\ge t\mid M_{<i}=\bar{m}_{<i},J_{<i}=\bar{j}_{<i}\right]\ge1-t=\Pr\left[U^{(i)}\ge t\right]
\]
for every choice of $\bar{m}_{<i}$ and $\bar{j}_{<i}$ and every
$t\in\left[0,1\right]$. Let $\delta^{(i)}=\log\frac{1}{U^{(i)}}$.
As the function $\log\frac{1}{x}$ is monotonically decreasing, we
can get that 
\[
\Pr\left[\log\frac{1}{p^{(i)}}\ge t\mid M_{<i}=\bar{m}_{<i},J_{<i}=\bar{j}_{<i}\right]\leq\Pr\left[\delta^{(i)}\ge t\right].
\]
for every choice of $\bar{m}_{<i}$ and $\bar{j}_{<i}$. Note that
for all real $a_{1}\dots a_{i-1}$ it holds that
\[
\Pr\left[\log\frac{1}{p^{(i)}}\ge t\mid p^{(1)}=a_{1},\dots,p^{(i-1)}=a_{i-1}\right]
\]
is a convex combination of the probabilities $\Pr\left[\log\frac{1}{p^{(i)}}\ge t\mid M_{<i}=\bar{m}_{<i},J_{<i}=\bar{j}_{<i}\right]$
for different $\bar{m}_{<i}$ and $\bar{j}_{<i}$, and thus
\[
\Pr\left[\log\frac{1}{p^{(i)}}\ge t\mid p^{(1)}=a_{1},\dots,p^{(i-1)}=a_{i-1}\right]\leq\Pr\left[\delta^{(i)}\ge t\right].
\]
By applying Fact \ref{fct:Dominance-of-sum} we get that 
\[
\Pr\left[\sum_{i=1}^{C}\log\frac{1}{p^{(i)}}\ge t\right]\le\Pr\left[\sum_{i=1}^{C}\delta^{(i)}\ge t\right].
\]
The left-hand side of the equation is just $\Pr\left[\KPrt\ge t\right]$.
In the right-hand side, the random variable $\delta^{(i)}$ is distributed
like the exponential distribution $\text{Ex}\left(\ln2\right)$ as
\[
\Pr\left[\delta^{(i)}\leq t\right]=\Pr\left[\log\frac{1}{U^{(i)}}\leq t\right]=\Pr\left[U^{(i)}\ge2^{-t}\right]=1-2^{-t}.
\]
As the Erlang random variable is a sum of exponential random variables,
we get that $\sum_{i=1}^{C}\delta^{(i)}$ is distributed like $\text{Erl}\left(C,\ln2\right)$,
and thus we complete the proof.
\end{myproof}
As a result of the above claim, it holds that in order to bound the
probability that $\text{\ensuremath{\KPrt>\KCnsPart C+\KpDeltaCns\Delta}}$,
it suffices to bound the probability that $\text{Erl}(C,\ln2)>\KCnsPart C+\KpDeltaCns\Delta$.
For convenience, we denote $t\defeq\KCnsPart C+\KpDeltaCns\Delta$
and $\lambda\defeq\ln2$. 
\begin{align*}
\Pr\left[\text{Erl}(C,\lambda)>t\right] & =e^{-\lambda\cdot t}\sum_{i=0}^{C-1}\frac{1}{i!}\cdot\left(\lambda t\right)^{i}.
\end{align*}
By our choice of $t$, it easy to see that $\frac{1}{(i+1)!}\cdot\left(\lambda t\right)^{i+1}$
is larger than $\frac{1}{i!}\cdot\left(\lambda t\right)^{i}$ by a
factor of at least~$2$. Therefore 
\begin{align*}
\sum_{i=0}^{C-1}\frac{1}{i!}\cdot\left(\lambda t\right)^{i} & \leq\sum_{i=0}^{C-1}\left(\frac{1}{2}\right)^{C-i}\frac{1}{C!}\cdot\left(\lambda t\right)^{C}\\
 & <\frac{1}{C!}\cdot\left(\lambda t\right)^{C}\cdot\sum_{i=1}^{\infty}\left(\frac{1}{2}\right)^{-i}\\
 & =\frac{1}{C!}\cdot\left(\lambda t\right)^{C}.
\end{align*}
Then 
\begin{align*}
\Pr\left[\text{Erl}(C,\lambda)>t\right] & \leq2^{-t}\cdot\left(\lambda t\right)^{C}\cdot\frac{1}{C!}\\
 & \leq2^{-t}\cdot\left(\lambda t\right)^{C}\cdot\left(\frac{e}{C}\right)^{C} & \text{(\ensuremath{C!\ge}\ensuremath{\left(\frac{C}{e}\right)^{C}})}.
\end{align*}
Substituting $t=\KCnsPart C+\KpDeltaCns\Delta$ we get
\begin{align*}
\Pr\left[\text{Erl}(C,\lambda)>\KCnsPart C+\KpDeltaCns\Delta\right] & \leq2^{-\KCnsPart C-\KpDeltaCns\Delta}\cdot\left(\KCnsPart\lambda C+\KpDeltaCns\lambda\Delta\right)^{C}\cdot\left(\frac{e}{C}\right)^{C}\\
 & =2^{-\KCnsPart C-\KpDeltaCns\Delta}\cdot\KCnsPart^{C}\cdot\left(\lambda e\right)^{C}\cdot\left(1+\frac{\KpDeltaCns\Delta}{\KCnsPart\cdot C}\right)^{C}\\
 & \leq2^{-\KCnsPart C-\KpDeltaCns\Delta}\cdot\KCnsPart^{C}\cdot\left(\lambda e\right)^{C}\cdot e^{C+\frac{\KpDeltaCns\Delta}{\KCnsPart}} & (\ensuremath{\left(1+x\right)^{y}\leq e^{x\cdot y}})\\
 & =2^{-\KCnsPart C-\KpDeltaCns\Delta}\cdot2^{\left(\log5\right)\cdot C}\cdot2^{\left(\log(\lambda e)\right)\cdot C}\cdot2^{\left(\log e\right)\cdot C+\left(\log e\right)\cdot\frac{\KpDeltaCns}{\KCnsPart}\cdot\Delta}\\
 & =2^{-\left(\KCnsPart-\log(\KCnsPart\lambda e^{2})\right)C-\left(\KpDeltaCns-\frac{\KpDeltaCns\cdot\log e}{\KCnsPart}\right)\cdot\Delta}. & \text{}
\end{align*}
As $\KCnsPart-\log(\KCnsPart\lambda e^{2})\ge0$ and $\KpDeltaCns-\frac{\KpDeltaCns\cdot\log e}{\KCnsPart}\ge1$
we have that 
\[
\Pr\left[\KPrt>\KCnsPart C+\KpDeltaCns\Delta\right]\leq\Pr\left[\text{Erl}_{C,\lambda}>\KCnsPart C+\KpDeltaCns\Delta\right]\leq2^{-\Delta}.
\]

\begin{myproof}[Proof of \ref{fct:Dominance-of-sum}]
We will prove by induction. The base case is trivial as $\sum_{i=1}^{1}B_{i}=B_{1}$.
Assume that $\Pr\left[\sum^{n-1}B_{i}\ge x\right]\ge\Pr\left[\sum^{n-1}A_{i}\ge x\right]$.
Then,
\begin{align*}
\Pr\left[\sum^{n}B_{i}\ge x\right] & =\E_{b_{1},\ldots,b_{n-1}\gets B_{1},\ldots,B_{n-1}}\left[\Pr\left[B_{n}\ge x-\sum^{n-1}b_{i}\mid B_{1}=b_{1},\ldots,B_{n-1}=b_{n-1}\right]\right]\\
 & \ge\E_{b_{1}\dots b_{n-1}\gets B_{1}\dots B_{n-1}}\left[\Pr\left[A_{n}\ge x-\sum^{n-1}b_{i}\right]\right]\\
 & =\Pr\left[A_{n}+\sum^{n-1}B_{i}\ge x\right]\\
 & =\E_{a_{n}\gets A_{n}}\left[\Pr\left[\sum^{n-1}B_{i}\ge x-a_{n}\right]\right]\\
 & \ge\E_{a_{n}\gets A_{n}}\left[\Pr\left[\sum^{n-1}A_{i}\ge x-a_{n}\right]\right]\\
 & =\Pr\left[\sum^{n}A_{i}\ge x\right].\qedhere
\end{align*}
\end{myproof}

\subsection{\label{subsec:Query-complexity-rand-tree}Query complexity of the
decision tree}

The following analysis of the query complexity of the randomized decision
tree is very similar to the analysis of the deterministic decision
tree. As in the analysis of deterministic case we define our potential
function to be 
\[
\Dm\left(X_{\free\left(\rho\right)},Y_{\free\left(\rho\right)}\right)=\Dm\left(X_{\free\left(\rho\right)}\right)+\Dm\left(Y_{\free\left(\rho\right)}\right).
\]
The main difference between the analysis of the deterministic and
randomized cases is as follows. While in the deterministic case we
could bound the increase in the deficiency caused by sending a message
$m$ by $\left|m\right|$, in the randomized setting this does not
hold. Nevertheless, $\KMsg$ bounds the increase in deficiency. Another
difference is that Step \ref{enu:partion-of-X} can increase the deficiency
by extra term of~$\log\frac{1}{p_{\ge j}}$. Due to Step \ref{enu:halting-msg-and-prt}
of the tree, it holds that $\KMsg+\KPrt\leq\KCns C+\KDeltaCns\Delta$
(otherwise the tree halts). 

With the latter issue resolved, the rest of the analysis is similar
to the deterministic case. We prove that any query decreases the deficiency
by at least $\Omega\left(\Delta\right)$ per queried bit. Thus, we
get an upper bound of $O\left(\frac{C+\Delta}{\Delta}\right)=O\left(\frac{C}{\Delta}+1\right)$
on the number of queries. We now analyze a single iteration of the
tree. Without loss of generality, we assume that it is Alice's turn
to speak. We go through the iteration step by step. We start by noting
that Steps \ref{enu:adding-to-K_prt} and \ref{enu:halting-msg-and-prt}
do not change the deficiency, and therefore are ignored in the following
list. 
\begin{itemize}
\item In Step \ref{enu:condition on uniformity and revealing}, the tree
conditions on an event with probability $1-2^{-\frac{1}{10}\Delta}\geq\frac{1}{2}$
by \ref{lem:main-lemma} with $\gamma=\frac{1}{10}$. Therefore, the
deficiency increases at by most $1$ by \ref{fact:deficiency-cond}.
\item In Step \ref{enu:sample-msg}, the tree samples a message and then
conditions on sending it. Therefore, the conditioning increases the
deficiency by at most $\log\frac{1}{p_{m}}$ by \ref{fact:deficiency-cond}. 
\item In Step \ref{enu:partion-of-X}, the tree chooses a partition class
$\mathcal{X}^{j}$ of $X$. By \ref{lem:density-restoring-partition}
we know that the deficiency increases by at most $\left(b-\sigma\cdot\Delta\right)\cdot\left|I_{j}\right|+\log\frac{1}{p_{\ge j}}$.
For the rest of the analysis we will denote the probability $p_{\ge j}$
and set $I_{j}$ at the $i$-th step by $p^{(i)}$ and $I^{(i)}$
respectively.
\item In Step \ref{enu:query-I_j}, the tree decreases the size of $\free\left(\rho\right)$
by $\left|I^{(i)}\right|$. It holds that $\Dm\left(Y_{\free\left(\rho\right)}\right)$
does not increase by \ref{fact:deficiency-monotone}, while $\Dm\left(X_{\free\left(\rho\right)}\right)$
decreases by $b\cdot\left|I^{(i)}\right|$ as $X_{I^{(i)}}$ is fixed
to $x_{I^{(i)}}$. Altogether we get that the deficiency decreases
by at least $b\cdot\left|I^{(i)}\right|$.
\item In Step \ref{enu:condion-on-input}, the tree conditions $Y$ on $g^{I^{(i)}}\left(x_{I^{(i)}},Y_{I^{(i)}}\right)=\rho_{I}$.
As $x$ is safe, we know that the probability of this event is at
least $2^{-\left|I^{(i)}\right|-1}\ge2^{-2\left|I^{(i)}\right|}$,
and therefore the deficiency increases by at most $2\left|I^{(i)}\right|$
by \ref{fact:deficiency-cond}.
\item In Step \ref{enu:restoring-Y}, the tree conditions on the event $\mathcal{E}$
such that $\Pr\left[\mathcal{E}\right]\geq\frac{1}{2}$. Thus, the
deficiency increases at most by $1$ by \ref{fact:deficiency-cond}.
\end{itemize}
Steps \ref{enu:condition on uniformity and revealing} and \ref{enu:restoring-Y}
can be executed at most $C$ times and therefore contribute at most
$2C$ to the deficiency. Step \ref{enu:sample-msg} contributes $\sum_{m\in\pi}\log\frac{1}{p_{m}}=\KMsg$,which
is at most $C+\Delta$ due to Step \ref{enu:adding-to-K_prt}. Similarly,
the contribution of the term $\log\frac{1}{p_{\ge j}}$ from Step
\ref{enu:partion-of-X} is $\sum\log\frac{1}{p^{(i)}}=\KPrt$ , which
is at most $\KCnsPart C+\KpDeltaCns\Delta$ due to Step \ref{enu:halting-msg-and-prt}.
We note that in fact when the tree halts at Steps \ref{enu:adding-to-K_prt}
or \ref{enu:halting-msg-and-prt} the counters $\KMsg$ or $\KPrt$
may be bigger then $C+\Delta$ and $\KCnsPart C+\KpDeltaCns\Delta$
respectively, but such an iteration does not make queries and therefore
it does not affect the analysis.

We turn to bound the change in deficiency that is caused by Step~\ref{enu:partion-of-X}
(without the $\log\frac{1}{p^{(i)}}$), and Steps \ref{enu:query-I_j}
and~\ref{enu:condion-on-input}. Assume that the tree queries $I$
in some iteration. In Steps \ref{enu:partion-of-X}, and \ref{enu:condion-on-input}
the deficiency increases by at most 
\[
\left(b-\sigma\cdot\Delta\right)\cdot\left|I^{(i)}\right|+2\left|I^{(i)}\right|,
\]
and Step \ref{enu:query-I_j} decreases the deficiency by at least
$b\cdot\left|I^{(i)}\right|$. Altogether, we get that the deficiency
decreases by at least 
\begin{align*}
b\cdot\left|I^{(i)}\right|-\left(b-\sigma\cdot\Delta\right)\cdot\left|I^{(i)}\right|-2\left|I^{(i)}\right| & =\Delta\cdot\left|I^{(i)}\right|\left(\sigma-\frac{2}{c}\right)\\
 & =\frac{1}{10}\cdot\Delta\cdot\left|I^{(i)}\right|\in\Omega\left(\Delta\cdot\left|I^{(i)}\right|\right). & \text{(\ensuremath{\sigma=\frac{1}{10}+\frac{\sigmaCRandCns}{c}})}
\end{align*}
Summing over all iterations of the simulation we get that
\begin{align*}
\Dm\left(X_{\free\left(\rho\right)}\right)+\Dm\left(X_{\free\left(\rho\right)}\right) & \leq\MaxDefTimes C+\KDeltaCns\Delta-\sum_{i}\frac{\Delta}{10}\cdot\left|I^{(i)}\right|\\
 & \leq\MaxDefTimes C+\KDeltaCns\Delta-\frac{\Delta}{10}\cdot\sum_{i}\left|I^{(i)}\right|.
\end{align*}
By \ref{fact:deficiency-nonnegative}, $\Dm\left(X_{\free\left(\rho\right)}\right)+\Dm\left(X_{\free\left(\rho\right)}\right)\geq0$
therefore we get that 
\begin{align*}
\frac{\Delta}{10}\cdot\sum_{i}\left|I^{(i)}\right| & \leq\MaxDefTimes C+\KDeltaCns\Delta\\
\sum_{i}\left|I^{(i)}\right| & \leq\MaxDefTimesTen\left(\frac{C}{\Delta}+1\right)\in O\left(\frac{C}{\Delta}+1\right).
\end{align*}
The term $\sum_{i}\left|I^{(i)}\right|$ is the equal to query complexity
of the tree, and thus we complete the proof. 

\section{\label{sec:Counterexample}The tightness of the main lemma of \cite{CFKMP19}}

Our work expands upon the results of \cite{CFKMP19} and generalizes
them into a wider regime of parameters. Despite the similarities between
our work and their work, a crucial difference between the two works
is that we use a weaker notion of safe values. To motivate the need
for a weaker notion of safety, we show that the conclusion of the
main lemma of \cite{CFKMP19} cannot hold when~$\Delta\ll b$. We
first recall the notion of dangerous values of \cite{CFKMP19} and
restate their main lemma in our terms. 
\begin{restated}{\ref{def:cfkmp-dangerous}}
Let $Y$ be a random variable taking values from $\Lambda^{n}$.
We say that a value $x\in\Lambda^{n}$ is \emph{leaking} if there
exists a set $I\subseteq\left[n\right]$ and an assignment $z_{I}\in\B^{I}$
such that 
\[
\Pr\left[g^{I}\left(x_{I},Y_{I}\right)=z_{I}\right]<2^{-\left|I\right|-1}.
\]
Let $\sigma_{Y},\varepsilon>0$, and suppose that $Y$ is $\sigma_{Y}$-sparse.
We say that a value $x\in\Lambda^{n}$ is $\varepsilon$\emph{-sparsifying}
if there exists a set $I\subseteq\left[n\right]$ and an assignment
$z_{I}\in\B^{I}$ such that the random variable $\mbox{\ensuremath{Y_{\left[n\right]-I}\mid g^{I}\left(x_{I},Y_{I}\right)=z_{I}}}$
is not $(\sigma_{Y}+\varepsilon)$-sparse. We say that a value $x\in\Lambda^{n}$
is $\varepsilon$\emph{-dangerous} if it is either leaking or $\varepsilon$-sparsifying.
\end{restated}

\begin{remark}
We note that even when $\varepsilon\ge1$, \ref{def:cfkmp-dangerous}
remains non-trivial, since the sparsity is measured with respect to~$\Delta$
rather than~$b$. On the other hand, from the perspective of the
simulation, a value $x$ that is $1$-dangerous is ``useless''.
The reason is that the discrepancy of~$g$ (and in particular, \ref{discrepancy-XOR-extractor,discrepancy-XOR-sampling})
cannot give any effective bound on the biases of $\varepsilon$-sparse
distributions for $\varepsilon\ge1$. 
\end{remark}

\begin{restated}{\ref{main-cfkmp}}[Main lemma of \cite{CFKMP19}]
There exists universal constants $h,c$ such that the following
holds: Let $b$ be some number such that $b\geq c\cdot\log n$ and
let $\gamma,\varepsilon,\sigma_{X},\sigma_{Y}>0$ be such that $\varepsilon\geq\frac{4}{\Delta}$,
and $\sigma_{X}+\sigma_{Y}\leq1-\frac{h\cdot b\cdot\log n}{\Delta^{2}\cdot\varepsilon}-\gamma$.
Let $X,Y$ be $\left(\rho,\sigma_{X},\sigma_{Y}\right)$-structured
random variables. Then, the probability that $X_{\free\left(\rho\right)}$
takes a value that is $\varepsilon$-dangerous for $Y_{\free\left(\rho\right)}$
is at most $2^{-\gamma\Delta}$. 
\end{restated}

It can be seen that the lemma is only applicable when $\Delta=\Omega(\sqrt{b\cdot\log n})$.
In this section we show this is almost optimal. The main result of
this section is that in the case where $\Delta\in O\left(\sqrt{b}\right)$,
the conclusion of the main lemma of \cite{CFKMP19} completely fails:
that is, it may happen that $X_{\free\left(\rho\right)}$ takes only
values that are $2$-dangerous for $Y_{\free\left(\rho\right)}$.
As noted in Remark 6.1, such values are useless for the simulation.
\begin{proposition}
\label{prop:exmple}For every $b\geq1000$ and $n\ge\left\lfloor 3\sqrt{b}\right\rfloor +1$
the following holds: There exist an inner function $\text{\ensuremath{g:\B^{b}\times\B^{b}\to\B}}$
with $\Delta\left(g\right)\in\Theta\left(\sqrt{b}\right)$ and two
random variables $X,Y$ over $\left(\B^{b}\right)^{n}$ that are $\frac{2}{\Delta}$-sparse
such that every $x\in\text{supp}\left(X\right)$ is $2$-dangerous
for $Y$. 
\end{proposition}

The rest of this section is dedicated to proving \ref{prop:exmple}.
We start by introducing some notation. Let $b,n$ be as in the proposition
and let $\Lambda=\B^{b}$. We denote by $I$ the set~$\left[\left\lfloor 3\sqrt{b}\right\rfloor \right]$
and define $d\defeq\left\lfloor \frac{1}{3}\sqrt{b}\right\rfloor $.
For each value $y\in\Lambda^{n}$, we view the last block~$y_{n}\in\Lambda$
as consisting of $\left|I\right|$ substrings of length $d$ and of
the remaining $b-d\left|I\right|$ bits. We call each such string
of $d$ bits a \emph{cell,} and denote the $i$-th cell by $y_{n,i}$
for every $i\in I$. Note that the definition of cells applies only
to the last block~$y_{n}$. For each $i\in I$, we refer to the first
$d$~bits of the $i$-th block $y_{i}$ as the \emph{prefix} of~$y_{i}$,
and denote it by $y_{i}^{\mathrm{pre}}$. Let $U$ be the uniform
distribution over $\Lambda^{n}$, and for each $i\in I$, let $\mathcal{A}_{i}$
be the event that $\left\langle U_{i}^{\mathrm{pre}},U_{n,i}\right\rangle =1$.

We now choose the function~$g$ to be $g(v,w)=\left\langle v^{\mathrm{pre}},w^{\mathrm{pre}}\right\rangle $
for $v,w\in\Lambda$. It hold that $\text{\ensuremath{\frac{d}{2}\leq\Delta\leq d+1}}$,
where the lower bound is by Lindsey's lemma and the upper bound is
trivial. We choose the random variable $X$ to be the uniform distribution
over $\Lambda^{n}$. We define $Y$ to be equal to the random variable
$U$ conditioned on the events~$\mathcal{A}_{i}$ for every $i\in I$.
In order to prove \ref{prop:exmple} we need to show that $X,Y$ are
$\frac{2}{\Delta}$-sparse and that every $x\in\text{supp}\left(X\right)$
is $2$-dangerous for $Y$. The variable $X$ is trivially $0$-sparse.
We turn to prove that $Y$ is $\frac{2}{\Delta}$-sparse. 
\begin{claim}
$Y$ is $\frac{2}{\Delta}$-sparse
\end{claim}

\begin{myproof}
For any set $S$, we prove that the deficiency $\Dm\left(Y_{S}\right)\leq2\left|S\right|=\frac{2}{\Delta}\cdot\Delta\cdot\left|S\right|$,
and this would imply that $Y$ is $\frac{2}{\Delta}$-sparse. Let
$U$ be the uniform distribution over $\Lambda^{n}$. In order to
bound $\Dm\left(Y_{S}\right)$, we start by expressing $\Pr\left[Y_{S}=y_{S}\right]\cdot2^{-b\left|S\right|}$
as follows
\begin{align*}
\frac{\Pr\left[Y_{S}=y_{S}\right]}{2^{b\left|S\right|}} & =\frac{\Pr\left[U_{S}=y_{S}\mid\forall i\in I:\mathcal{A}_{i}\right]}{\Pr\left[U_{S}=y_{S}\right]} & \text{(\ensuremath{Y=U\mid\forall i\in I:U\in\mathcal{A}_{i}})}\\
 & =\frac{\Pr\left[\forall i\in I:\mathcal{A}_{i}\mid U_{S}=y_{S}\right]}{\Pr\left[\forall i\in I:\mathcal{A}_{i}\right]}. & \text{(Bayes' formula)}
\end{align*}
We note that $\Pr\left[\forall i\in I:\mathcal{A}_{i}\right]=\left(\frac{1-2^{-d}}{2}\right)^{\left|I\right|}$:
this holds as each event $\mathcal{A}_{i}$ has probability~$\frac{1-2^{-d}}{2}$
and they are all independent. We turn to bound the probability in
the above numerator. The events~$\mathcal{A}_{i}$ are independent
even conditioned on $U_{S}=y_{S}$, and therefore we can bound their
probabilities separately. In general, the probability of $\mathcal{A}_{i}$
conditioned on $U_{S}=y_{S}$ can be as large as~$1$ (for example,
if both $i$ and~$n$ are in~$S$, and $\left\langle y_{i}^{\mathrm{pre}},y_{n,i}\right\rangle =1$).
Nevertheless, it is easy to see that for each $i\in I\backslash S$,
the probability of $\mathcal{A}_{i}$ conditioned on $U_{S}=y_{S}$
is at most~$\frac{1}{2}$ even if $n\in S$. Therefore we can get
a bound as follows
\[
\frac{\Pr\left[\forall i\in I:\mathcal{A}_{i}\mid U_{S}=y_{S}\right]}{\Pr\left[\forall i\in I:\mathcal{A}_{i}\right]}\leq\frac{2^{-\left|I\backslash S\right|}}{\left(\frac{1-2^{-d}}{2}\right)^{\left|I\right|}}=2^{\left|S\right|}\cdot\left(1-2^{-d}\right)^{-\left|I\right|}.
\]
The second term can be bounded as follows 
\[
\left(1-2^{-d}\right)^{-\left|I\right|}\leq e^{\left|I\right|2^{-d}}\leq e^{3\sqrt{b}\cdot2^{-\left\lfloor \frac{1}{3}\sqrt{b}\right\rfloor }},
\]
which is at most $2$ for $b\ge1000$. Therefore
\begin{align*}
\Dm\left(Y_{S}\right) & =\log\max_{y_{S}}\frac{\Pr\left[Y_{S}=y_{S}\right]}{2^{b\left|S\right|}}\leq1+\left|S\right|\leq2\left|S\right|.\qedhere
\end{align*}
\end{myproof}
It remain to prove that every $x\in\text{supp}\left(X\right)$ is
$2$-dangerous for $Y$. In order to do so, we use the following claim
on the distribution of $Y$. 
\begin{claim}
\label{clm:Prob_Y_ni_lower_bound}Let $y_{n}$ be such that $y_{n,i}\ne\overline{0}$
for every~$i$. Then $\Pr\left[Y_{n}=y_{n}\right]\ge2^{-b}$.
\end{claim}

\begin{myproof}
First, observe that the event $Y_{n,i}=y_{n,i}$ is independent of
the event $\mathcal{A}_{j}$ for every $j\neq i$ and thus it holds
that
\begin{align*}
\Pr\left[Y_{n,i}=y_{n,i}\right] & =\Pr\left[U_{n,i}=y_{n,i}\mid\mathcal{A}_{i}\right]\\
 & =\frac{\Pr\left[\mathcal{A}_{i}\mid U_{n,i}=y_{n,i}\right]\Pr\left[U_{n,i}=y_{n,i}\right]}{\Pr\left[\mathcal{A}_{i}\right]} & \text{(Bayes' rule).}
\end{align*}
As $y_{n,i}\neq\bar{0}$ it easy to see that $\Pr\left[\mathcal{A}_{i}\mid U_{n,i}=y_{n,i}\right]\ge\Pr\left[\mathcal{A}_{i}\right]$,
and therefore
\[
\frac{\Pr\left[\mathcal{A}_{i}\mid U_{n,i}=y_{n,i}\right]\Pr\left[U_{n,i}=y_{n,i}\right]}{\Pr\left[\mathcal{A}_{i}\right]}\ge\Pr\left[U_{n,i}=y_{n,i}\right]=2^{-d}.
\]
Finally by combining this inequality for all cells and and the fact
that the remaining $b-d\cdot\left|I\right|$ bits are uniformly distributed
and indipendet, we get the desired result.
\end{myproof}
We finally show that all values in $\text{supp}\left(X\right)$ are
$2$-dangerous. It holds that every $x$ such that $x_{i}^{\mathrm{pre}}=\overline{0}$
for some~$i\in I$ is leaking as $\Pr\left[g^{n}\left(x,Y\right)=1^{n}\right]=0$,
and thus such an $x$ is dangerous. It remains to handle the case
where $x_{i}^{\mathrm{pre}}\neq0^{d}$ for all $i\in I$. Let $x\in\Lambda^{n}$
be such a value. We show that every such~$x$ is $2$-sparsifying.
Specifically, we show that there exists a value $y_{n}$ such that~$\Pr\left[Y_{n}=y_{n}\mid g^{I}\left(x_{I},Y_{I}\right)=1^{\left|I\right|}\right]$
is too high. We choose $y_{n}$ be equal to the concatenation of all
prefixes $x_{i}^{\mathrm{pre}}$ appended by $b-d\left|I\right|$
zeros. For this choice of $y_{n}$ it hold that

\begin{align*}
Y_{n}=y_{n} & \Rightarrow\forall i\in I:\left\langle y_{n,i},Y_{i}^{\mathrm{pre}}\right\rangle =1 & \text{(definition of \ensuremath{Y})}\\
 & \Rightarrow\forall i\in I:\left\langle x_{i}^{\mathrm{pre}},Y_{i}^{\mathrm{pre}}\right\rangle =1 & \text{(definition of \ensuremath{y_{n,1}})}\\
 & \Rightarrow g^{I}\left(x_{I},Y_{I}\right)=1^{I}.
\end{align*}
Thus we have that
\begin{align*}
 & \Pr\left[Y_{n}=y_{n}\mid g^{I}\left(x_{I},Y_{I}\right)=1^{\left|I\right|}\right]\\
= & \frac{\Pr\left[Y_{n}=y_{n}\right]\cdot\Pr\left[g^{I}\left(x_{I},Y_{I}\right)=1^{\left|I\right|}\mid Y_{n}=y_{n}\right]}{\Pr\left[g^{I}\left(x_{I},Y_{I}\right)=1^{\left|I\right|}\right]} & \text{(Bayes' formula)}\\
= & \frac{\Pr\left[Y_{n}=y_{n}\right]}{\Pr\left[g^{I}\left(x_{I},Y_{I}\right)=1^{\left|I\right|}\right]} & \text{(\ensuremath{Y_{n}=y_{n}\Rightarrow g^{I}\left(x,Y\right)=1^{I}}).}
\end{align*}
We note that the events $g\left(x_{i},Y_{i}\right)=1$ are independent
for each $i\in I$, and that each of them occurs with probability
at most $\leq\frac{1}{2\cdot\left(1-2^{-d}\right)}$, thus
\begin{align*}
\Pr\left[g^{I}\left(x_{I},Y_{I}\right)=1^{\left|I\right|}\right] & \leq\left(\frac{1}{2\cdot\left(1-2^{-d}\right)}\right)^{\left|I\right|}.
\end{align*}
Therefore we get 
\begin{align*}
\Pr\left[Y_{n}=y_{n}\mid g^{I}\left(x_{I},Y_{I}\right)=1^{\left|I\right|}\right] & \geq\Pr\left[Y_{n}=y_{n}\right]\cdot\left(2-2^{1-d}\right)^{\left|I\right|}.
\end{align*}
It holds that $\Pr\left[Y_{n}=y_{n}\right]\ge2^{-b}$ by \ref{clm:Prob_Y_ni_lower_bound}
as $y_{n,i}=x_{i}^{\mathrm{pre}}\neq\bar{0}$ for all $i$. Thus,
we get that 
\begin{align*}
\Pr\left[Y_{n}=y_{n}\mid g^{I}\left(x_{I},Y_{I}\right)=1^{\left|I\right|}\right] & \ge2^{-b}\cdot\left(2-2^{1-d}\right)^{\left|I\right|}\\
 & =2^{\left|I\right|-b}\cdot\left(1-2^{-d}\right)^{\left|I\right|}\\
 & \ge2^{\left|I\right|-b}\cdot\left(1-2^{-\left\lfloor \frac{1}{3}\sqrt{b}\right\rfloor }\right)^{3\sqrt{b}}\\
 & \ge2^{\left|I\right|-b-1} & \text{(\ensuremath{b\ge1000}).}
\end{align*}
Hence, the distribution $Y_{n}\mid g^{I}\left(x_{I},Y_{I}\right)=1^{\left|I\right|}$
is not $\frac{\left|I\right|-1}{\Delta}$-sparse and thus $x$ is
$\frac{\left|I\right|-3}{\Delta}$-sparsifying. As a result we can
see that all values of $x$ are $\left(\frac{\left|I\right|-2}{\Delta}\right)$-dangerous.
We have that
\begin{align*}
\left(\frac{\left|I\right|-3}{\Delta}\right) & \geq\frac{\left\lfloor 3\sqrt{b}\right\rfloor -3}{\left\lfloor \frac{1}{3}\sqrt{b}\right\rfloor +1} & \text{(\ensuremath{\Delta\leq d+1},\ensuremath{d=\left\lfloor \frac{1}{3}\sqrt{b}\right\rfloor },\ensuremath{\left|I\right|=\left\lfloor 3\sqrt{b}\right\rfloor })}\\
 & \geq\frac{3\sqrt{b}-4}{\frac{1}{3}\sqrt{b}+1} & \text{(\ensuremath{x\geq\left\lfloor x\right\rfloor \ge x-1})}\\
 & =9\left(1-\frac{13}{3\sqrt{b}+9}\right)\\
 & \geq2 & \text{(\ensuremath{b\geq9})},
\end{align*}
as required.
\begin{remark}
We note that the choice of the constant $2$ in \ref{prop:exmple}
is arbitrary. For any constant $\varepsilon$ one can construct an
example such that all values are $\varepsilon$-dangerous, where $\Delta\in\Theta\left(\sqrt{\frac{b}{\varepsilon}}\right)$
and $n\ge b=\Omega(\varepsilon)$.
\end{remark}

\section{\label{sec:Discrepancy-on-product}Discrepancy with respect to product
distributions.}

Discrepancy is commonly defined with respect to an underlying distribution
as follows.
\begin{definition}
Let $\mu$ be a distribution over $\Lambda\times\Lambda$ and let
$g:\Lambda\times\Lambda\to\B$ be a function. Let $\left(V,W\right)\in\Lambda\times\Lambda$
be a pair of random variables that are distributed according to $\mu$.
Given a combinatorial rectangle $R\subseteq\Lambda\times\Lambda$,
the \emph{discrepancy of~$g$ with respect to~$R$} (and $\mu$),
denoted $\disc_{\mu,R}(g)$, is defined as follows:
\[
\disc_{\mu,R}(g)=\left|\Pr_{\mu}\left[g(V,W)=0\text{ and }(V,W)\in R\right]-\Pr_{\mu}\left[g(V,W)=1\text{ and }(V,W)\in R\right]\right|.
\]
The \emph{discrepancy of~$g$} with respect to $\mu$, denoted $\disc_{\mu}(g)$,
is defined as the maximum of $\disc_{\mu,R}(g)$ over all combinatorial
rectangles~$R\subseteq\Lambda\times\Lambda$. We define 
\[
\Delta_{\mu}(g)\defeq\log\frac{1}{\disc_{\mu}\left(g\right)}
\]
\end{definition}

Note that the definition of discrepancy that we used throughout this
paper is the special case of the above definition for the uniform
distribution. A natural question is to ask whether our main result
(\ref{main-theorem}) can be generalized to other distributions. We
show that the theorem holds with respect to every \emph{product} distribution~$\mu$.
\begin{theorem}
\label{generalized-main-theorem}There exists a universal constant~$c$
such that the following holds: Let $\cS$ be a search problem that
takes inputs from~$\B^{n}$, and let $g:\Lambda\times\Lambda\to\B$
be an arbitrary function such that $\Delta_{\mu}(g)\ge c\cdot\log n$
for some product distribution $\mu$. Then 
\[
\dcc(\cS\circ g^{n})\in\Omega\left(\ddt(\cS)\cdot\Delta_{\mu}(g)\right),
\]
and for every $\beta>0$ it holds that
\[
\rcc_{\beta}(\cS\circ g^{n})\in\Omega\left(\left(\rdt_{\beta^{\prime}}(\cS)-O(1)\right)\cdot\Delta_{\mu}(g)\right),
\]
where $\beta^{\prime}=\beta+2^{-\Delta_{\mu}(g)/50}$.
\end{theorem}

We start by defining a simple kind of reduction between two communication
problems,
\begin{definition}
We say that a function $g^{\prime}:\D^{\prime}\times\D^{\prime}\to\cO$
is reducible to~$g:\D\times\D\to\cO$ if there exist functions~$r_{A},r_{B}:\D^{\prime}\to\D$
such that for every $x^{\prime},y^{\prime}\in\D^{\prime}$ it holds
that $g^{\prime}\left(x^{\prime},y^{\prime}\right)=g\left(r_{A}\left(x\right),r_{B}\left(y\right)\right)$. 
\end{definition}

It easy to see that if $g'$ is reducible to $g$ then $\dcc\left(g^{\prime}\right)\leq\dcc\left(g\right)$
and $\rcc_{\beta}\left(g^{\prime}\right)\leq\rcc_{\beta}\left(g\right)$. 
\begin{claim}
Given a search problem $S:\B^{n}\to\cO$ and functions $g^{\prime}:\D^{\prime}\times\D^{\prime}\to\B$
that is reducible to $g:\D\times\D\to\B$ it hold that $S\circ\left(g^{\prime}\right)^{n}$
is reducible to $S\circ g^{n}$. 
\end{claim}

\begin{myproof}
Let $r_{A}^{\prime},r_{B}^{\prime}$ the functions that reduce $g^{\prime}$
to $g$, and let $r_{A}=\left(r_{A}^{\prime}\right)^{n}$and let $r_{B}=\left(r_{B}^{\prime}\right)^{n}$.
Then, it easy to see that 
\[
\left(S\circ g^{n}\right)\circ\left(r_{A}^{\prime}\times r_{B}^{\prime}\right)^{n}=S\circ\left(g\circ\left(r_{A}^{\prime}\times r_{B}^{\prime}\right)\right)^{n}=S\circ\left(g^{\prime}\right)^{n}\qedhere
\]
\end{myproof}
To prove \ref{generalized-main-theorem}, we use the following lemma
that relates discrepancy of a function $g$ with respect to the some
product distribution $\mu$ to the discrepancy with respect to the
uniform distribution for some function $g^{\prime}$ that is reducible
to $g$. 
\begin{lemma}
\label{clm:matrix-disc-prod}Let $g:\Lambda\times\Lambda\to\B$ be
a function and let $\mu=\mu_{X}\times\mu_{Y}$ be a product distribution.
Then, for every constant $\varepsilon>0$ there exists a function~$g^{\prime}:\D^{\prime}\times\D^{\prime}\to\B$
reducible to~$g$ such that $\disc_{U}\left(g^{\prime}\right)\leq\disc_{\mu}\left(g\right)+\varepsilon$,
where $U$ is the uniform distribution over $\Lambda^{\prime}\times\Lambda^{\prime}$. 
\end{lemma}

We prove \ref{clm:matrix-disc-prod} later in this section. We turn
to prove \ref{generalized-main-theorem} by applying \ref{main-theorem}
with a inner function $g^{\prime}$ that is constructed by \ref{clm:matrix-disc-prod}.
\begin{myproof}[Proof of \ref{generalized-main-theorem} from \ref{clm:matrix-disc-prod}]
Let $c'$ be the universal constant $c$ from \ref{main-theorem}.
We choose the universal constant $c$ to be equal to $\max\left(c^{\prime}+1,2\right)$.
Let $n$, $\mu$, $\cS$, and~$g$ be as in the theorem. Let $\varepsilon=\disc_{\mu}\left(g\right)$
and let $g^{\prime}$ be the function obtained from \ref{clm:matrix-disc-prod}
for $g$ and~$\mu$. As $g^{\prime}$ is reducible to $g$, it hold
that $S\circ g^{n}$ is reducible to $S\circ\left(g^{\prime}\right){}^{n}$.
Therefore, it hold that $\dcc(\cS\circ\left(g^{\prime}\right){}^{n})\leq\dcc(\cS\circ g^{n})$
and $\rcc_{\beta}(\cS\circ\left(g^{\prime}\right){}^{n})\leq\rcc_{\beta}(\cS\circ g^{n})$.

We note that the discrepancy of $g^{\prime}$ is bounded from above
as follows 
\begin{align*}
\disc_{U}\left(g^{\prime}\right) & \leq\disc_{\mu}\left(g\right)+\disc_{\mu}\left(g\right)\leq2n^{-c}\leq n^{-c^{\prime}}
\end{align*}
In other words, it holds that $\Delta_{U}\left(g^{\prime}\right)\ge\Delta_{\mu}\left(g\right)-1\ge c'\cdot\log n$.
Hence, we can apply \ref{main-theorem} on $S\circ g^{\prime}$ and
get
\[
\dcc\left(\cS\circ g^{n}\right)\geq\dcc\left(\cS\circ\left(g^{\prime}\right){}^{n}\right)\in\Omega\left(\ddt(\cS)\cdot\Delta_{U}\left(g^{\prime}\right)\right)=\Omega\left(\ddt(\cS)\cdot\Delta_{\mu}(g)\right),
\]
and for every $\beta>0$ it holds that
\[
\rcc_{\beta}\left(\cS\circ g^{n}\right)\geq\rcc_{\beta}\left(\cS\circ\left(g^{\prime}\right){}^{n}\right)\in\Omega\left(\left(\rdt_{\beta^{\prime}}(\cS)-O(1)\right)\cdot\Delta_{U}\left(g^{\prime}\right)\right)=\Omega\left(\left(\rdt_{\beta^{\prime}}(\cS)-O(1)\right)\cdot\Delta_{\mu}(g)\right),
\]
where $\beta^{\prime}=\beta+2^{-\Delta(g)/50}$. 
\end{myproof}
It only remain to prove \ref{clm:matrix-disc-prod}. In order to prove
the lemma, we first introduce a notion of a \emph{canonical rectangle
}with respect to function $g$ and then we then show that the discrepancy
is maximized with respect to such rectangles. 
\begin{definition}
Let $g:\D\times\Lambda\to\B,g^{\prime}:\Lambda^{\prime}\times\Lambda^{\prime}$
be functions such that $g^{\prime}$ is reducible to $g$ and let
$r_{A},r_{B}:\D'\to\D$ be the reductions. A rectangle $R^{\prime}\subseteq\D^{\prime}\times\D^{\prime}$
is \emph{canonical} if and only if there exists a rectangle $R=A\times B\subseteq\D\times\D$
such that $R^{\prime}=r_{A}^{-1}(A)\times r_{B}^{-1}(B)$. 
\end{definition}

\begin{claim}
\label{clm:disc-max-by-canonical}Let $g:\D\times\D\to\B$ and $g':\D'\times\D'\to\B$
be functions such that $g'$ is reducible to~$g$. The discrepancy
of $g^{\prime}$ with respect to any product distribution $\mu$ is
maximized by a canonical rectangle. 
\end{claim}

\begin{myproof}
We show that for every rectangle $R^{\prime}=M\times N$ of $g^{\prime}$,
there exists a canonical rectangle of $g^{\prime}$ with at least
the same discrepancy and this will imply the claim. The discrepancy
of a rectangle~$R^{\prime}$ can be written as follow 
\[
\disc_{\mu,R^{\prime}}(g^{\prime})=\left|\sum_{\left(x^{\prime},y^{\prime}\right)\in R}\left(-1\right)^{g^{\prime}\left(x^{\prime},y^{\prime}\right)}\Pr_{\left(V,W\right)\leftarrow\mu}\left[\left(V,W\right)=\left(x^{\prime},y^{\prime}\right)\right]\right|.
\]
Assume without loss of generality that the above sum inside the absolute
value is positive. It hold that each row $x^{\prime}\in M$ either
has a positive contribution to the sum or not. If the contribution
is positive then we add all the rows $v^{\prime}$ such that $r_{A}\left(v^{\prime}\right)=r_{A}\left(x^{\prime}\right)$.
For each such row $v^{\prime}$ it hold that $g^{\prime}\left(x^{\prime},y^{\prime}\right)=g^{\prime}\left(v^{\prime},y^{\prime}\right)$
for every $y^{\prime}\in N$ and thus the contributions to the sum
from $x^{\prime}$ and $v^{\prime}$ are equal. As a result we get
that adding $v^{\prime}$ increases the discrepancy. Similarly, if
the contribution of a row $x^{\prime}$ is not positive we remove
this row from $M$, and the new rectangle has at least the same discrepancy
as the one with the row $x^{\prime}$. The same can be done for the
columns. Therefore we get a canonical rectangle that has at least
the same discrepancy as $R^{\prime}$. 
\end{myproof}
We also use the following folklore fact 
\begin{fact}[Folklore]
\label{clm:rational-distrbution-approx}For every distribution $\nu$
over finite set $\Lambda$ and $\varepsilon>0$ there exists a distribution~$\nu^{\prime}$
over $\Lambda$ such that all the probabilities that $\mu'$ assigns
are rational and $\left|\nu^{\prime}-\nu\right|\leq\varepsilon$.
\end{fact}

Finally, we prove \ref{clm:matrix-disc-prod}. Let $\mu=\mu_{X}\times\mu_{Y}$
be a product distribution. The high-level idea of the proof is that
we choose $\D'$ and $r_{A},r_{B}:\D^{\prime}\to\D$ such that for
every $x\in\D$ it holds that $\mu_{X}(x)\approx\frac{\left|r_{A}^{-1}(x)\right|}{\left|\D'\right|}$
(the same is done for $r_{B}$ and $\mu_{Y}$). As the discrepancy
is maximized by canonical rectangles, we get that for each input $x$
the set $r_{A}^{-1}\left(x\right)$ is either contained in the rows
of the rectangle or disjoint from them. Therefore when calculating
the discrepancy we get that the contribution of~$x$ is multiplied
by about $\Pr\left[\mu_{X}=x\right]$ as in the definition of discrepancy
with respect to $\mu$. 
\begin{myproof}[Proof of \ref{clm:matrix-disc-prod}]
Let $g$, $\mu=\mu_{X}\times\mu_{Y}$, and $\varepsilon$ be as in
the lemma. Let $\varepsilon^{\prime}=\frac{\varepsilon}{4}$. Let
$\mu_{X}^{\prime}$ (respectively, $\mu_{Y}^{\prime}$) be some distribution
that is $\varepsilon^{\prime}$-close to $\mu_{X}$ (respectively,~$\mu_{Y}$)
and is rational, whose existence guaranteed by \ref{clm:rational-distrbution-approx}.
As $\mu_{X}^{\prime}$ and $\mu_{Y}^{\prime}$ are rational then exists
an integer $l>0$ such that~$l\cdot\mu_{X}^{\prime}\left(x\right)$
and~$l\cdot\mu_{Y}^{\prime}\left(y\right)$ are integers for all
$x,y\in\D$. We choose $\Lambda^{\prime}=\left[l\right]$, and choose
the function $r_{A}$ such that for every $x\in\Lambda$, the function
$r_{A}$ maps $l\cdot\mu_{X}^{\prime}\left(x\right)$ values to $x$.
The function $r_{B}$ is chosen in the same way with respect to $\mu_{Y}^{\prime}$. 

We prove that $\disc_{U}(g')\le\disc(g)+\varepsilon$. Let $R^{\prime}=M\times N$
be rectangle that maximizes the discrepancy of $g^{\prime}$, and
recall that we can assume $R^{\prime}$ is a canonical rectangle by
\ref{clm:disc-max-by-canonical}. Let $R=r_{A}\left(M\right)\times r_{B}\left(N\right)$
be the rectangle of~$g$ that corresponds to~$R'$. Then it holds
that
\begin{align*}
\disc_{U}\left(g^{\prime}\right) & =\disc_{U,R^{\prime}}\left(g^{\prime}\right)\\
 & =\left|\sum_{\left(x^{\prime},y^{\prime}\right)\in R^{\prime}}\left(-1\right)^{g^{\prime}\left(x^{\prime},y^{\prime}\right)}\Pr\left[U=\left(x^{\prime},y^{\prime}\right)\right]\right|\\
 & =\left|\sum_{\left(x,y\right)\in R}\sum_{\left(x^{\prime},y^{\prime}\right)\in r_{A}^{-1}\left(x\right)\times r_{B}^{-1}\left(y\right)}\left(-1\right)^{g^{\prime}\left(x^{\prime},y^{\prime}\right)}\frac{1}{l^{2}}\right|\\
 & =\left|\sum_{\left(x,y\right)\in R}\left(-1\right)^{g\left(x,y\right)}\sum_{\left(x^{\prime},y^{\prime}\right)\in r_{A}^{-1}\left(x\right)\times r_{B}^{-1}\left(y\right)}\frac{1}{l^{2}}\right|\\
 & =\left|\sum_{\left(x,y\right)\in R}\left(-1\right)^{g\left(x,y\right)}\frac{\left|r_{A}^{-1}\left(x\right)\times r_{B}^{-1}\left(y\right)\right|}{l^{2}}\right|\\
 & =\left|\sum_{\left(x,y\right)\in R}\left(-1\right)^{g\left(x,y\right)}\cdot\mu_{X}'(x)\cdot\mu_{Y}'(y)\right|.\\
 & =\left|\sum_{\left(x,y\right)\in R}\left(-1\right)^{g\left(x,y\right)}\Pr_{\left(V,W\right)\leftarrow\mu^{\prime}}\left[\left(V,W\right)=\left(x,y\right)\right]\right|\\
 & =\disc_{\mu^{\prime},R}\left(g\right).
\end{align*}
Next, we show that as $\mu$ and $\mu^{\prime}$ are $\varepsilon^{\prime}$-close
then $\disc_{\mu^{\prime},R}\left(g\right)$ is only bigger then $\disc_{\mu,R^{\prime}}\left(g\right)$
by $\varepsilon$ as follows\quad{}
\begin{align*}
\disc_{U}\left(g^{\prime}\right)= & \disc_{\mu^{\prime},R}\left(g\right)\\
= & \left|\Pr_{\left(V,W\right)\leftarrow\mu^{\prime}}\left[g(V,W)=0\text{ and }(V,W)\in R\right]\right.\\
 & \left.-\Pr_{\left(V,W\right)\leftarrow\mu^{\prime}}\left[g(V,W)=1\text{ and }(V,W)\in R\right]\right|\\
\leq & 4\varepsilon^{\prime}+\left|\Pr_{\left(V,W\right)\leftarrow\mu}\left[g(V,W)=0\text{ and }(V,W)\in R\right]\right. & \text{(\ensuremath{\mu^{\prime}} is \ensuremath{2\varepsilon^{\prime}}-close to \ensuremath{\mu})}\\
 & \left.-\Pr_{\left(V,W\right)\leftarrow\mu}\left[g(V,W)=1\text{ and }(V,W)\in R\right]\right|\\
= & \varepsilon+\disc_{\mu,R}\left(g\right)\\
\leq & \varepsilon+\disc_{\mu}\left(g\right) & \qedhere
\end{align*}
\end{myproof}
\bibliographystyle{alpha}
\bibliography{refs}

\end{document}